\documentclass[twocolumn,floatfix,,amsmath,amssymb,longbibliography]{revtex4}
\usepackage{amsmath,amssymb,color,latexsym,natbib,graphicx,bm}
\usepackage{lpic} \usepackage{subfigure} \usepackage[normalem]{ulem}

\bibpunct{[}{]}{,}{n}{,}{,} 
 
\def\ADD#1{{#1}} 

\def\DEL#1{{\textcolor{green}{}}} 

\newcommand{\vomega}{\mbox{\boldmath $\omega$}}
\newcommand\BV{{Brunt-V\"ais\"al\"a frequency}}
\newcommand{\be}{\begin{equation}}  \newcommand{\ee}{\end{equation}}

\begin{document}
\title{Generation of turbulence through frontogenesis in sheared stratified flows}

\author{N.E. Sujovolsky$^1$, P.D. Mininni$^1$, and A. Pouquet$^{2,3}$}   
\affiliation{$^1$ Departamento de F\'{\i}sica, Facultad de Ciencias Exactas y Naturales, Universidad de Buenos Aires and IFIBA, CONICET, Buenos Aires 1428, Argentina. \\
                  $^2$NCAR, P.O. Box 3000, Boulder, Colorado 80307-3000, USA. \\
                  $^3$Laboratory for Atmospheric and Space Physics, CU, Boulder, Colorado 80309-256, USA.}

\begin{abstract}
The large-scale structures in the ocean and the atmosphere are in geostrophic balance, and a conduit must be found to channel the energy to the small scales where it can be dissipated. In turbulence this takes the form of an energy cascade, whereas one possible mechanism in a balanced flow at large scales is through the formation of fronts, a common occurrence in geophysical dynamics. We show in this paper that an iconic configuration in laboratory and numerical experiments for the study of turbulence, that of the \ADD{Taylor-Green} or von K\'arm\'an swirling flow, can be suitably adapted to the case of fluids with large aspect ratios, leading to the creation of an imposed large-scale vertical shear. To this effect we use direct numerical simulations of the Boussinesq equations without net rotation and with no small-scale modeling, \ADD{and with this idealized Taylor-Green set-up. Various grid spacings are used, up to $2048^2\times 256$ spatial points. The grids are always isotropic, with box aspect ratios of either $1:4$ or $1:8$}. We find that \ADD{when shear and stratification are comparable}, the imposed shear layer resulting from the forcing leads to the formation of multiple fronts and filaments which destabilize and further evolve into a turbulent flow in the bulk, with a sizable amount of dissipation and mixing, and with a cycle of front creation, instability, and development of turbulence. \ADD{The results depend on the vertical length scales for shear and for stratification, with stronger large-scale gradients being generated when the two length scales are comparable.}
\end{abstract} \maketitle

\section{Introduction} \label{S:intro}

The large scales of the atmosphere and the oceans are in geostrophic balance, an equilibrium between pressure gradients, Coriolis and gravity forces due to the presence of both rotation and stratification, leading to a flow that can sustain inertia-gravity waves at smaller scales \cite{charney_71}. The large-scale geostrophy can be broken through resonances between the waves, although finite-size effects can weaken such resonances due to the resulting discretization in wave numbers \cite{kartashova_08}. However, the energy which is injected in the system, e.g., through solar radiation, tides, or large-scale temperature gradients, has to find a way to small-scale dissipation, and how this process takes place remains a puzzle in atmospheric and oceanic dynamics. \ADD{As an example, it was shown in \cite{kafiabad_16} that a large-scale balanced flow in the presence of both rotation and stratification remains balanced at low Rossby number $\textrm{Ro}$, with the ageostrophic part of the flow being weak but increasing as $\textrm{Ro}$ increases.  Typically, only a fraction of the energy cascades to small scales, with co-existing constant energy and enstrophy fluxes to small scales \cite{deusebio_13}. Such fluxes can also be diagnosed with third-order structure functions, as measured in numerical modeling and in observations of atmospheric flows \cite{deusebio_14b, king_15a}. Nevertheless, numerous studies show that the ageostrophic component of the flow develops for buoyancy Reynolds numbers sufficiently high, with a threshold around 10 (see for example \cite{pouquet_17j} and references therein).}

\ADD{Evidence of dissipative processes} in the ocean, and of the development of three-dimensional mixing, has been 
\ADD{demonstrated} by remarkable visualizations of oceanic surface motions, which have been achieved for quite a while using plankton as markers. For example, it has been observed that the phytoplankton density increases significantly at the passage of a hurricane within chlorophyll-{\it a} filaments and eddies  (see, e.g., \cite{davis_04}), and they represent a signature of upwelling and lateral mixing that both occur in a turbulent flow \cite{rossi_09, gruber_11}. These mesoscale and sub-mesoscale geostrophic as well as ageostrophic motions, affecting nutrients and organic matter filamentary structures in coastal systems, are crucial for halieutic management and the fishing industry \cite{shulman_15}. Ocean tracers can also be studied using large-scale models of such flows, and it was shown that their dynamics depend on flow regimes and on their interaction with the structures and turbulent eddies \cite{smith_16}.

Filaments in the upper layers of the oceans are commonplace, either at the borders of large eddies \cite{papenberg_10}, or as an ensemble of parallel structures (see \cite{mcwilliams_16} for a recent review). Their typical scale of a few kilometers makes them part of what are called sub-mesoscale eddies, between the large scales in geostrophic balance and the turbulent small scales that have presumably recovered homogeneity and isotropy. They presently attract a lot of attention, since ``eddy-resolving'' numerical models are now able to reach such small scales\ADD{, and since they are believed to play an important role in the departure of the flow from a balanced state.}

As departure from geostrophic balance develops and turbulent eddies strengthen, nonlinear coupling through advection leads naturally, through a turbulent cascade, to the formation of intense localized structures. In the absence of rotation or stratification, these take the form of shocks in the one-dimensional Burgers equation, or fronts for the passive scalar \cite{celani_04}, whereas in three-dimensional homogeneous isotropic turbulence it is vorticity filaments which prevail in fluids \cite{ishihara_09}, and current sheets and flux ropes in magnetohydrodynamics \cite{oieroset_11, mininni_06b}. Fronts are also a well-known feature of atmospheric flows \cite{hoskins_82}; as they are embedded in a turbulent environment, they affect for example the growth of rain droplets and the overall cloud system \cite{grabowski_13}. \ADD{Stratified flows can also develop strong intermittency \cite{rorai_14}: quiescent and strongly mixed regions are often juxtaposed with sharp edges between them, and with internal gravity waves propagating outward from the turbulent regions thus redistributing the energy \cite{maffioli_14}. Such waves can become unstable through resonant harmonic generation even in the absence of perturbation \cite{liang_17}. It was also recently shown experimentally that the confinement of secondary waves is instrumental in the destabilization process, the waves then displaying high horizontal vorticity \cite{brouzet_17}.}

It was clear starting with the pioneering numerical work of Herring and M\'etais \cite{metais_89, metais_94} that stratified turbulence was different from homogeneous isotropic turbulence, for example through the formation of intense vertical layers as already advocated in \cite{lilly_83}, or because the decay of energy in the absence of external forcing is slow. \ADD{Spectral scaling in these flows is thus quite complex:} one has to consider the spectra of the kinetic, potential, and total energy, and their dependence in the vertical and horizontal directions. Only in some cases (e.g., of sufficiently strong stratification) can one expect spectral laws to clearly emerge, as shown using two-point closures of turbulence \cite{cambon_04}, \ADD{and more recently with direct numerical simulations (DNS) (see, {\it e.g.}, \cite{lindborg_05, lindborg_06, waite_09, rorai_15, maffioli_16}).} It is known that  angular spectra show diverse power laws \cite{rorai_15}, and it is only when one particular angle in spectral space becomes dominant that a scaling can be identified straightforwardly. In the decay case, it was also concluded in \cite{metais_94} that there may well be a lack of universality in these flows. These studies laid the ground for further numerical investigations in which, today, scales are better resolved.

The development of turbulence in the atmosphere and the oceans has recently been considered in high resolution numerical simulations and in theoretical studies of rotating and stratified flows, or of purely stratified flows using the Boussinesq approximation. As mentioned above, in both cases secondary instabilities play a crucial role in the organization of the flow and in the distribution of kinetic and potential energy among scales, as demonstrated for example by a careful analysis of the evolution of a simple large-scale vortex \cite{augier_12}: The Kelvin-Helmholtz instability first sets in and destabilizes scales down to the buoyancy wavenumber $k_B=U_0/N$, with $U_0$ a typical r.m.s.~velocity and $N$ the \BV. The associated buoyancy scale $L_B = 2\pi/k_B$ also corresponds to the typical thickness of stratified layers in the vertical direction \cite{billant_01}. Further instabilities of these structures lead to an excitation down to a scale at which isotropy is recovered. Depending on the strength of rotation, this can be the Zeman scale $\ell_{ze}=2\pi/k_{ze}$ (the scale at which Coriolis and advection forces balance \cite{3072}), with $k_{ze} = (f^3/\epsilon_V)^{1/2}$, $f$ the Coriolis frequency, and $\epsilon_V$ the kinetic energy dissipation rate, or the Ozmidov scale $\ell_{oz}=2\pi/k_{oz}$ (the scale at which buoyancy and advection forces balance), with $k_{oz}=(N^3/\epsilon_V)^{1/2}$. Whether these scales are properly resolved or not may well alter the efficiency of mixing and the properties of stratified turbulence, as advocated in \cite{brethouwer_07, ivey_08, waite_11}.

\ADD{This mixing can be quantified for example by the ratio of the buoyancy flux due to internal waves to the kinetic energy dissipation, or as sometimes advocated, by the ratio of potential energy to total energy dissipation \cite{venayagamoorthy_16}. Mixing can be shown to depend in a simple manner on the small parameter of the problem, namely the Froude number \cite{maffioli_16b, marino_15p, pouquet_17p, pouquet_17j}. The amount of mixing in stratified flows has also been predicted using tools emanating from statistical mechanics  \cite{venaille_17}; it is shown to depend on the background density gradient and on the global Richardson number, with an irreversible increase of potential energy which becomes equal to the kinetic energy. Mixing is also routinely measured in the ocean (see, e.g., \cite{ivey_08, klymak_08, vanharen_16, clement_17}), and can vary by several orders of magnitude depending on whether the flow is quiescent or dominated by instabilities, together with the fact that oceanic jets can form strong barriers to mixing \cite{david_17}. The resulting parametrization schemes in global weather and climate codes are thus quite complex, one issue being what is the minimum number of dimensionless parameters that is necessary to trigger the inclusion of such eddy transport coefficients \cite{mater_14, mater_14b, karimpour_15}, including in the presence of fronts \cite{bachman_17}.} At even smaller scales, quasi-isotropy, Kolmogorov scaling, and strong mixing are thought to recover down to the Kolmogorov length scale $\eta$ at which dissipation prevails. However, it was shown in \cite{smyth_00b} that the anisotropy of dissipation may persist in this range of scales at least in the presence of an imposed anisotropic shear.

In this paper we examine stably stratified turbulence \ADD{in the absence of rotation,} and search for front-like and filamentary structures \ADD{in an idealized setting, using computations in a box with an aspect ratio $A_r$, for the highest Reynolds number run, of $1:8$ (where $A_r$ is defined as the ratio of the vertical to horizontal length scales). Note that in this case geostrophic balance of the large scales does not apply, but it can be replaced by a so-called cyclostrophic balance, which is due to the balance between pressure gradients and the centrifugal force engendered by strong vorticity; it thus can take place even in the absence of Coriolis force like at the Equator or in sub-mesoscales.}

\ADD{We also consider the effect of varying the background stratification with fixed aspect ratio, and of varying the aspect ratio by doing simulations with $A_r$ of $1:4$. In all cases,} the fluid is forced with a configuration which can be viewed as quasi two-dimensional with a super-imposed strong vertical shear; \ADD{it is called the Taylor-Green (TG) flow and was first proposed in \cite{taylor_37}.} This is a paradigmatic flow which has been \ADD{at the core of several numerical studies of non-dissipative neutral and conducting fluids in search of potential singular structures \cite{brachet_83, brachet_05, brachet_13}, of studies of the dynamo problem for the generation of magnetic fields by fluid motions in magnetohydrodynamics \cite{nore_97, ponty_05}, and which was also considered for the study of stratified turbulence \cite{riley_03}. The TG flow mimics a laboratory configuration of two counter-rotating co-axial disks stirring a fluid which is called the von K\'arm\'an swirling flow. This configuration has been used experimentally to detect vortex filaments in hydrodynamic turbulence \cite{douady_91} and for turbulent dynamo experiments in liquid sodium and gallium \cite{odier_98,lee_10}.}

\begin{table*}
\begin{ruledtabular}
\begin{tabular}{lccccccccccccccc}
Run & $n_\perp$ & $n_z$ & $N$ & $\textrm{Fr}$ & $\textrm{Re}$ & ${\cal R}_B$ & $\left<Ri_g\right>$ & $L_z/L_B$ & $\ell_{oz}/\ell_{min}$ & $L_0/\lambda_0$ & $r_E$ &  $r_\epsilon$  & $\beta$ & $L_\perp/L_\parallel$ & $\lambda_\perp/\lambda_\parallel$ \\
\hline \hline
A8 & 2048 & 256 & 8 & 0.03 & 40000 & 36 & 730 & 0.98 & 51 & 4.0 & 0.19 & 0.27 & 1.05 & 7.0 & 0.85 \\
\hline
B8 & 1024 & 128 & 8 & 0.03 & 15000 & 14 & 390 & 0.98 & 29 & 2.6 & 0.15 & 0.29 & 1.20 & 7.1 & 1.09 \\
${\bf B8^\ast}$ & 1024 & 128 & 8 & 0.03  & 15000 & 14 & 680 & 1.04  & 27 & 2.4 & 0.16  & 0.25 & 1.40 & 7.4 & 1.19 \\
C8 & 1024 & 128 & 16 & 0.01 & 17000 & 3.8 & 370 & 1.61 & 11 & 2.3 & 0.16 & 0.33 & 1.13 & 8.7 & 1.52 \\
\hline
D4 & 768 & 192 & 4 & 0.05 & 10000 & 25 & 120 & 1.02 & 44 & 3.1 & 0.13 & 0.27 & 1.10 & 3.8 & 1.06 \\
E4 & 768 & 192 & 8 & 0.03 & 13000 & 13 & 200 & 1.82 & 17 & 2.8 & 0.10 & 0.23 & 0.90 & 4.6 & 1.39 \\
F4 & 768 & 192 & 16 & 0.01 & 14000 & 3.1 & 280 & 3.33 & 6 & 2.1 & 0.05 & 0.15 & 0.35 & 6.0 & 3.18 \\
\end{tabular}
\end{ruledtabular}
\caption{\ADD{List of the runs with $n_\perp$ and $n_z$ grid points in the horizontal and the vertical direction. Runs A8 to C8 have an aspect ratio $A_r=n_z/n_\perp=1/8$, whereas runs D4 to F4 have $A_r=1/4$. Run ${\bf B8^\ast}$ has the same parameters as run B8, but is forced both in the velocity and in the temperature, taken to be balanced up to third order. $N$ is the \BV, $\textrm{Fr}$, $\textrm{Re}$ and ${\cal R}_B$ are respectively the Froude, Reynolds and buoyancy Reynolds numbers, and $\left<Ri_g\right>$ is the mean value of the local gradient Richardson number. In the ratio $L_z/L_B$, $L_z=2\pi/A_r$ is the box height and also the characteristic length of the shear, and $L_B$ is the buoyancy length scale; $\ell_{oz}/\ell_{min}$ is the ratio of the Ozmidov length scale to the minimally resolved scale of the run, namely $\ell_{min}=2\pi/n_\perp$. These ratios have to be equal or larger than unity for the buoyancy and the Ozmidov scales to be resolved by the run. $L_0/\lambda_0$ is the ratio of the isotropic integral scale to the isotropic Taylor scale. The quantity $r_E=E_P/(E_V+E_P)$ measures the ratio of potential to total energy, $r_\epsilon = \epsilon_P/(\epsilon_V+\epsilon_P)$ is the ratio of potential energy dissipation rate to total energy dissipation rate, and $\beta = \epsilon_V/\epsilon_V^{Kol}$ measures the effective to dimensional (Kolmogorovian) kinetic energy dissipation rate. Finally, $L_\perp/L_\parallel$ and $\lambda_\perp/\lambda_\parallel$ give respectively the ratios of integral and Taylor perpendicular to parallel scales, quantifying respectively the anisotropy at large and at small scales. All quantities were averaged in time over the turbulent steady state of each simulation.}}
\label{t:runs}
\end{table*}

From the flow starting at rest, we shall consider the formation of fronts and the subsequent development of turbulence, as well as the large-scale and small-scale cyclic dynamics that ensues. In the next section we present the equations and a brief description of the flow \ADD{and of all the runs analyzed in this paper.} We then move on to describe in detail the temporal evolution of the flow \ADD{at the highest Reynolds number} in Sec.~\ref{S:TIME}, starting from the flow at rest and until turbulence develops. In Sec.~\ref{S:SS} we show that fronts and filament-like structures easily form in this configuration, using visualizations of the flow and studying its spatial structures. Section \ref{S:FOURIER} presents energy spectra and discusses flow anisotropies, while \ADD{Section \ref{S:PARAM} studies the effect of varying the Reynolds number (and the spatial resolution), the \BV, and the aspect ratio of the domain.} Finally, we give our conclusions in Sec.~\ref{S:CONCLU}.

\section{Formulation of the problem}

\subsection{The Boussinesq equations}

The incompressible Boussinesq equations are written for the velocity field ${\bf u}$, with Cartesian components ${\bf u} = (u,v,w)$, and for the temperature fluctuations $\theta$ as:
\begin{eqnarray} 
\partial_t {\mathbf u} +{\mathbf u} \cdot \nabla {\mathbf u}  &=&  -\nabla P - N \theta\  e_z + \nu \Delta {\mathbf u} + {\bf F} \ , \\       \label{eq:mom}
\partial _t \theta\  +{\mathbf u} \cdot \nabla \theta\  &=& N w + \kappa \Delta \theta\  \ , \\  \label{eq:temp}
 \nabla \cdot {\bf u} &=&0 \ .
\end{eqnarray}
Here $P$ is the pressure, ${\bf F}$ is a mechanical forcing function to be defined later, $\nu$ the viscosity, and $\kappa$ the diffusivity, with a unit Prandtl number $\textrm{Pr}=\nu/\kappa=1$. The square of the Brunt-V\"ais\"al\"a frequency is  $N^2=-(g/\theta) (d\bar \theta /dz)$, with $d\bar \theta /dz$ the imposed background stratification, assumed to be linear, and  $g$ the acceleration due to gravity. No modeling of the small scales is included. 

\begin{figure}
\includegraphics[width=7cm]{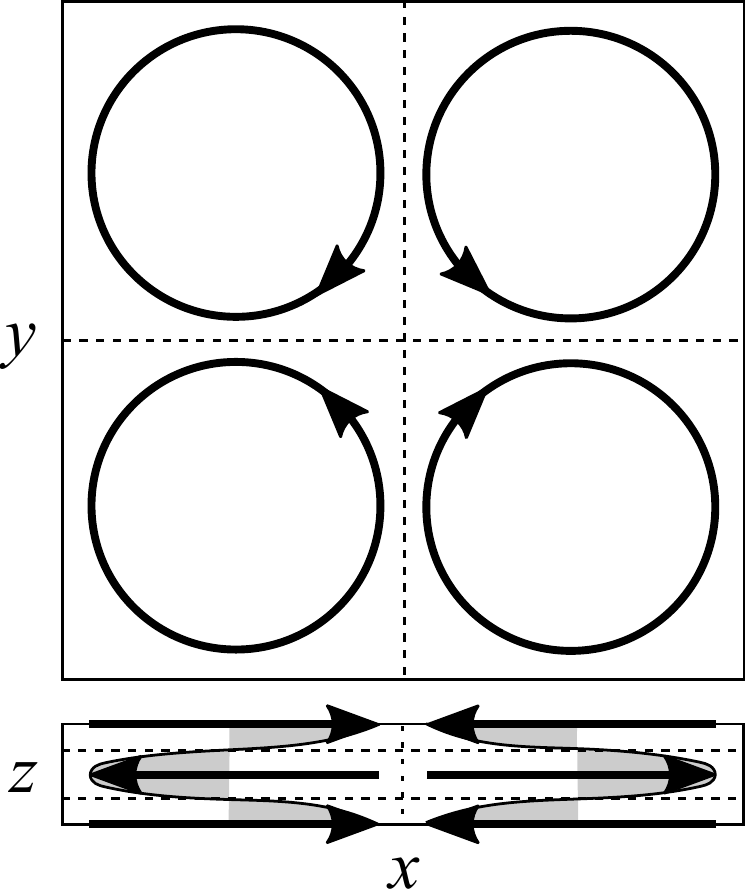}
\caption{Sketch of the Taylor-Green forcing with a strong vertical shear corresponding to $k_z=8$ in a box with aspect ratio 
\ADD{$1:8$,} see Eq.~(\ref{TGf8}). {\it Top}: a horizontal cut of the domain, with sides $L_x=L_y=2\pi$. {\it Bottom}: a vertical cut, with $L_z=L_x/8=\pi/4$. Dashed lines indicate each von K\'arm\'an cell. In the bottom figure, the horizontal lines also correspond to the regions of strongest vertical shear and zero forcing amplitude.}
\label{f:TG}
\end{figure}

In the absence of dissipation and forcing ($\nu=\kappa=0, \ {\bf F}=0$) these equations conserve the total (kinetic plus potential) energy, 
\be E_T = \frac{1}{2} \left<|{\bf u}|^2 + \theta ^2\right> = E_V+E_P \ , \ee
and the pointwise potential vorticity 
\be P_V= -N\omega_z +  \vomega \cdot \nabla \theta\ ,  \label{pveq} \ee
with $\vomega=\nabla \times {\bf u}$ the vorticity, with Cartesian components $\vomega = (\omega_x, \omega_y, \omega_z)$.

\begin{figure}   
\includegraphics[width=8.7cm]{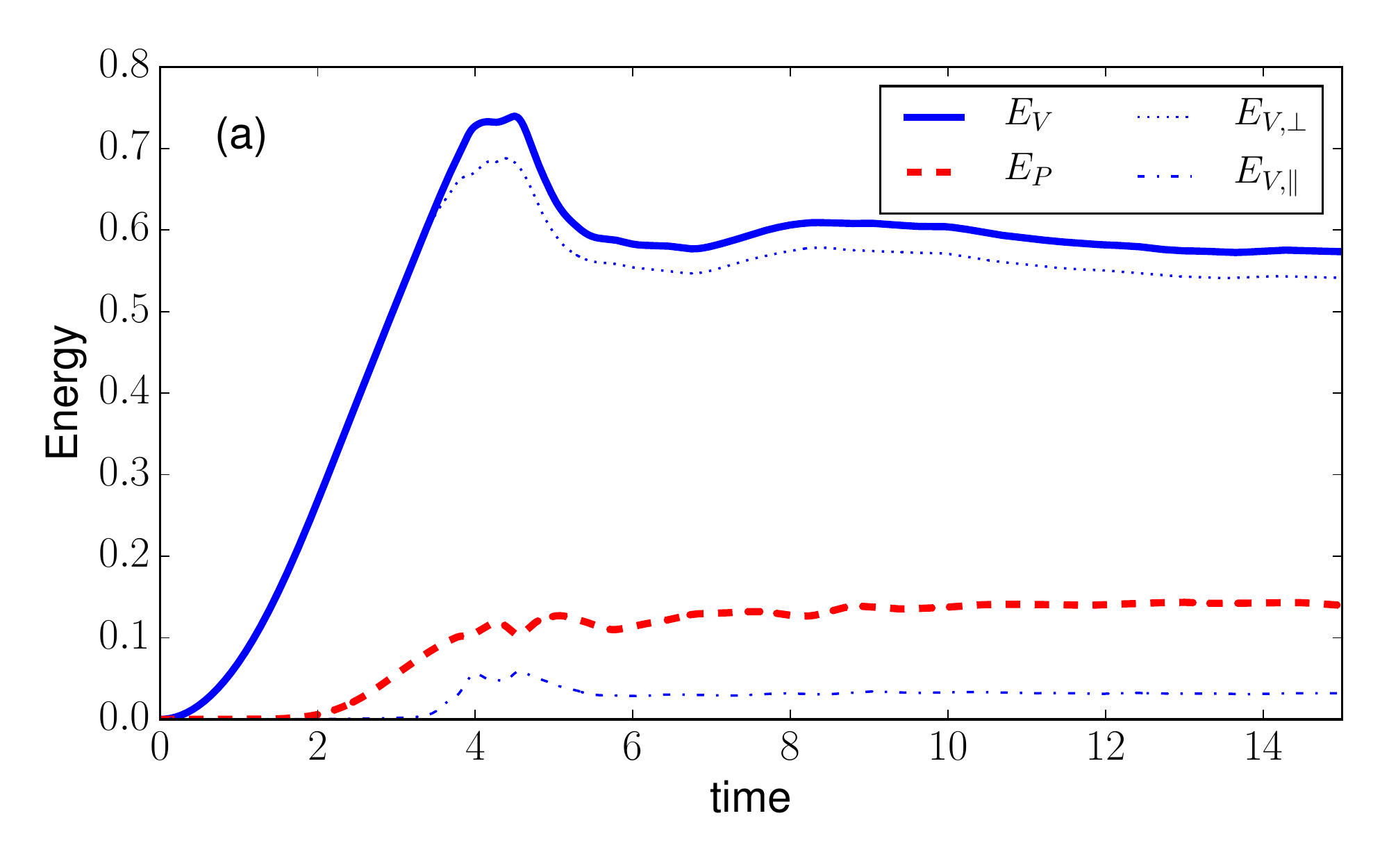}
\includegraphics[width=8.7cm]{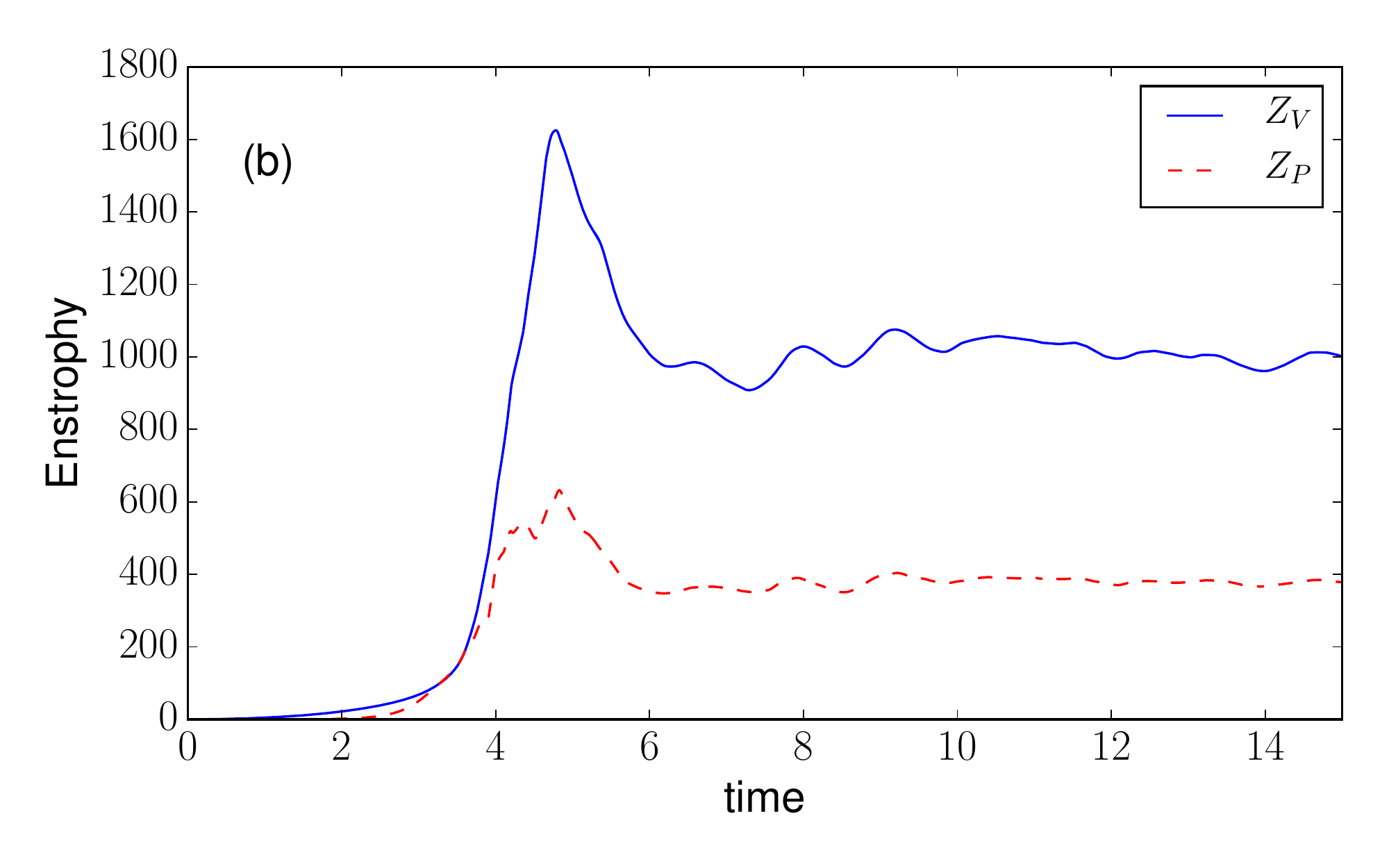}
\includegraphics[width=8.7cm]{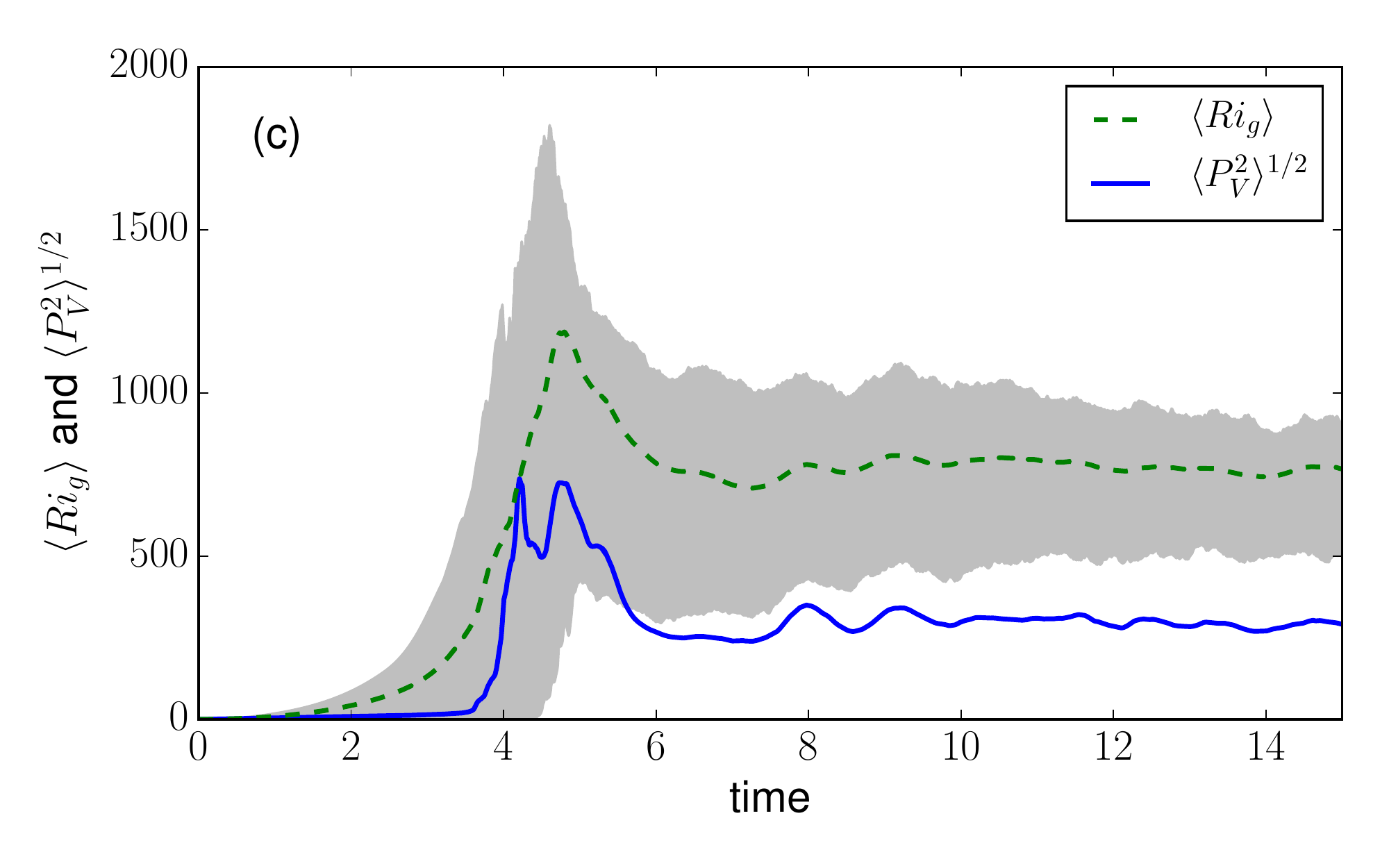}
\caption{({\it Color online}) Temporal evolution \ADD{for run A8} of: (a) Energy, with different partitions; kinetic ($E_V$) and potential ($E_P$), as well as kinetic energy in perpendicular ($E_{V,\perp}$) and in parallel motions ($E_{V,\parallel}$). (b) Kinetic and potential enstrophies $Z_V$ and $Z_P$ (dissipation rates are \ADD{given by $2\nu Z_V$ and $2\kappa Z_P$ respectively}). (c) Mean gradient Richardson number, and r.m.s.~value of the potential vorticity. The shaded region indicates minimum and maximum values of $Ri_g$ when averaged in horizontal planes.}
\label{f:T} 
\end{figure}

The  three  dimensionless parameters of the problem are the Reynolds number
\be \textrm{Re}=\frac{U_\perp L_\perp}{\nu} , \ee
where $U_\perp$ and $L_\perp$ are respectively the characteristic velocity and integral scale of the flow in the horizontal direction, the Froude number
\be \textrm{Fr}=\frac{U_\perp}{L_\perp N} ,\ee
and the Prandtl number defined above. The buoyancy Reynolds number ${\cal R}_B=\textrm{Re} \, \textrm{Fr}^2$ is also an important dimensionless number of the problem, as it measures the amount of small-scale turbulence present in the flow: for ${\cal R}_B>1$, the Ozmidov scale is larger than the Kolmogorov dissipation scale and strong quasi-isotropic mixing can be recovered at small scales.

Mixing and small-scale turbulence in stratified flows is also often quantified by the Richardson number. In our case, as there will be no imposed uniform shear, the Richardson number \ADD{can simply be taken as being proportional to} $1/\textrm{Fr}^2$. A local gradient Richardson number can be also defined pointwise as \cite{rosenberg_15}
\be Ri_g = N(N-\partial_z \theta)/(\partial_z u_\perp)^2, \label{eq:rich} \ee
where $u_\perp = (u^2+v^2)^{1/2}$. When strong temperature gradients develop, this Richardson number becomes small, and in fact can become negative, with a classical transitional value of $1/4$ for \ADD{local shear instabilities} to be able to develop.

\subsection{The code for an elongated box}

To solve numerically these equations we use the GHOST code ({\it Geophysical High-Order Suite for Turbulence}), first developed for incompressible magnetohydrodynamics (MHD) \cite{gomez_05}. It was then extended to have many solvers in both two and three space dimensions: the Navier-Stokes (and Euler) equations, with or without passive scalars, with or without solid-body rotation, gravity, or compressibility, as well as the shallow water equations, the surface quasi-geostrophic equations, Hall-MHD, and more recently the Gross-Pitaevskii equations \cite{clark_16}. It also includes several formulations for subgrid scale modeling. Here we use no subgrid scale modeling and perform instead direct numerical simulations, with all relevant spatial and temporal scales explicitly resolved.

GHOST is a pseudo-spectral code with a Fourier decomposition of the basic fields, periodic boundary conditions, and adjustable-order Runge-Kutta methods to evolve fields in time. It is parallelized using a hybrid method with both MPI and OpenMP \cite{hybrid_11}; it scales linearly over 100,000 processors and it now has CUDA capability to run in GPUs, with simulations done using up to 6250 GPUs \cite{rosenberg_15}. Furthermore, a new version of the code has been developed for this work that allows for non-cubic boxes. Both the lengths of the domain in the three space directions $(L_x, L_y, L_z)$, and the number of grid points in each direction$(n_x, \ n_y, \ n_z)$ can be different, with a total number of grid points $n^3 \equiv n_x n_y n_z$. This new version of the code has been tested for conservation of the global invariants in each set of equations, and of Parseval's theorem, both in single and double precision, with all versions of FFTs and MPI libraries supported by the code.

We chose in this work to set $L_x=L_y=2\pi$ in the horizontal plane (in dimensionless units, typical sizes for the domain with physical dimensions are discussed in Sec.~\ref{sec:dimensions}). To obtain an isotropic grid at small scales this implies $n_x=n_y=n_\perp$ and we chose, \ADD{for the highest resolved flow}, $n_\perp=2048$ \ADD{(see run A8 in Table \ref{t:runs})}. Thus, the wave numbers in the horizontal direction are integers starting with $k_{\perp,\textrm{min}}=1$ and with unit increments up to $k_{\perp,\textrm{max}}=n_\perp/3$ (\ADD{$\approx 683$ for the highest resolution}) assuming a classical $2/3$ dealiasing rule; this corresponds to a horizontal grid spacing of $\Delta_x=\Delta_y=\Delta_\perp=2\pi/n_\perp$ ($=2\pi/2048\approx 3\times 10^{-3}$ in dimensionless units \ADD{for the best resolved case}). 

The second choice made in this work is to set $L_z=L_x \, A_r$, with an aspect ratio $A_r$ of either \ADD{$1:4$ or $1:8$.} The latter, which results in $L_z=\pi/4$, is the case considered for the run at the highest resolution \ADD{(see Table \ref{t:runs}). Three other runs with the same aspect ratio were performed at lower Reynolds numbers (runs B8, ${\bf B8^\ast}$, and C8 in Table \ref{t:runs}) using a resolution of $1024^2 \times 128$ grid points, while three simulations with aspect ratio of $1:4$ were performed with resolutions of $768^2 \times 192$ grid points (runs D4, E4, and F4 in Table \ref{t:runs}). Note that these choices result in all cases in the grid resolution} to be the same in the three directions. Although it is not the only possibility in GHOST, the importance of having an isotropic grid is emphasized in \cite{waite_04} as it allows for an unbiased simulation of the small scales which can thus recover isotropy beyond the Ozmidov scale. Thus, in our simulation $\Delta_z = \Delta_\perp$ and $n_z=n_\perp/A_r$, which results in $k_{z,\textrm{min}}=1/A_r$ \ADD{(either $8$ or $4$). Thus, the simulations with aspect ratio $1:4$ will allow for cases with lesser vertical shear.} Also, $k_{z,\textrm{max}}=k_{\perp,\textrm{max}}=n_\perp/3$. 

Note that as a result of \ADD{these choices for the two sets of runs,} $k_z$ increases by increments of \ADD{4 or} 8: the resolution in wave numbers is scarce at large scales in the vertical direction. The physical implication is that, in elongated boxes such as the ones considered here, the (exact) resonant condition between three waves becomes more difficult to satisfy, and such resonant interactions, at the basis of weak turbulence coupling, are scarcer. Thus the waves can be expected to be less efficient at transferring energy to the small scales: any large-scale 
 balance should be stronger in such a domain, compared to a flow in a cubic box with the same physical dimensionless parameters. Indeed, finite-size effects in wave turbulence are well known \cite{kartashova_08}, although the analysis in wave turbulence theory is generally confined to a cubic box in which only the overall length of the box is considered. However, it should also be noted that the temporal evolution of the pointwise total energy is directly linked to the nonlinear advection terms, as well as pressure gradients, dissipation and forcing, so that nonlinear coupling must also take place in such flows even when resonant triadic interactions are scarce. Indeed, even for isotropic boxes, in rotating and stratified flows for which parameters are chosen to enforce the absence of resonances \cite{smith_02}, one can still find nonlinear interactions leading to direct or inverse cascades \ADD{including} when stratification is strong, \ADD{that is, for Froude or buoyancy Reynolds number small enough to be in a regime dominated by waves \cite{ivey_08} (or regime I in \cite{pouquet_17j}, see also \cite{marino_13i,marino_15w})}. The aspect ratio is also known to alter the strength of direct and inverse cascades when compared to cubic boxes even in the case of homogeneous isotropic turbulence \cite{celani_10}, as well as for geophysical flows with either rotation \cite{deusebio_14}, or stratification \cite{sozza_15}. In spite of this, we can expect the geometric (physical) constraint considered to lead to a more resilient large-scale balance than for a fluid in a cubic box.

Given the choices made for the spatial resolutions, we set in the following the viscosity and diffusivity $\nu=\kappa=1.2\times 10^{-4}$ \ADD{at the highest resolution (run A8), $\nu=\kappa \approx 3.2\times 10^{-4}$ for runs B8, ${\bf B8^\ast}$, and C8, and 
$\nu=\kappa \approx 4 \times 10^{-4}$ for the remaining runs; the \BV\ for all runs is given in Table \ref{t:runs}.}

We will also consider in the following the isotropic, perpendicular, and parallel integral and Taylor scales, defined as
\be
L_\alpha  = \frac{2\pi}{E_V} \int \frac{E_V(k_\alpha)}{k_\alpha} dk_\alpha \ , \alpha=0, \ \perp \ , \parallel \ , 
\label{eq:scales} \ee
and similarly 
\be
\lambda_\alpha= 2\pi \left[ E_V \bigg/ \int  k_\alpha^2 E_V(k_\alpha) dk_\alpha \ \right]^{1/2}  \ , \alpha={0, \ \perp, \ \parallel} \ ,
\label{eq:scalesT} \ee
where the subindex $\alpha = 0$ stands for isotropic quantities. The spectra $E_V(k_\alpha)$ are the reduced energy spectra defined below. In all cases, the integral scale is characteristic of the energy-containing eddies (and thus also called the energy-containing scale), and the Taylor scale is the scale at which the dissipation would equal the dissipation of the actual flow if all its energy were to be concentrated at only one scale. Thus, the Taylor scale is a characteristic scale lying in the inertial range, and the smaller this scale, the more developed the turbulence. \ADD{Similarly, the buoyancy and Ozmidov scales, $L_B$ and $\ell_{oz}$, are defined (for the dimensionless box of length $2\pi$) as:
\be
 L_B=2\pi {U_\perp}/{N} \ , \ \ell_{oz}= 2\pi ({\epsilon_V}/{N^3})^{1/2} \ ,
 \label{eq:Oz} \ee
under the assumption that a Kolmogorov $\sim k^{-5/3}$ spectrum is recovered at scales sufficiently smaller than $\ell_{oz}$. Table \ref{t:runs} gives the following ratios of these length scales for all the runs: $L_z/L_B$ and $\ell_{oz}/\ell_{min}$ (indicating, when larger or equal to unity, that the buoyancy and Ozmidov scales are resolved in the simulation, and where $\ell_{min}$ is the minimal resolved scale in the run), $L_0/\lambda_0$ (which estimates the scale separation dynamically), $L_\perp/L_\parallel$ and $\lambda_\perp/\lambda_\parallel$ (which can be considered respectively as estimations of the anisotropy at large and at small scales in the flows), and
\begin{equation}
r_E=E_P/E_T, \,\,\,\,\,\,\, r_\epsilon=\epsilon_P/\epsilon_T, \,\,\,\,\,\,\, \beta=\epsilon_V/\epsilon_V^{Kol},
 \label{eq:ratios} \end{equation} 
with $\epsilon_T=\epsilon_V+\epsilon_P$ the total energy dissipation rate and $\epsilon_P$ the dissipation rate of potential energy, and where $\epsilon_V^{Kol}=U_\perp^3/L_\perp$ is the dimensional (Kolmogorovian) estimate of the kinetic energy dissipation rate. Note that $\beta$ is observed to vary linearly with Froude number \cite{pouquet_17j}, although its actual level does not seem constant for a given $Fr$, the geometry of the flow and the presence or not of strong shear likely playing a role.}

\begin{figure}
\includegraphics[width=8.7cm]{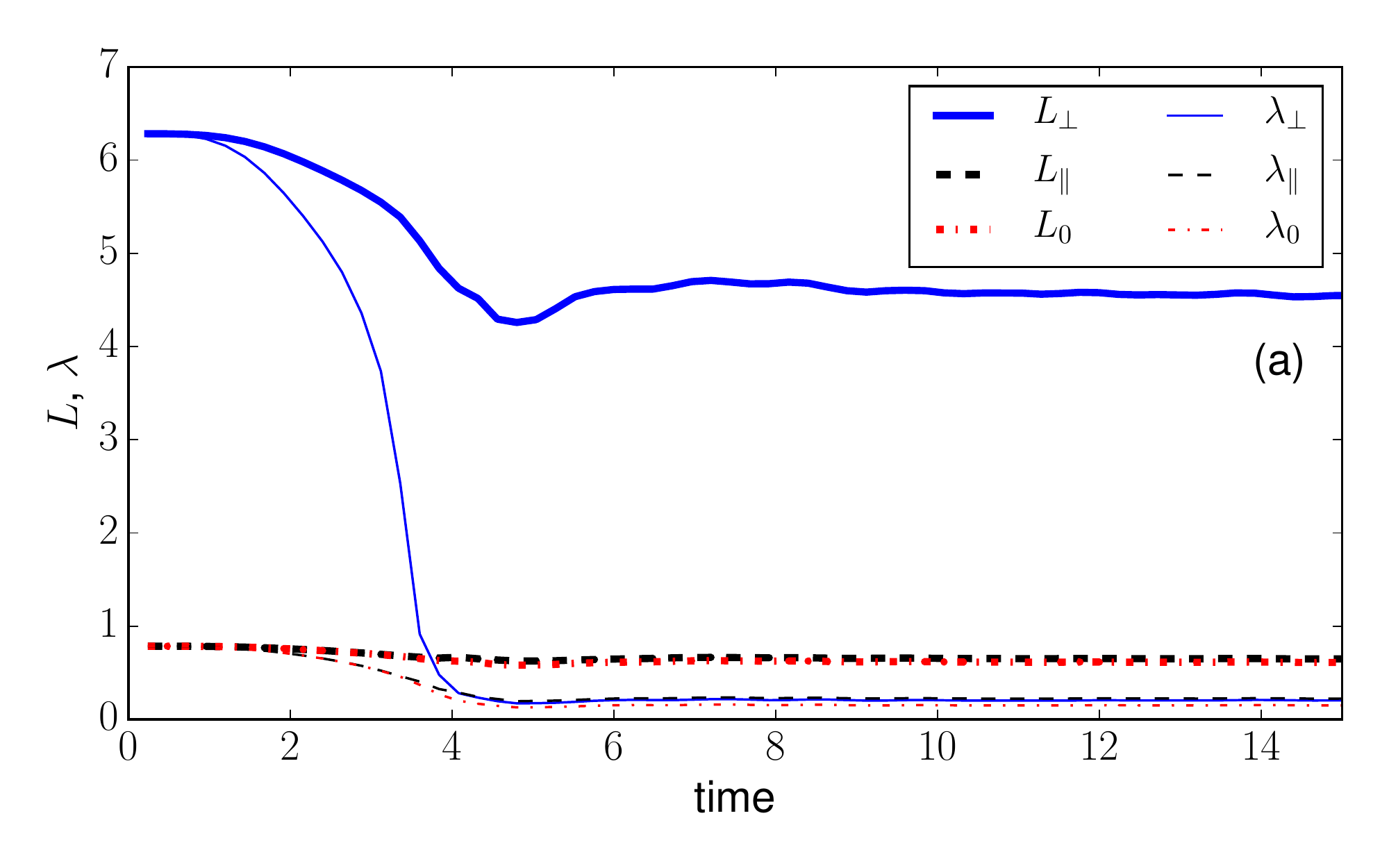}
\includegraphics[width=8.7cm]{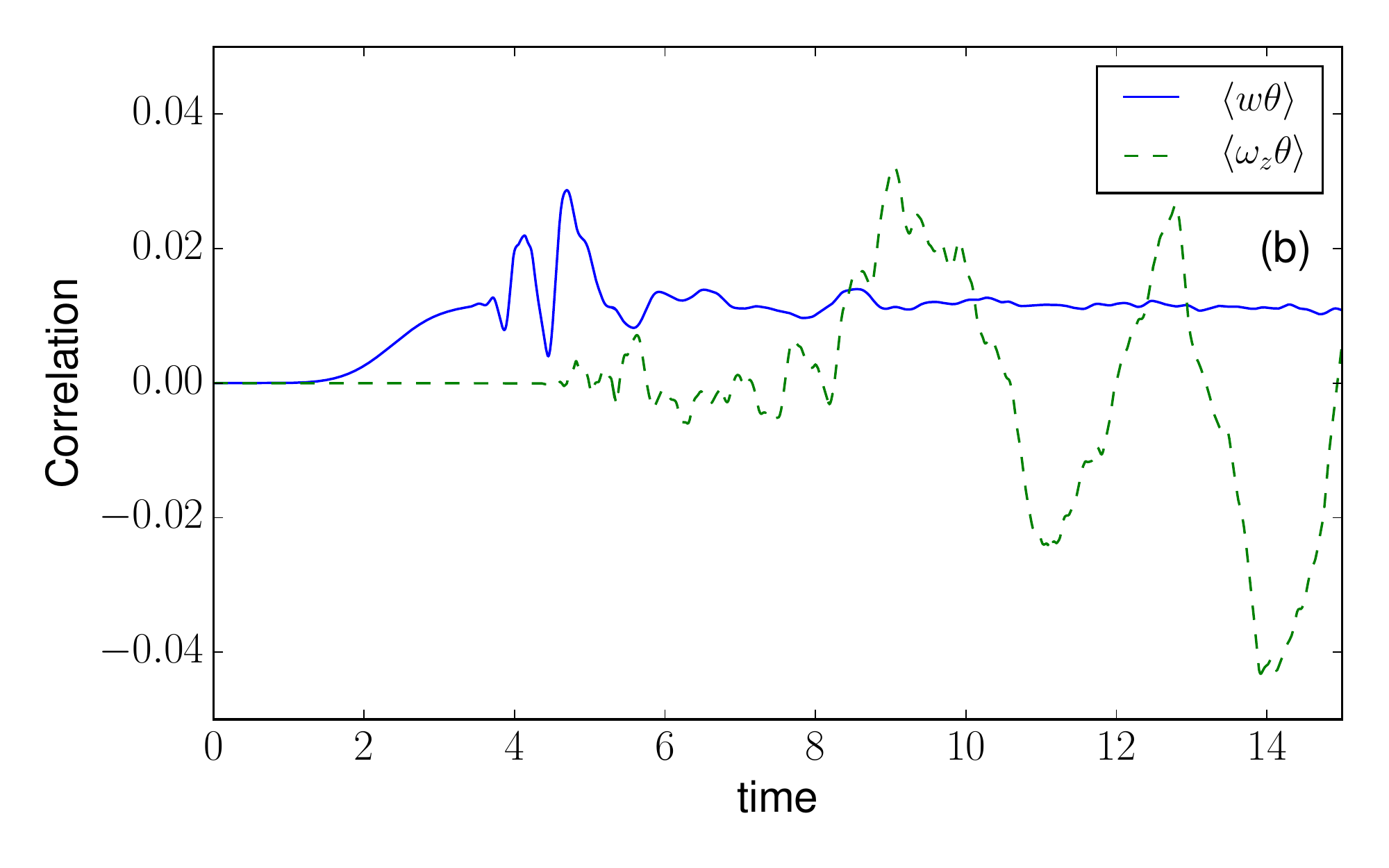}
\caption{({\it Color online}) (a) Time evolution of integral ($L$) and Taylor ($\lambda$) scales \ADD{in run A8}: isotropic ($L_0$ and $\lambda_0$), perpendicular ($L_\perp$ and $\lambda_\perp$), and parallel ($L_\parallel$ and $\lambda_\parallel$). (b) Vertical flux of temperature fluctuations $\langle w \theta \rangle$, and correlation between vertical vorticity and temperature fluctuations $\langle \omega_z \theta \rangle$.}
\label{f:T2}
\end{figure}

Finally, the above expressions are defined using isotropic and anisotropic Fourier spectra for the fields, which in the elongated box are built from the correlation functions in Fourier space, or equivalently, from the power spectral densities. As an example, for the kinetic energy using the velocity correlation function in Fourier space $U({\bf k})$ (see, e.g., \cite{3072}), we can define the axisymmetric spectrum
\be 
e_V(k_\perp,k_\parallel) =  e_V(k, \theta) = \int U({\bf k}) k \sin \theta \textrm{d}\phi \ ,
\label{etheta}  \ee 
with $\theta$ the co-latitude in Fourier space, $\phi$ the longitude in Fourier space, $k_\perp=(k_x^2 + k_y^2)^{1/2}$, $k_\parallel = k_z$,  and $k=(k_\perp^2 + k_\parallel^2)^{1/2}$. From this spectrum one can also define reduced isotropic, perpendicular, and parallel spectra respectively as
\begin{eqnarray}
E_V(k) &=& \int e_V(k, \theta) k \textrm{d}\theta , \label{eq:iso} \\
E_V(k_\perp) &=& \int e_V(k_\perp,k_\parallel) \textrm{d}k_\parallel , \label{eq:para} \\
E_V(k_\parallel) &=& \int e_V(k_\perp,k_\parallel) \textrm{d}k_\perp \label{eq:per} .
\end{eqnarray}
Similar definitions hold for the  potential energy spectra built on the temperature fluctuations.

\begin{figure*}
\includegraphics[width=18cm,trim=0 15 0 25,clip]{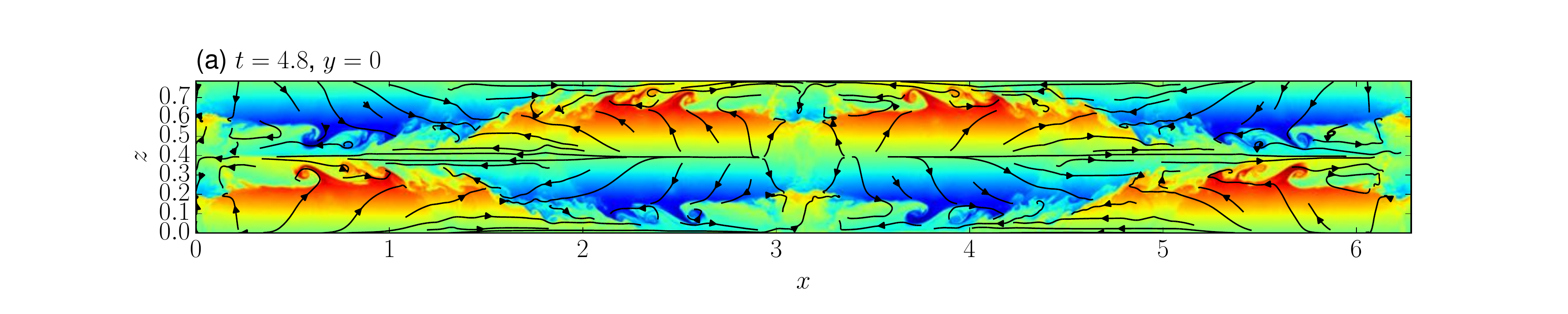} 
\includegraphics[width=18cm,trim=0 15 0 25,clip]{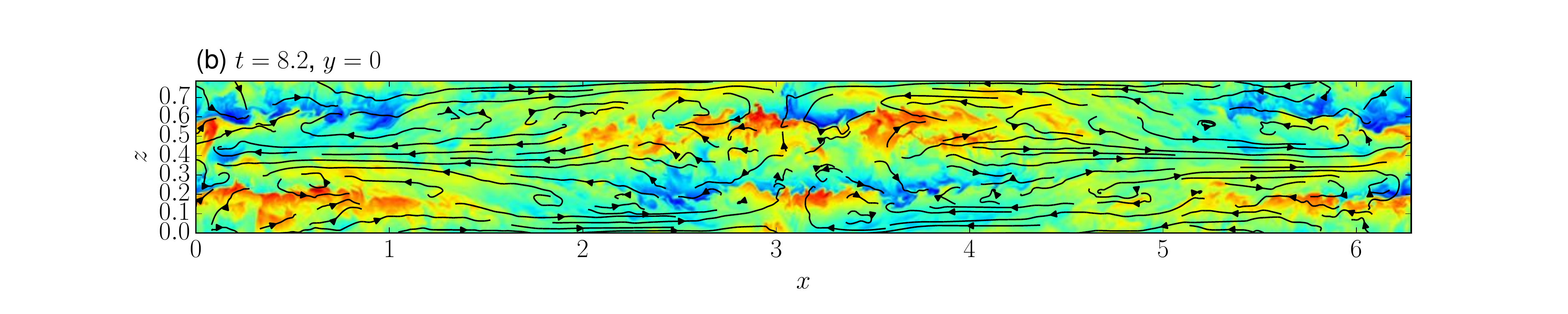}   
\includegraphics[width=18cm,trim=0 15 0 25,clip]{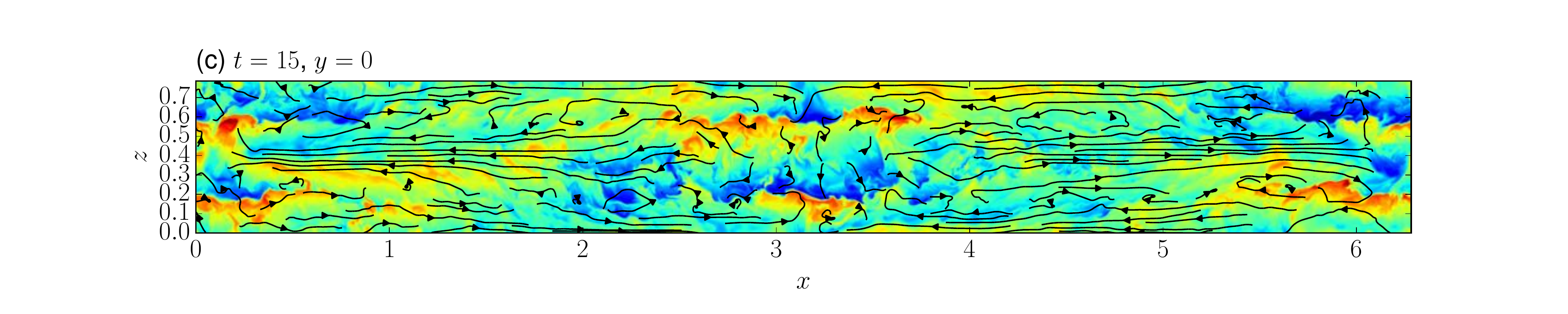}    
\caption{({\it Color online}) Vertical $(x,z)$ slices at $y=0$ of temperature fluctuations \ADD{for run A8} at (a) $t=4.8$, (b) $t=8.2$, and (c) $t=15$, with super-imposed velocity vectors. \ADD{The same color map applies to all three snapshots; dark blue corresponds to $\theta = -2$ (in units of velocity), and red to $\theta = 2$.} Note at large-scales the Taylor-Green flow as sketched in Fig.~\ref{f:TG} (bottom), with the flow at $z=\pi/8$ going from the center of the box to the boundary, and vice-versa at $z=0$ and $\pi/4$. At late times a circulation develops, with cold downdrafts and hot updrafts (see $x=0$, $\pi$ and $2\pi$).}
\label{f:V}
\end{figure*}

\begin{figure}   
\includegraphics[width=4.25cm,trim=15 0 15 0,clip]{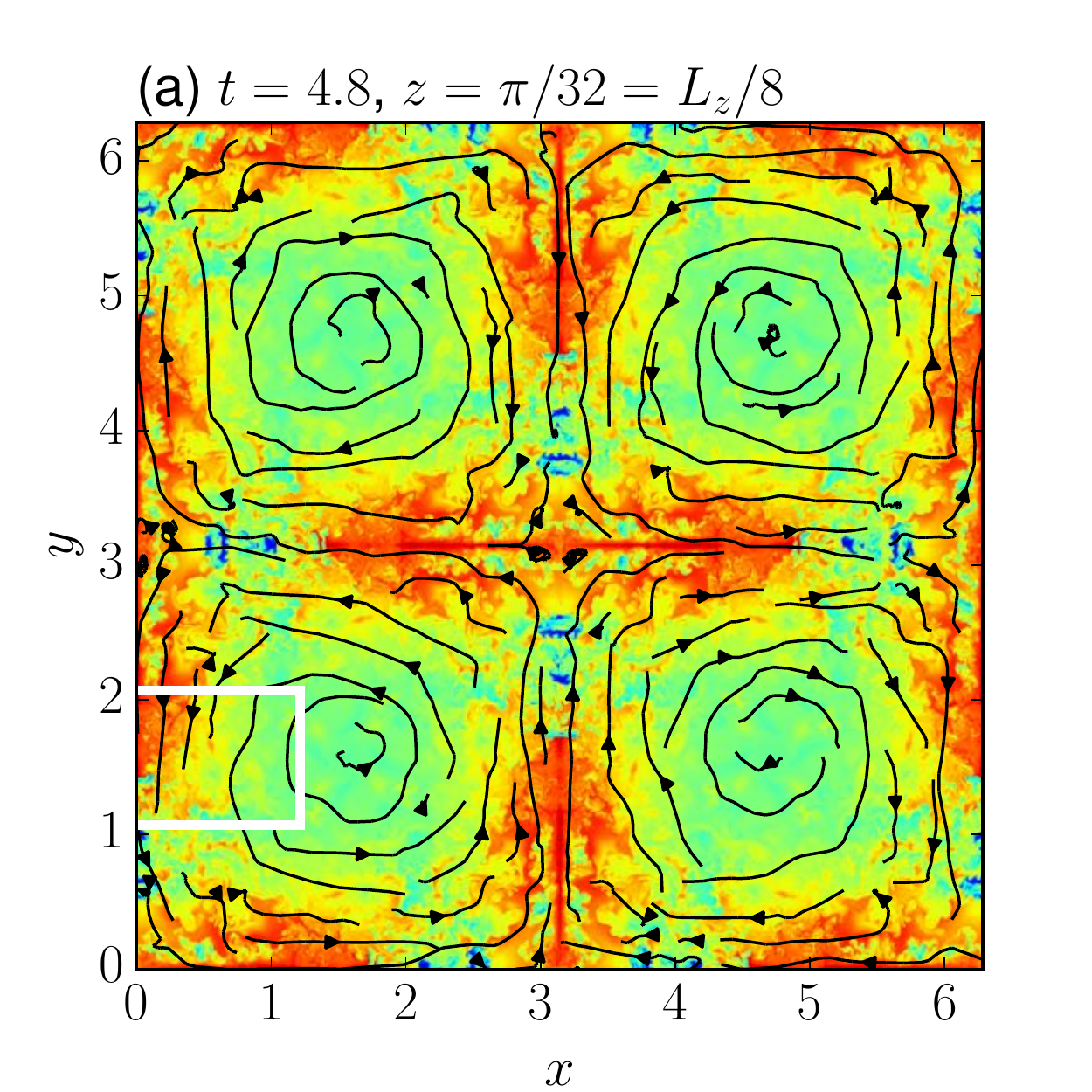}
\includegraphics[width=4.25cm,trim=15 0 15 0,clip]{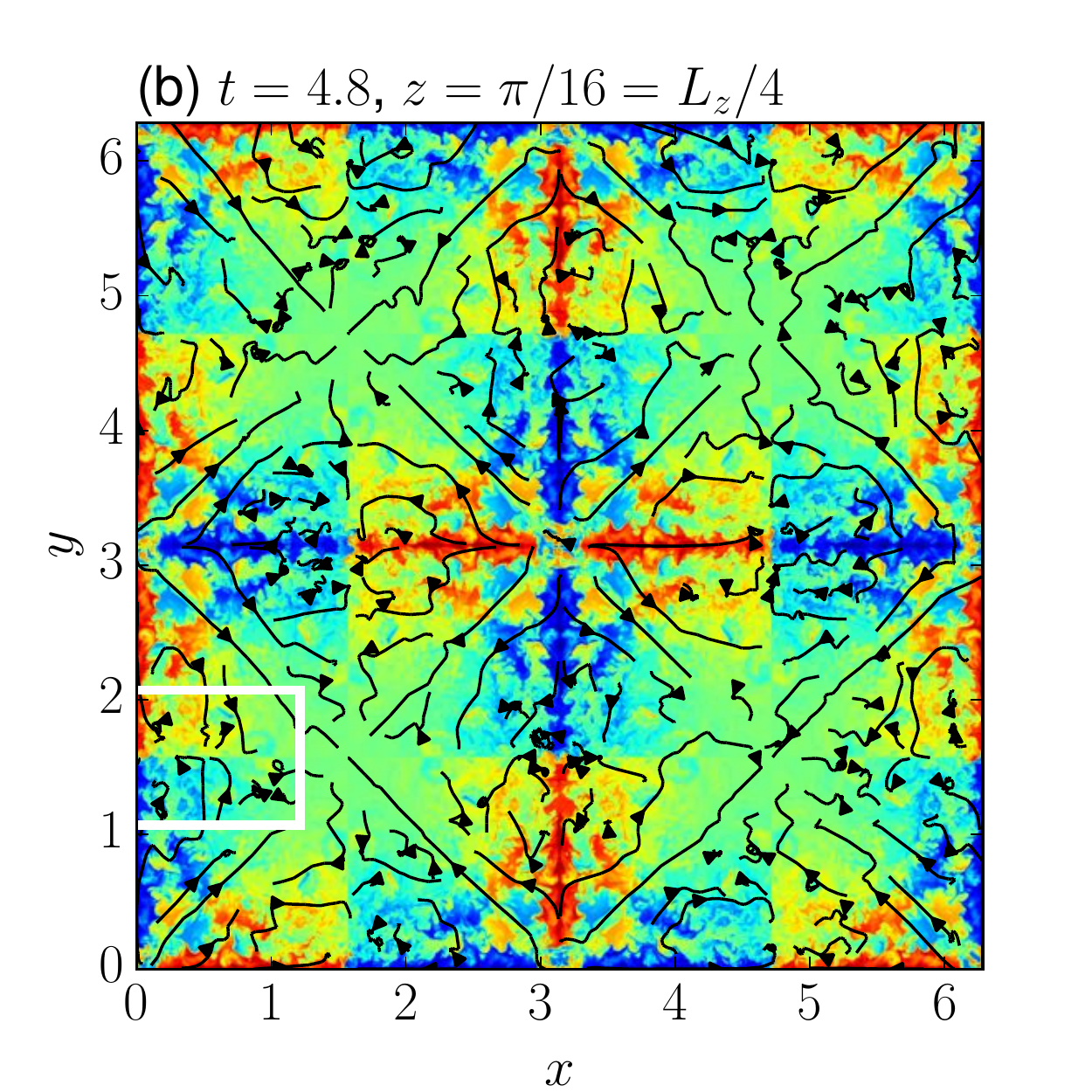}
\includegraphics[width=4.25cm,trim=15 0 15 0,clip]{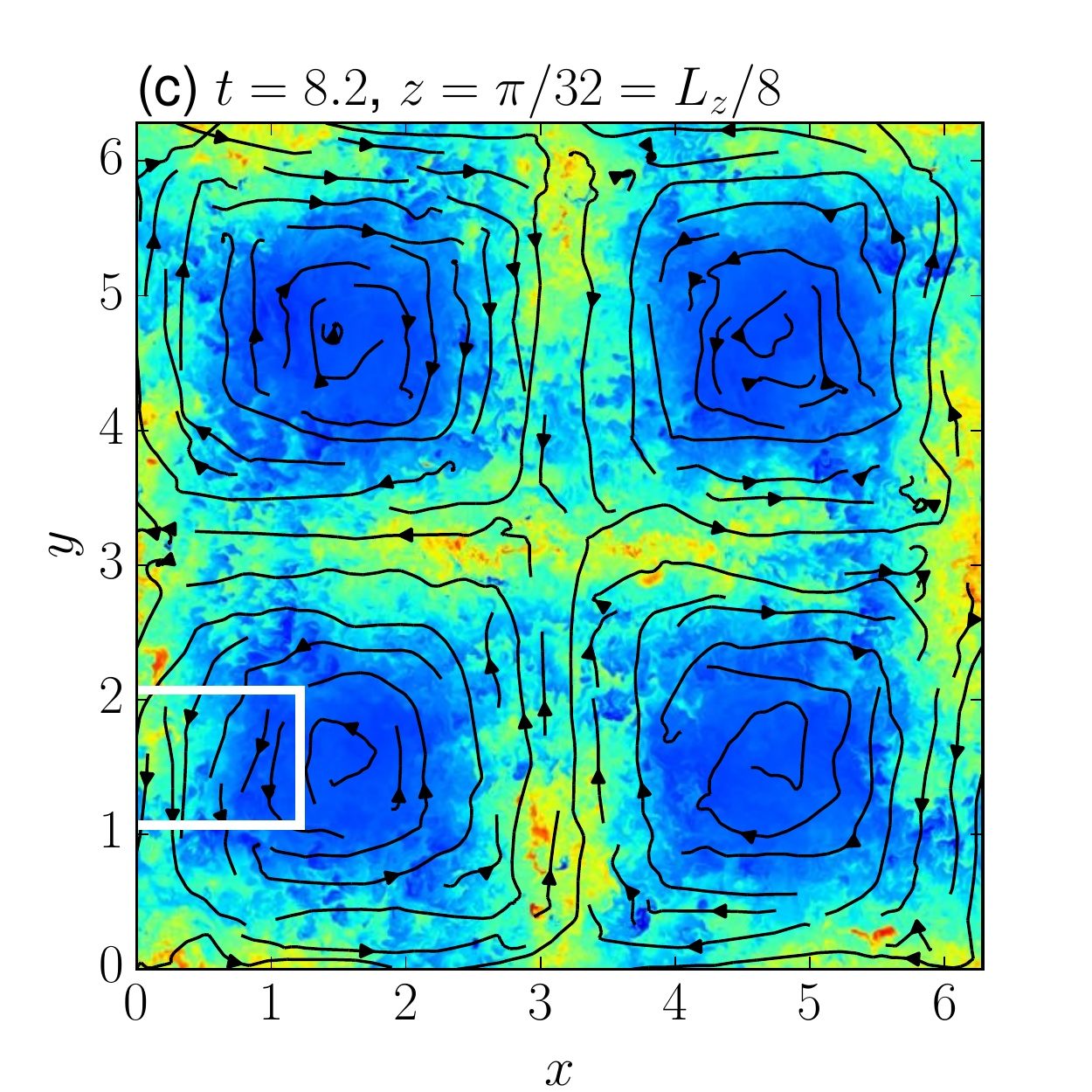}
\includegraphics[width=4.25cm,trim=15 0 15 0,clip]{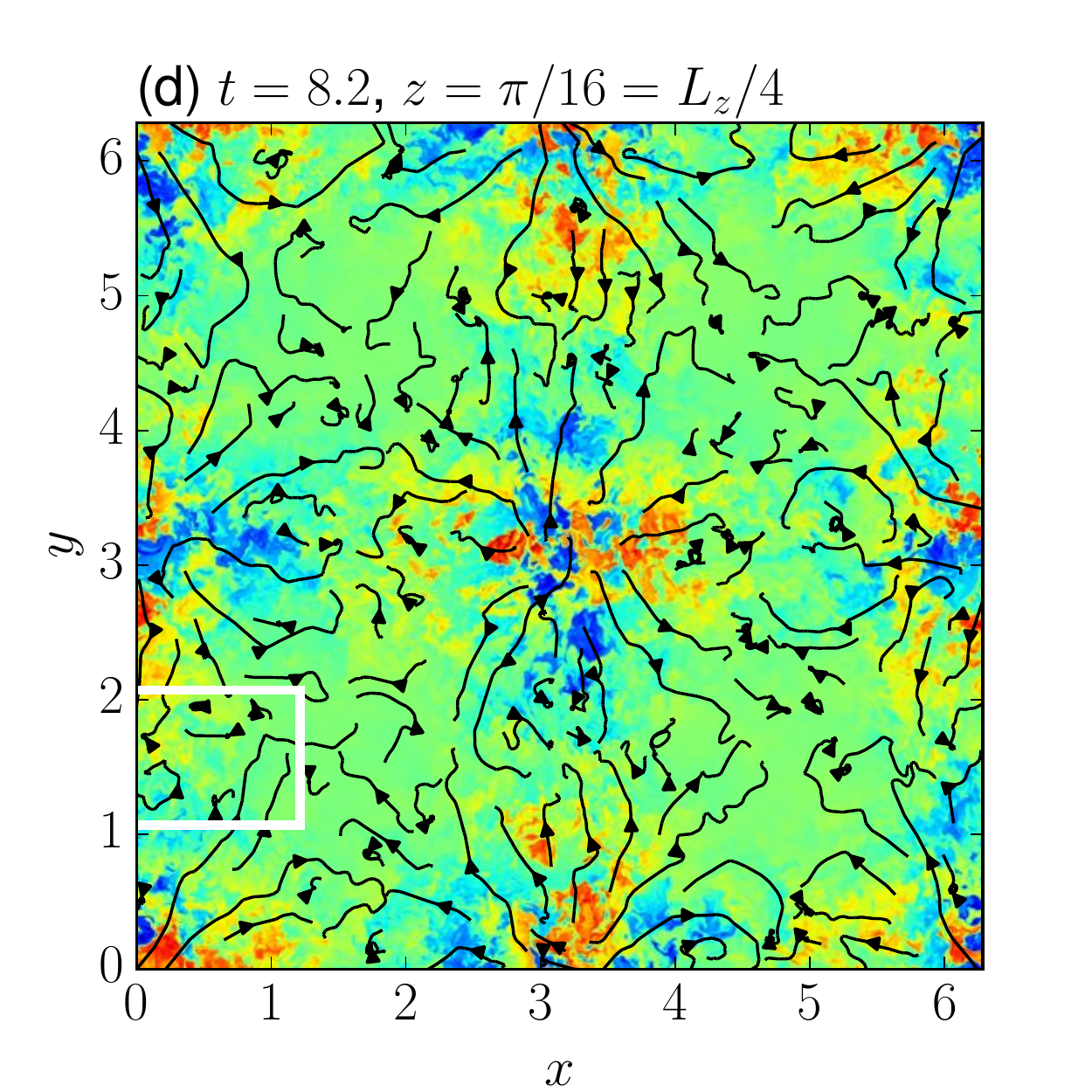}   
\caption{({\it Color online)} Two-dimensional horizontal cuts, \ADD{for run A8,} of temperature fluctuations, with super-imposed velocity fields at various times and at different heights: (a) $t=4.8$ at $z=L_z/8$, (b) same time at $z=L_z/4$ (i.e., at the shear layer, where forcing is zero), (c) $t=8.2$ at $z=L_z/8$, and (d) same time at $z=L_z/4$. The white boxes indicate a region with fronts near the shear layer, where perspective volume renderings will be performed (see Fig.~\ref{f:3DT}). \ADD{The colormap is the same as in Fig.~\ref{f:V}.}}
\label{f:CT}
\end{figure}

\subsection{The forcing} 

\begin{figure*}  
\includegraphics[width=15cm]{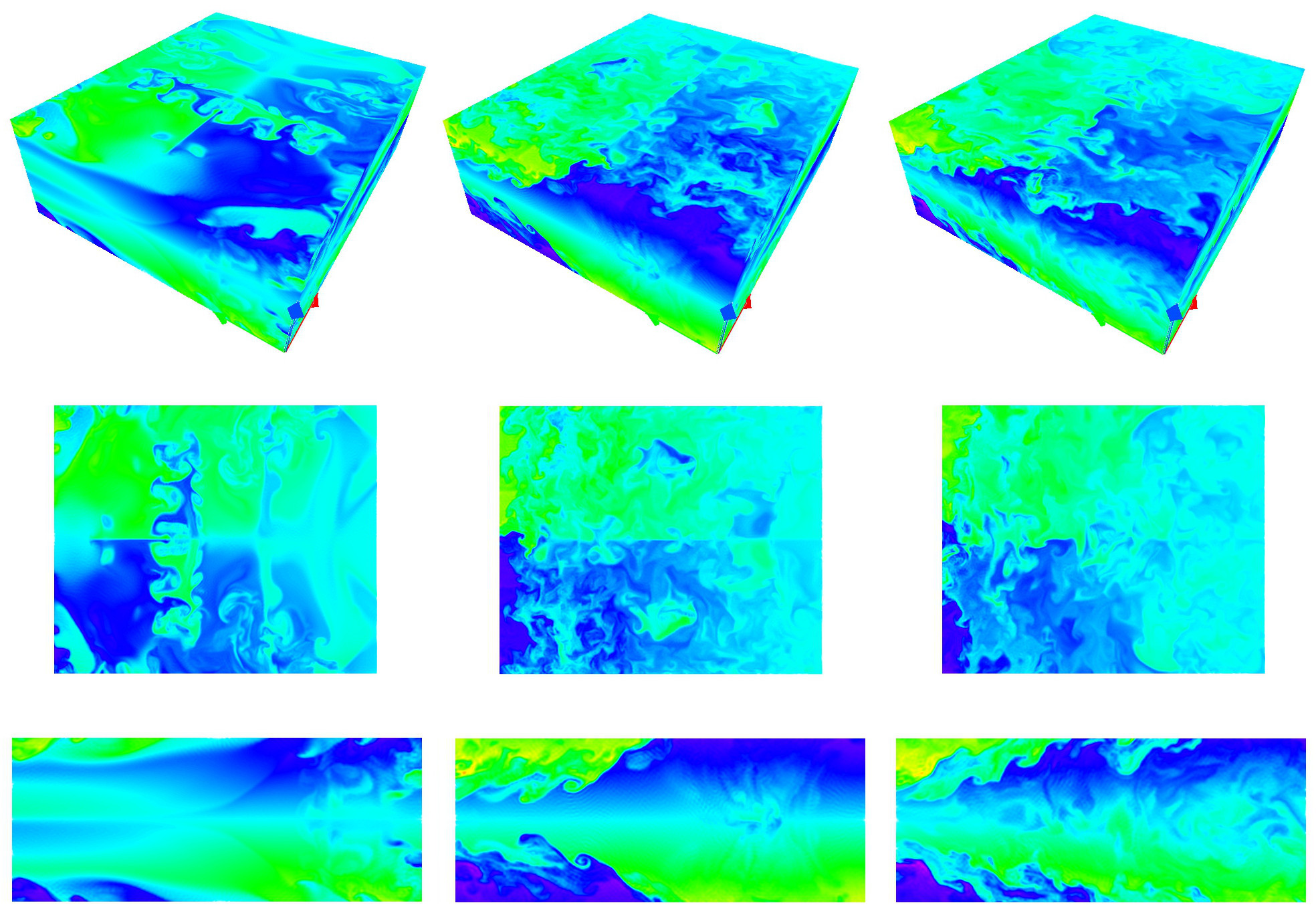}
\includegraphics[width=15cm]{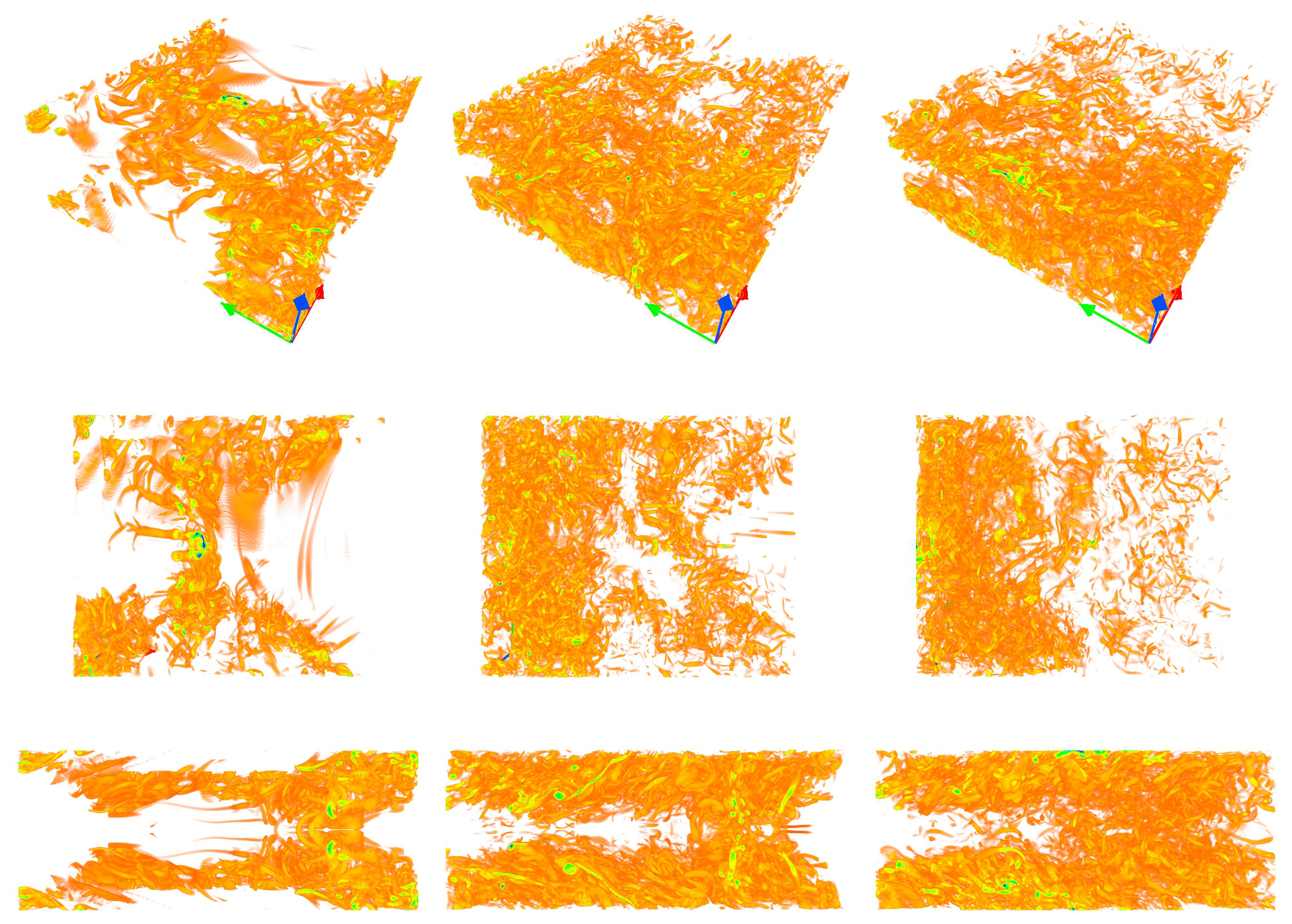}
\caption{({\it Color online}) Three-dimensional renderings, \ADD{for run A8,} of temperature fluctuations (first three rows) and of vorticity intensity (last three rows), at $t=4.3$ when the first fronts are created (left column), at $t=4.8$ (middle column), and $t=5.2$ (right column) when turbulence has developed. Red, green, and blue arrows are respectively the $x$, $y$ and $z$ directions. For both quantities, top row is a perspective volume rendering of the region indicated in white in Fig.~\ref{f:CT}, middle row is a top view, and bottom row is a size view. Observe the vertically slanted destabilizing fronts \ADD{for $t\approx 4.8$ and $\approx 5.2$.} Also note, in the vorticity at $t=4.8$ (bottom row\ADD{, middle column),} the creation of pairs of Kelvin-Helmholtz like vortices that feed the turbulence \ADD{on the left side of the domain. Color maps are linear, from $-2$ (light yellow) to $2$ (dark blue) for the temperature; for the vorticity only regions with intensity larger than $3\sigma$ are shown, where $\sigma$ is the variance.}}
\label{f:3DT}
\end{figure*}

\ADD{For all runs but run B8$^\ast$, the forcing} is only incorporated in the momentum equation, and initial conditions for both the velocity and temperature fluctuations are zero. \ADD{Run ${\bf B8^\ast}$ has the same initial conditions but a balanced forcing, and will be discussed in Sec.~\ref{S:PARAM}.} In all cases, the mechanical forcing is based on the Taylor-Green (TG) vortex \cite{taylor_37}, and is applied only to the horizontal components of the velocity. \ADD{As stated in the Introduction,} the TG vortex is a classical flow in the study of turbulence. It consists of two counter-rotating vortices, with a shear layer in between, and mimics many experimental configurations in the laboratory. In isotropic periodic domains, the TG flow is written as:
\begin{eqnarray}
u^{TG} &=& u_0 \sin (k_0x) \cos (k_0y) \cos (k_0z) \nonumber  \ , \\
v^{TG} &=& -u_0 \cos (k_0x) \sin (k_0y) \cos (k_0z) \nonumber  \ , \\
w^{TG} &=& 0 \ , \label{TGf}
\end{eqnarray}
with $k_0$ a characteristic wavenumber. The TG flow was also used in studies of stratified flows \cite{riley_03}, where it was shown that its vertical characteristic scale decreases with time, as predicted, e.g., in \cite{lilly_83, babin_97}, and that regions with strong shear are prone to many instabilities leading to the development of small-scale turbulence.

The TG flow has strong differential rotation as well as point-wise helicity, defined as the correlation between velocity and vorticity, $H_V({\bf x})= {\bf u}({\bf x}) \cdot \vomega({\bf x})$, although on average, because of the symmetries of the flow, it has no global helicity. Furthermore, for homogeneous isotropic turbulence, this flow develops in time a vertical velocity in the form of a recirculation, as shown \ADD{analytically} in \cite{taylor_37} using an expansion in time to fourth order. This secondary flow is created by the vertical pressure gradient. However, in the stratified case such a recirculation has to fight against gravity and, as shown later, an energetically more favorable secondary flow develops.

We use the TG flow as mechanical forcing and adapt it to the particular elongated geometry chosen here, by stating that the forcing must fill the box both in the horizontal plane and in the vertical direction. This results in \begin{eqnarray}
F_x^{TG} &=& F_0 \sin (x) \cos (y) \cos (z/A_r) \nonumber  \ , \\
F_y^{TG} &=& -F_0 \cos (x) \sin (y) \cos (z/A_r) \nonumber  \ , \\
F_z^{TG} &=& 0 \ , \label{TGf8}
\end{eqnarray}
a forcing we label \ADD{TGz8 for $A_r=1/8$, and TGz4 for $A_r=1/4$.} The choice of $F_0$ is such that $U_\perp$ is of order unity in the turbulent steady state. This formulation of the forcing leads to the formation of elongated vortices, with an aspect ratio which is that of the box. Such a forcing has strong shear in the vertical  (corresponding to \ADD{$k_z=1/A_r$),} and is thus intended to mimic the vertical shear that is often encountered in the atmosphere or the ocean. The TGz8 forcing is depicted schematically in Fig.~\ref{f:TG} in two-dimensional $(x,y)$ and $(x,z)$ slices. The arrows indicate the direction of the forcing. On the top is a horizontal cut: for $k_x=k_y=1$ as depicted here, there are basically four circular cells to this flow. In the vertical, the box is 8 times smaller, as shown in the bottom of Fig.~\ref{f:TG}, and the cells are flattened; the TGz8 forcing is maximum (in absolute value) at $z_\textrm{max}=0$, $L_z/2$, and $L_z$, and is zero at $z_0=L_z/4$ and $3L_z/4$, the two planes where vertical shear is strongest (indicated by two dashed horizontal lines). The shaded region with a bell-like curve represents the amplitude of the flow, with the velocity going from zero at the intersection of the bell-like curve with the horizontal dashed lines, to its maximum absolute values in between these horizontal lines.

\ADD{\section{Temporal evolution of run A8 \label{S:TIME}}}

\ADD{In this section and in the following we analyze run A8, which has the highest Reynolds and buoyancy Reynolds numbers in a box with aspect ratio $1:8$ (and thus, with TGz8 forcing). The \BV\ is $8$, and thus the horizontal scales associated with the shear and with buoyancy are comparable. The effects of varying the \BV\ and the vertical shear are studied in Sec.~\ref{S:PARAM}.}

\subsection{Global quantities and correlations}

We first show in Fig.~\ref{f:T}(a) the temporal evolution of the energy in run A8 decomposed into its kinetic and potential components, $E_V$ and $E_P$, and the kinetic energy in perpendicular and parallel motions, respectively $E_{V,\perp} = \langle u_\perp^2 \rangle /2$ and $E_{V,\parallel} = \langle w^2 \rangle /2$. For all of them, after an initial increase starting from zero initial conditions, a maximum is reached around $t=4$ and, after a short relaxation, the flow settles into a statistical steady state with r.m.s.~velocity $U_0 \approx 1.1$ and r.m.s.~temperature fluctuations $\theta_0 \approx 0.5$; data is averaged between $t=5$ and 15; in the following all time averages are computed in this window. The kinetic energy in horizontal motions is dominant, and the ratio of potential to total energy for this run \ADD{can bee seen in Table \ref{t:runs}}.

An increase similar to that of energy is observed in run A8 for kinetic enstrophy \ADD{and its potential equivalent, both} shown in Fig.~\ref{f:T}(b), with $Z_V = \left< |\vomega|^2 \right>/2 = (\left< |\vomega_\perp|^2 \right> + \left< \omega_z^2 \right>)/2$, and $Z_P= \left< |{\bf \nabla} \theta|^2 \right>/2$.  With these definitions, the energy dissipation rates for each energy component are $\epsilon_V= 2 \nu Z_V$ and $\epsilon_P= 2 \kappa Z_P$. They display a sharper peak than the energies; for longer times, the $\epsilon_P/\epsilon_T$ ratio also converges to a slightly higher value than the ratio of energies (see Table \ref{t:runs}). This is indicative of a more efficient mixing in the small scales, as expected for a flow that has developed turbulent structures. 

In Fig.~\ref{f:T}(c) we also show the r.m.s.~potential vorticity $P_V$, and the gradient Richardson number $\langle Ri_g \rangle$, averaged over the entire domain for run A8. They display the same type of evolution as the enstrophies, except for a sharper double peak in $P_V$, the trace of which can be seen also in $Z_P$. This indicates that in the $P_V$ evaluation, the nonlinear term $\vomega \cdot \nabla \theta$ is dominant at the peak of dissipation; at later times the evolution of the r.m.s.~value of $P_V$ is quite similar to that of kinetic dissipation; the rather strong closeness of the two evolutions now suggest that the kinetic enstrophy is dominated by the vertical component of the vorticity. Concerning the averaged gradient Richardson number, although its values are large, note that fluctuations are also very large. In Fig.~\ref{f:T}(c) we indicate with a shaded area the minimum and maximum values of $Ri_g$ after averaging in horizontal planes, $\langle Ri_g \rangle_\perp$ (i.e., averaged over the $x$ and $y$ coordinates). After $t\gtrsim 4$, $\min\{\langle Ri_g \rangle_\perp\} >0$, i.e., all horizontal planes are (on average) stable against local shear and overturning instabilities. However, pointwise fluctuations of $Ri_g$ in each plane are still very large, and the flow has unstable points with $Ri_g < 0.25$ and with $Ri_g < 0$ at all times, as will be seen in the next sections.

\begin{figure}      
\includegraphics[width=8.5cm]{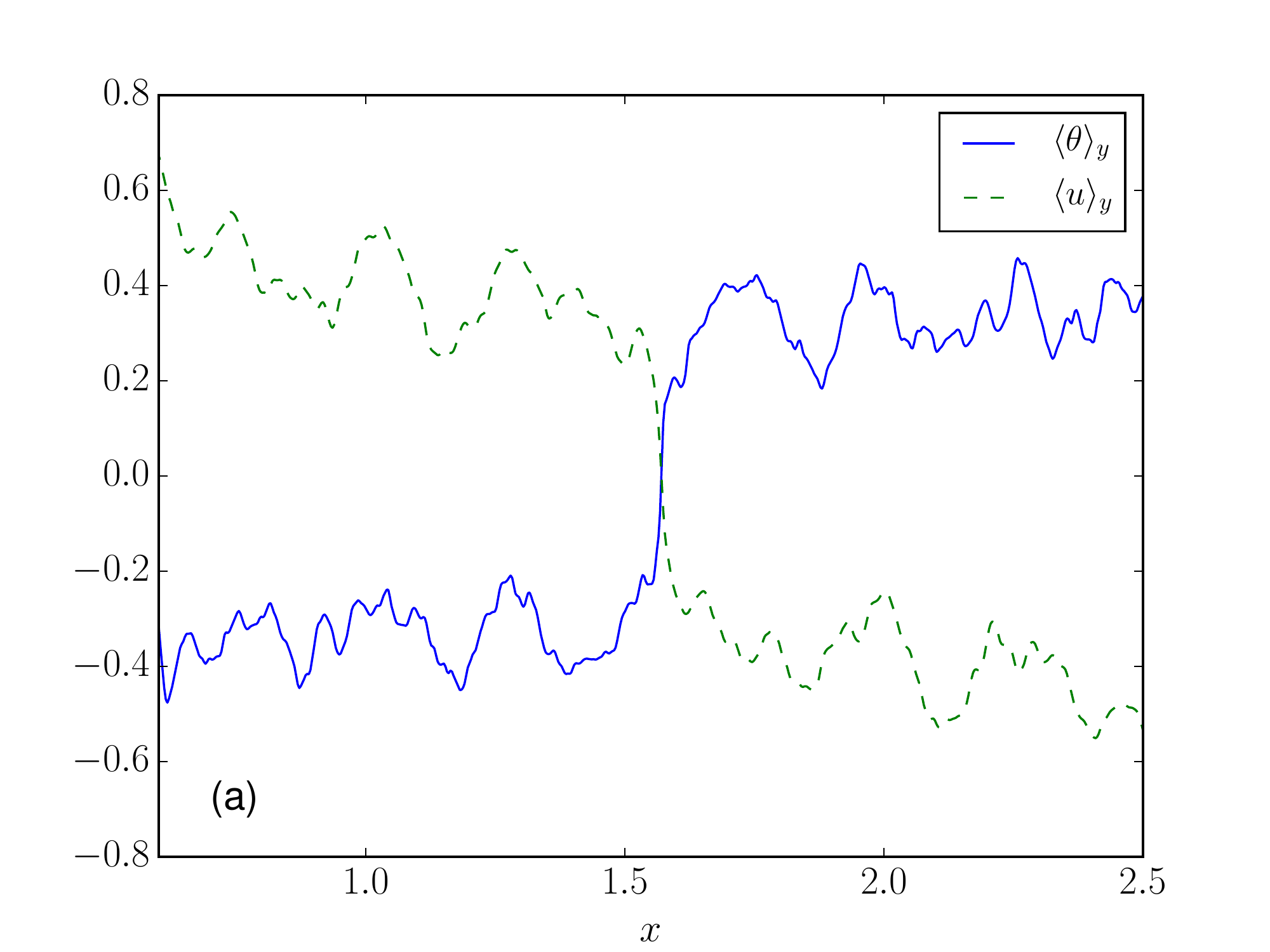}  
\includegraphics[width=8.5cm]{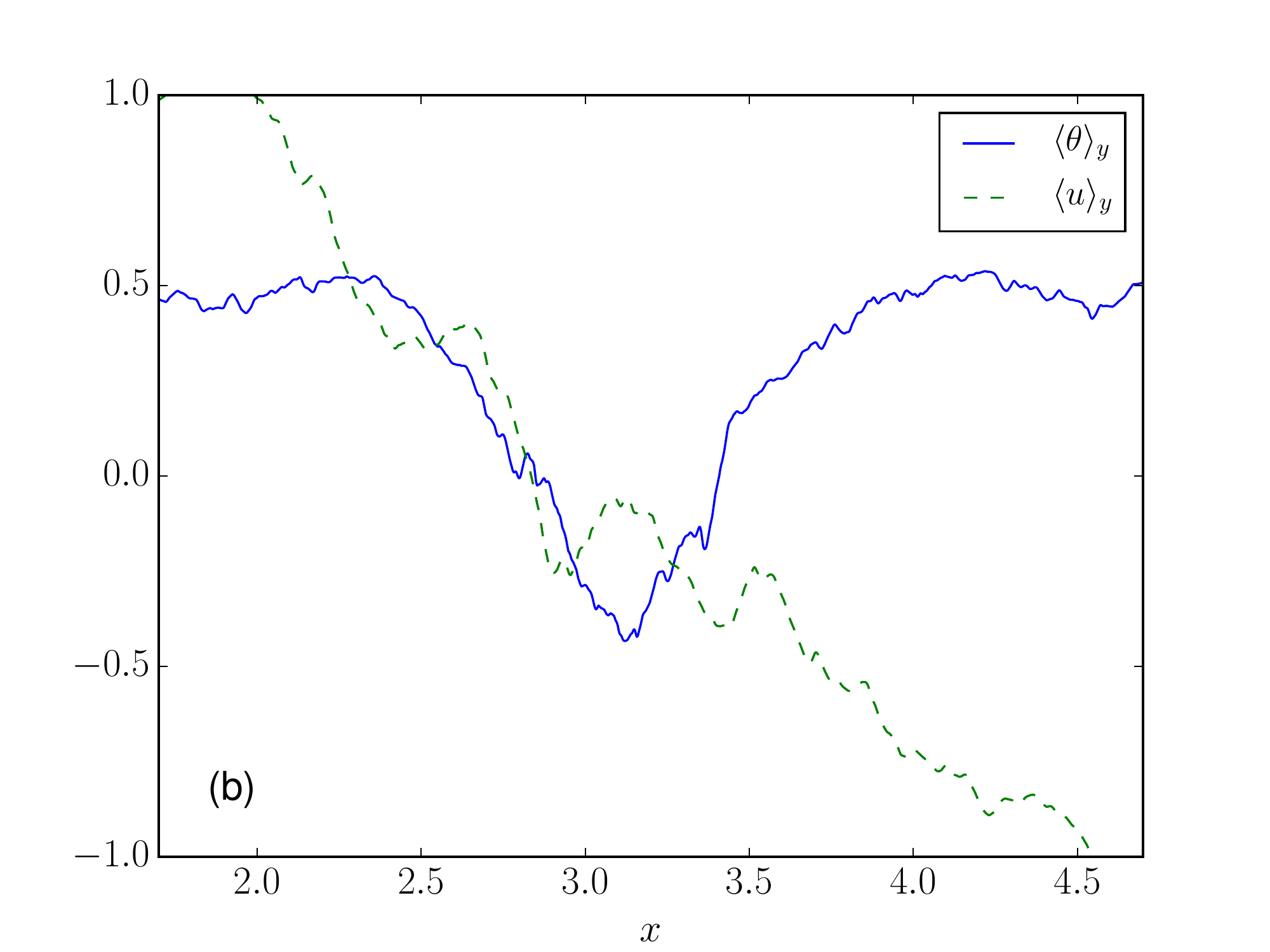}   
\caption{({\it Color online}) Instantaneous temperature and velocity ($u$) profiles, \ADD{for run A8,} averaged in the horizontal ($y$) direction in the vicinity of the structure, of (a) a front at early time ($t=4.8$), and (b) a cold filament-like structure at a later time ($t=8.2$).
} 
\label{f:P} 
\end{figure}

When examining the characteristic scales of the flow in run A8, we see in Fig.~\ref{f:T2}(a) that all scales are smaller than unity except for the perpendicular integral scale, since the TGz8 forcing is applied at $k_x=k_y=1$ and $k_z=8$. Between $t\approx 2$ and $4$, the Taylor scales decrease abruptly and become smaller than the integral scales, indicating the flow becomes unstable and develops small scale turbulence. After this transient, the flow is strongly anisotropic at large scales since $L_\perp$ and $L_\parallel$ are quite different (resulting both from the spectral anisotropy of the forcing and from the stratification), but isotropy seems to recover at small scales in the sense that $\lambda_\perp \approx \lambda_\parallel \approx 0.15$ (see Table \ref{t:runs}). Anisotropy will be studied in more detail in Sec.~\ref{S:FOURIER}.

Finally, Fig.~\ref{f:T2}(b) gives the temporal evolution of the vertical temperature flux, which is proportional to the vertical buoyancy flux $N \left< w \theta \right>$, and of the correlation between the vertical vorticity and temperature fluctuations $\left< \omega_z \theta \right>$. Zero initially, they both grow with time, although the onset of the cross correlation between $\omega_z$ and $\theta$ has a later departure with at first smaller fluctuations. We note that the buoyancy flux is always positive, indicating that upward currents are associated with lighter (warmer) fluid, and colder patches of fluid are correlated with downdrafts. This sign of the buoyancy flux is traditionally associated with the formation of front-like structures \cite{mcwilliams_16}, which are observed in this flow as discussed in the next section. 

Correlations between temperature and vertical vorticity in Fig.~\ref{f:T2}(b) undergo instead large semi-regular excursions around $\langle\omega_z \theta\rangle = 0$, on a time scale of the order of the horizontal large-scale turn-over time $T_\perp = L_\perp/U_\perp \approx 4.3$, \ADD{as will be confirmed for other runs with different \BV\ in Sec.~\ref{S:PARAM}.} As shown in the next section, these oscillations correspond to a cycle of (1) creation of front- and filament-like structures, and (2) dissipation through destabilization of these structures with creation of turbulence, as the flow continues to be fed energy through the Taylor-Green forcing. The $\langle\omega_z \theta\rangle$ correlation can also be linked to the pointwise conservation of potential vorticity as temperature fronts are pushed together by the coherent large-scale velocity, as will be observed in flow visualizations.

\subsection{Dimensionless numbers}\label{SS:PARAM}

With this data, we can now compute the dimensionless parameters for run A8, evaluated in a dynamical sense in the quasi-stationary regime; we find (see Table \ref{t:runs}):
$$\textrm{Re}\approx 40000, \  \textrm{Fr}\approx 0.03.$$
Of course, as in all direct numerical simulations, the Reynolds number is  low compared to geophysical flows but, as we shall see below, this flow is already in an efficient regime in which energy is strongly dissipated.

We can also deduce several derived parameters of interest, \ADD{averaged over the developed turbulent regime.} For example, the buoyancy Reynolds number in run A8 is ${\cal R}_B\approx 36$, and  the Taylor Reynolds number is $R_\lambda=U_\perp \lambda_\perp/\nu \approx 1700$. The Richardson number is $1/\textrm{Fr}^2 \approx 1100$, of the order of the averaged gradient Richardson number $\left<Ri_g\right>$ shown in Fig.~\ref{f:T} and in Table \ref{t:runs}. The buoyancy wavenumber and length scale are respectively $k_B\approx 8$  and $L_B\approx 0.8$ (note \ADD{that} the parallel integral scale of the flow is $L_\parallel \approx 0.65$). We also find that the Ozmidov and dissipation wave numbers can be estimated as $k_{oz}\approx 40$, $k_\eta = [\epsilon_V/\nu^3]^{1/4} \approx 650$, \ADD{with $k_\eta$ evaluated assuming Kolmogorov-like scaling, that is for ${\cal R}_B>1$, which is the case for all runs of Table \ref{t:runs}.} This gives $k_\eta/k_\textrm{max}\approx 0.9$, indicating that the dissipative range in this flow is reasonably resolved.

The kinetic energy dissipation rate measured directly from the run A8 is $\epsilon_V \approx 0.24$, while the potential energy dissipation rate is $\epsilon_P \approx 0.09\approx \epsilon_V/3$. Using Kolmogorov phenomenology, the kinetic energy dissipation rate can be  estimated as
\be \epsilon_V^\textrm{Kol}=U_\perp^3/L_\perp \approx 0.23 \ . \label{eq:kolm} \ee
This is quite comparable to the actually measured kinetic energy dissipation, indicative of a strongly turbulent flow which is efficient at dissipating all the available energy at small scales (see Table \ref{t:runs} for more details). \ADD{The level of turbulent dissipation observed here is strong for the $\textrm{Fr}$ and ${\cal R}_B$ considered in this run; typically, direct numerical simulations of stably stratified turbulence have $\epsilon_V$ smaller than $\epsilon_V^\textrm{Kol}$ up to ${\cal R}_B\approx 200$ \cite{pouquet_17j}. Strong effective dissipation rates have already been reported in the literature for flows at low Froude number: using hyper-viscosity (with a higher power in the vertical than in the horizontal, since stronger gradients are expected in the vertical), it was shown in \cite{lindborg_06} that $r_\epsilon=\epsilon_V/\epsilon_V^{Kol}\approx 1$; this is for effective high buoyancy Reynolds numbers ${\cal R}_B$, comparable to those in the ocean and in the atmosphere. This was also found using a normal Laplacian for the dissipation in \cite{pouquet_17p, pouquet_17j} for rotating stratified flows in the absence of forcing, but with $r_\epsilon\sim {\cal R}_B^{1/2}$ in the transitional regime up to ${\cal R}_B\approx 200$. Moreover, smaller dissipation rates can also be found in our runs, specially at lower values of ${\cal R}_B$ (see, e.g., run F4 in Table \ref{t:runs}), and the coefficient of proportionality in front of the scaling relationship in Eq.~(\ref{eq:kolm}) can also depend on the geometry of the flow. Thus, the values of $r_\epsilon$ observed in run A8 (as well as in other runs in Table \ref{t:runs}) indicate a very efficient mixing and dissipation of the flows at small scales at these $\textrm{Fr}$ and ${\cal R}_B$ numbers.}

\subsection{Typical dimensional values\label{sec:dimensions}}

Sub-mesoscale structures in the ocean have horizontal scales of 1-10 km. \ADD{To compare with dimensional quantities in the ocean, we must choose some set-up and dimensionalize all quantities using typical velocities and length scales. Here we do so, noting that the motivation is to see if the ordering of scales, and the orders of magnitudes, are reasonable, but keeping in mind that our geometrical set up is different from those in oceanic measurements. Any sub-mesoscale configuration with sufficient measurements can then be used to this end. We thus} consider the Kuroshio current, for which detailed measurements of turbulence and enhanced dissipation in fronts are available \cite{dasaro_11}. Mean horizontal velocities in this flow are $\approx 0.3$ m s$ ^{-1}$ \cite{Zuo_12}. Associating energy-containing structures in run A8 with these \ADD{observed} structures, dimensions can be obtained by multiplying dimensionless lengths by $L^* = 10 \, \textrm{km}/L_\perp = 10 \, \textrm{km} /4.5 \approx 2.2$ km, velocities by $U^* = 0.3 \, \textrm{m s}^{-1}/U_\perp = 0.3 \, \textrm{m s}^{-1}/1.03 \approx 0.29$ m s$ ^{-1}$, and times by $T^* = L^*/U^* \approx 7600$ s.

Based on these numbers, run A8 has a horizontal size $L_x = L_y \approx 14$ km, a vertical size $L_z \approx 1700$ m, r.m.s.~horizontal velocity $U_\perp \approx 0.3$ m s$^{-1}$, and r.m.s.~vertical velocity $U_\parallel \approx 7\times 10^{-2}$ m/s. This later value is comparable to those measured in sub-mesoscale ocean fronts \cite{dasaro_11}. The \BV\ is $N \approx 10^{-3}$ s$^{-1}$. The spatial resolution of the simulation is $\Delta x = \Delta y = \Delta z \approx 7$ m, clearly insufficient to realistically resolve the Kolmogorov dissipation scale of turbulence in the ocean. Thus, all turbulent length scales will be overestimated when compared with realistic values. However, the kinetic energy dissipation rate measured in run A8 is $\epsilon_V \approx 2.6 \times 10^{-6}$ W kg$^{-1}$. This value is directly comparable with measurements of energy dissipation rates in ocean fronts within the Kuroshio current \cite{dasaro_11}, which yield values between $1.6 \times 10^{-7}$ and $4.3 \times 10^{-6}$ W kg$^{-1}$.

\begin{figure}      
\includegraphics[width=8.7cm]{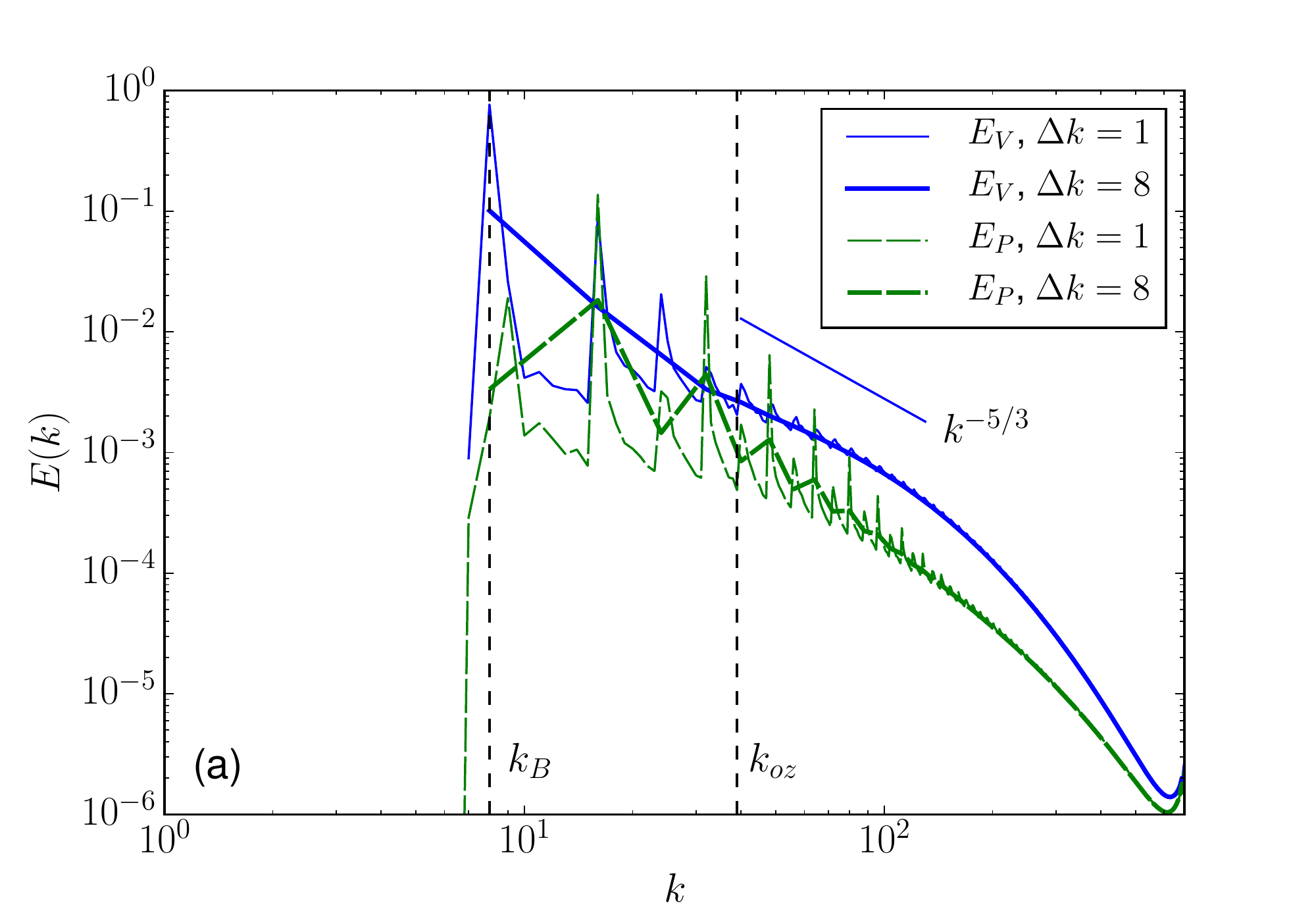} 
\includegraphics[width=8.7cm]{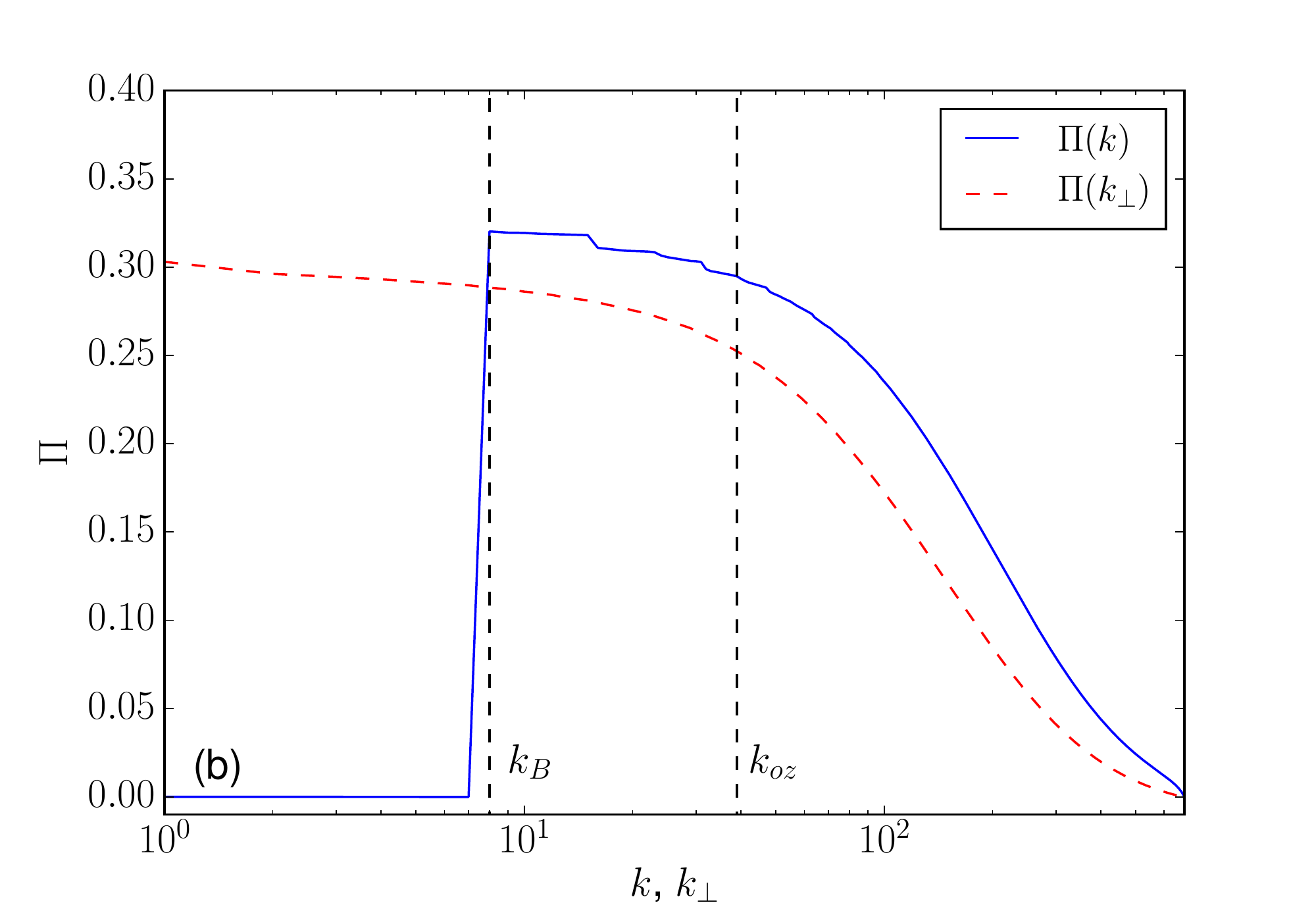}
\caption{({\it Color online}) (a) Total energy isotropic spectrum \ADD{for run A8}, separated into its kinetic ($E_V$) and potential ($E_P$) components. Thin and thick lines correspond  to different widths of the Fourier shells over which the data is averaged. The Kolmogorov slope is shown as a reference. (b) Isotropic total energy flux $\Pi(k)$, and perpendicular total energy flux $\Pi(k_\perp)$. The two vertical dashed lines indicate the buoyancy and Ozmidov wave numbers.}
\label{f:F}
\end{figure}

In the next section we will present visualizations of front- and filament-like structures in run A8. In these units, typical widths of the structures range between $700$ m and $1.5$ km. Velocity gradients in these structures are $|\partial_x u| \approx 1\times 10^{-4}$ to $3\times 10^{-4}$ s$^{-1}$. Measurements of fronts in the Kuroshio current \cite{dasaro_11} yield gradients of $\approx 10^{-5}$ s$^{-1}$. However, \ADD{as mentioned above}, it is important to note that the formation of structures in our simulation is driven by the Taylor-Green flow, and important ingredients for ocean modeling such as surface winds and the boundary layer are missing. \ADD{Thus, the dimensional values are only considered to give a better idea of scale separation and of typical strengths when comparing with turbulence in more realistic set ups.}

\ADD{\section{Spatial structures for run A8}} \label{S:SS}
 
The dynamics of the flows computed in this work is rather classical in terms of temporal evolution, except perhaps for the cyclic behavior in $\langle \omega_z \theta \rangle$ \ADD{observed in run A8 (and also observed in the other runs, as discussed later in Sec.~\ref{S:PARAM}).} But to what type of structures does such behavior correspond to?

We show in Fig.~\ref{f:V} vertical cuts of the temperature fluctuations at three different times for run A8, one close to the maximum of enstrophy, one after the maximum, and one at the latest time of the computation. Black lines correspond to instantaneous velocity field lines. Examining in Fig.~\ref{f:V}(a) the field at the earlier time in the vicinity of ($x\approx \pi/2, \ z\approx 0.25$), we observe hot fluid which is ascending to the left of the front, and to the right cold fluid which is descending. Hot fluid in this region is also pushed to the right, and cold fluid to the left. At this time the symmetries of the TGz8 forcing are evident, and thus one can observe three other such contrasting frontal configurations (one between each von K\'arm\'an cell, i.e., all lying in the vicinity of the shear layer; compare the flow with the sketch in the bottom of Fig.~\ref{f:TG}). This structure is reminiscent of a classical front, as  proposed in \cite{hoskins_72}. In all these fronts, hot and cold fluid elements are being pushed against each other by the flow, forming ever sharper fronts. As turbulence has not fully developed yet, the formation of the front-like structure is only arrested by the dissipation wavelength.

Figure \ref{f:CT} shows horizontal cuts of temperature fluctuations and instantaneous velocity field lines in two ($x,y$) planes at $z=L_z/8 \approx 0.1$ and at $z=L_z/4 \approx 0.2$ (at the shear layer, where the TGz8 forcing is zero), and at two times $t=4.8$ (before the maximum of enstrophy) and $t=8.2$ (once turbulence has fully developed), also for run A8. As in Fig.~\ref{f:V}, the large-scale flow generated by the TGz8 forcing is clearly seen. At early times in the \ADD{von} K\'arm\'an cell, see Fig.~\ref{f:CT}(a), cold fluid concentrates (and descends) in the center of each TG vortex, and hot fluid concentrates (and ascends) in the surroundings of the vortices. In the shear layer, see \ref{f:CT}(b), the hot and cold fluid meet in regions in which hot and cold are pushed against each other, creating sharp fronts as the one indicated by the white box; note that, as a result of the symmetries of the forcing, many other front-like structures can be be seen.

In the presence of gravity, the recirculation associated with the TG flow is quite different from the homogeneous and isotropic case with, unsurprisingly, descending cold and rising hot fluid parcels. This unbalanced secondary circulation, in the form of vertical motions, is  created first by pressure gradients. This can be seen by a Taylor expansion of the flow at early times in terms of a small time $dt$, by solving Eqs.~(\ref{eq:mom}) and (\ref{eq:temp}) iteratively as done in \cite{taylor_37}. To the lowest non-zero order, the velocity components and temperature fluctuations for TGz8 forcing are
\begin{eqnarray}
u &=& dt F_0 \sin(x) \cos(y) \cos(8z) , \\
v &=& -dt F_0 \cos(x) \sin(y) \cos(8z) , \\
w &=& -2 dt^2 F_0 [\cos(2x) + \cos(2y)] \sin(16z) , \\
\theta &=& -2 dt^3 N F_0 [\cos(2x) + \cos(2y)] \sin(16z) \label{eq:parity} ,
\end{eqnarray}
where viscous contributions were neglected. Note the flow is two-dimensional to the lowest order (only $u$ and $v$ are different from zero at order $dt$), but vertical displacements at twice the wave numbers of the large-scale flow arise from the pressure gradient term $-\partial_zp$ in Eq.~(\ref{eq:mom}) at order $dt^2$. This creates a circulation with vertical displacements in each K\'arm\'an cell, which in turn excites temperature fluctuations at order $dt^3$ and with the periodicity of the large-scale pattern seen in Fig.~\ref{f:CT}. Furthermore, potential vorticity $P_V$ is created by the forcing but it must be conserved pointwise by the equations. This implies immediately that density gradients that are quasi-aligned with the vorticity will be counter-balanced by vertical vorticity, and that the two may be correlated, see Fig.~\ref{f:T2}(b) (although it has also been argued by some authors that the formation of fronts and filaments can be an indication of non-conservation of $P_V$ \cite{mcwilliams_16}). Once the descending cold elements and rising hot elements are excited, frontogenesis can arise from the instability of the buoyancy field when submitted to a large scale horizontal shear \cite{hoskins_72, hoskins_82, molemaker_10b}, as observed in the atmosphere and the ocean and also found in direct numerical simulations (see, e.g., \cite{kimura_12, debruynkops_15}).

\begin{figure}
\includegraphics[width=8.7cm]{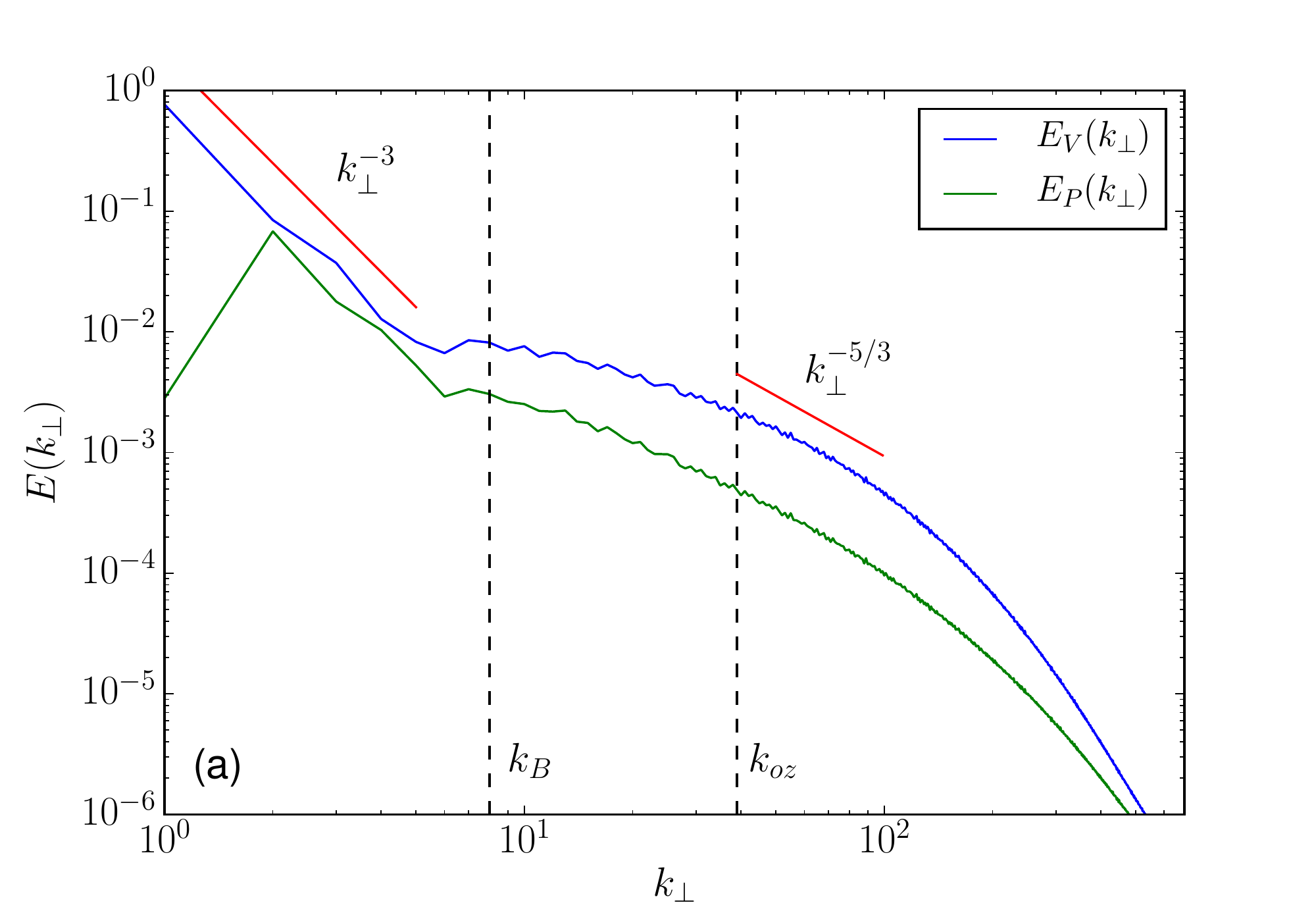} 
\includegraphics[width=8.7cm]{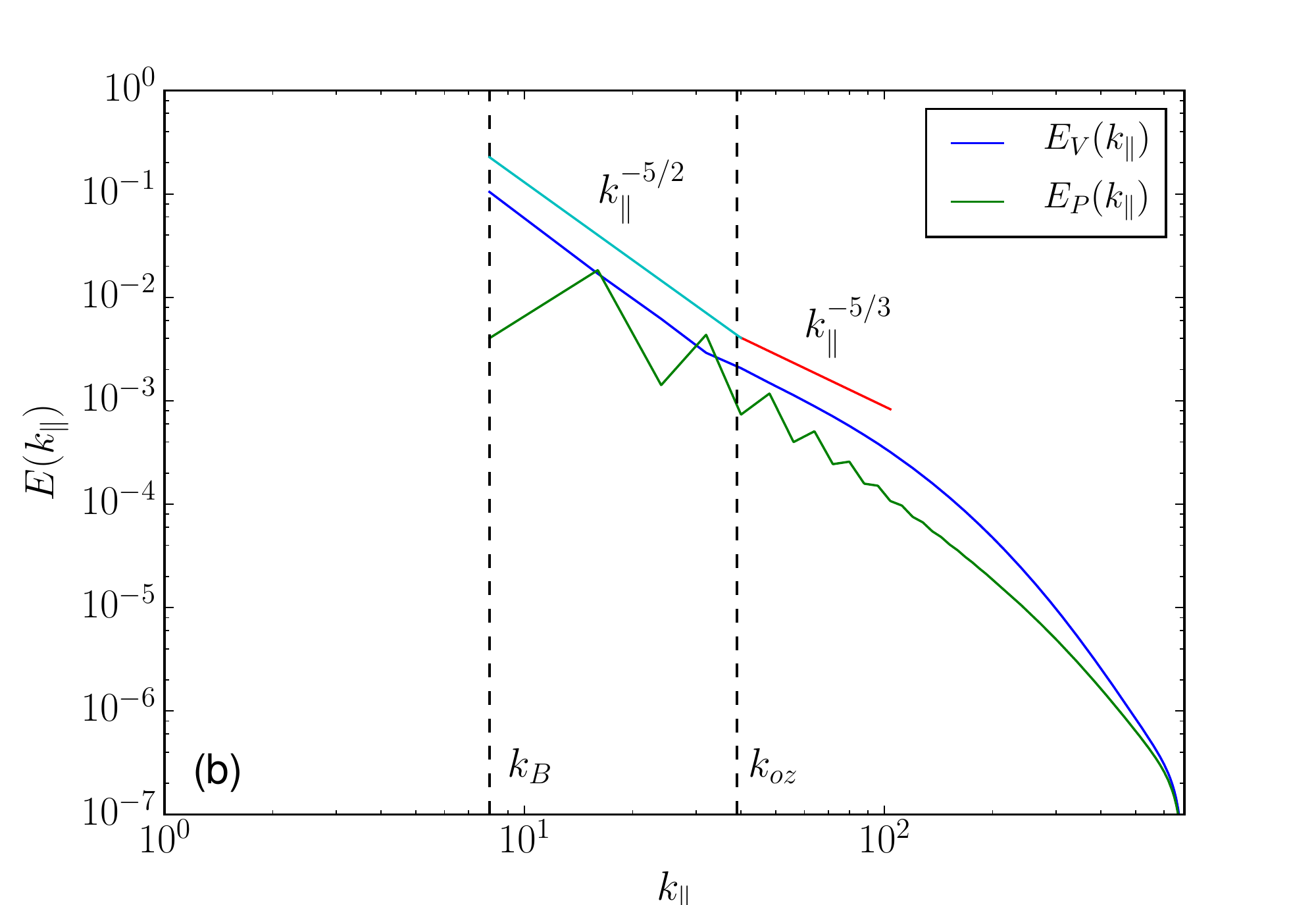}
\caption{Reduced spectra of kinetic ($E_V$) and potential ($E_P$) energy, defined in Eq.~(\ref{etheta}), \ADD{for run A8} as a function of (a) perpendicular and of (b) parallel wavenumbers. The two vertical dashed lines indicate the buoyancy and Ozmidov wave numbers. Several power laws are given as references.}
\label{f:SS}
\end{figure}

The evolution of the sharp front and the subsequent creation of turbulence in run A8 is shown in Fig.~\ref{f:3DT}, where we present perspective volume renderings of the temperature and of vorticity intensity in the subdomain indicated by the white boxes in Fig.~\ref{f:CT}, using the VAPOR software \cite{clyne_07}. Three different times are shown, respectively at $t=4.3$ (before the peak of enstrophy), $t=4.8$ and $t=5.2$ (after turbulence develops). At early times the gradient is very sharp, and gradients increase further as hot fluid is pushed against cold fluid. Note also that above this structure, the large-scale flow resulting from the TGz8 forcing moves fluid along the structure (see Fig.~\ref{f:CT}). The destabilization of the sharp gradient through shear instabilities at intermediate times (note the vortex sheets in light yellow at $t=4.8$ in the bottom row), further allows the development of turbulence, with no discernible structures beyond vortex filaments and which fill the bulk of the flow. This excitation of turbulence gives a path for energy dissipation, and for the arrest of the growth of the gradient. Finally, note that, overall, the geometry of the structure in  Fig.~\ref{f:3DT} is reminiscent of one of the mechanisms for the creation of fronts depicted in \cite{mcwilliams_16}.

After turbulence develops, we see in Figs.~\ref{f:V}(b) and (c) and in Figs.~\ref{f:CT}(c) and (d) that several such front-like structures are formed, and that the flow can make them almost collide into filament-like structures \cite{mcwilliams_16}, i.e., a succession of cold-hot-cold fluid or vice-versa . This is visible in Figs.~\ref{f:V}(b) and (c) near $x\approx \pi$, with the velocity indicating convergence of the two sharp gradients. These structures are deep in the third ($y$) direction, as seen in Fig.~\ref{f:CT}(c) (look for example at $x=\pi$, $y=0$ or $y=2\pi$). Once the system reaches the fully turbulent regime, these front- and filament-like structures are cyclically created by the large-scale flow, and dissipated by small-scale turbulence, giving rise to the cyclic behavior observed in Fig.~\ref{f:T2}(b). Note also that at these late times, the symmetries of the TGz8 forcing are broken, with each von K\'arm\'an cell still discernible but showing different small-scale features.

While the width of the sharp gradient at early times (before the development of turbulence) is controlled by viscous dissipative processes, the typical width of the structures afterwards is larger and \ADD{is} determined by the large-scale flow in which they are embedded, \ADD{as will be confirmed by varying $A_r$ and the \BV\ in Sec.~\ref{S:PARAM}}. This can also be seen in Fig.~\ref{f:P}, which shows, for the sharp gradient at early times and for a cold filament-like structure at late times (both along the $y$ direction), the averaged velocity profile $\left<u\right>_y$ (where the average in $y$ is done over the extension of the structure), and the averaged temperature profile $\left<\theta \right>_y$. In the case of the sharp gradient, both the velocity and the temperature change sign rapidly. In the case of the filament-like structure, temperature drops while the velocity changes sign more smoothly.

\begin{figure}
\includegraphics[width=8.7cm,trim=0 10 0 20,clip]{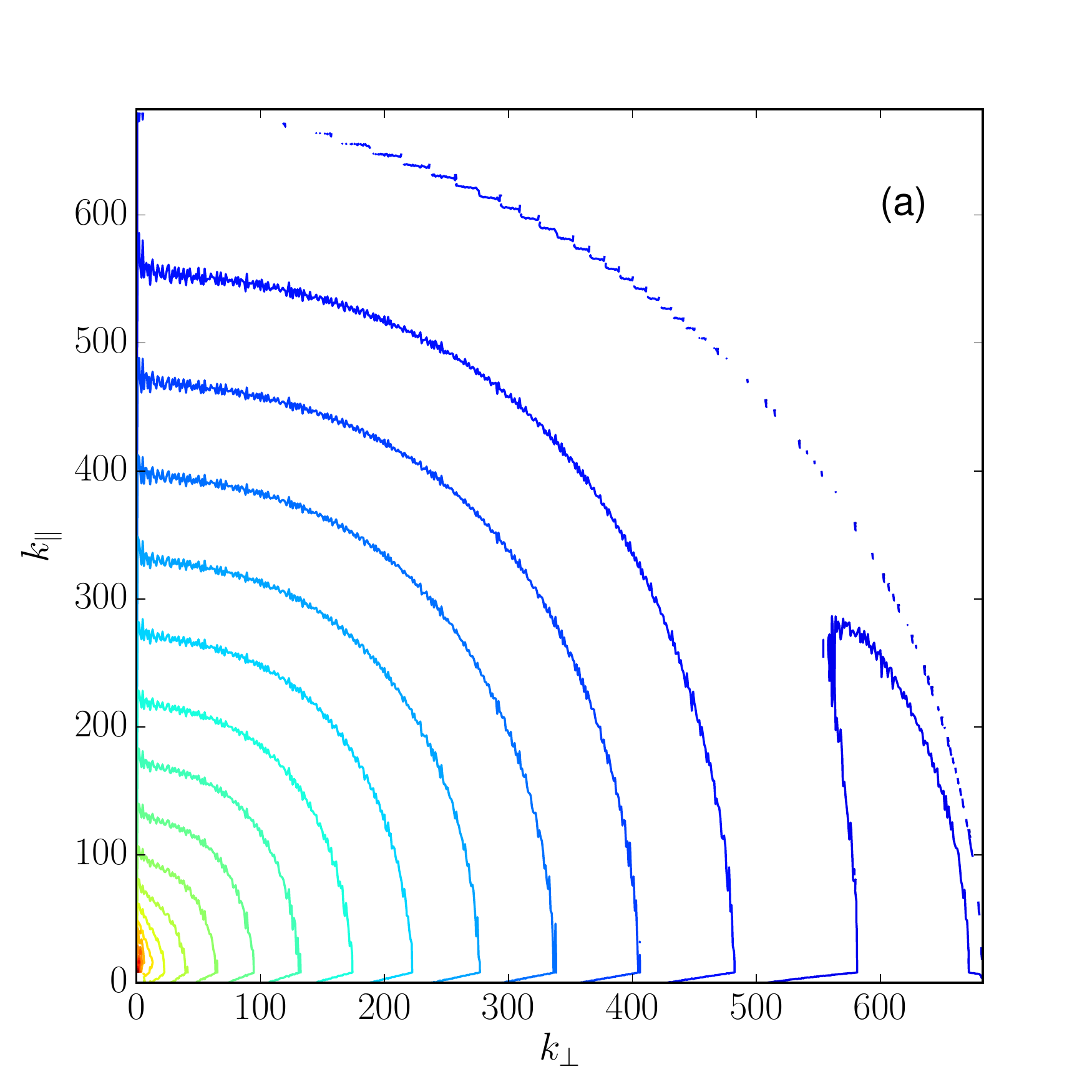} 
\includegraphics[width=8.7cm,trim=0 10 0 20,clip]{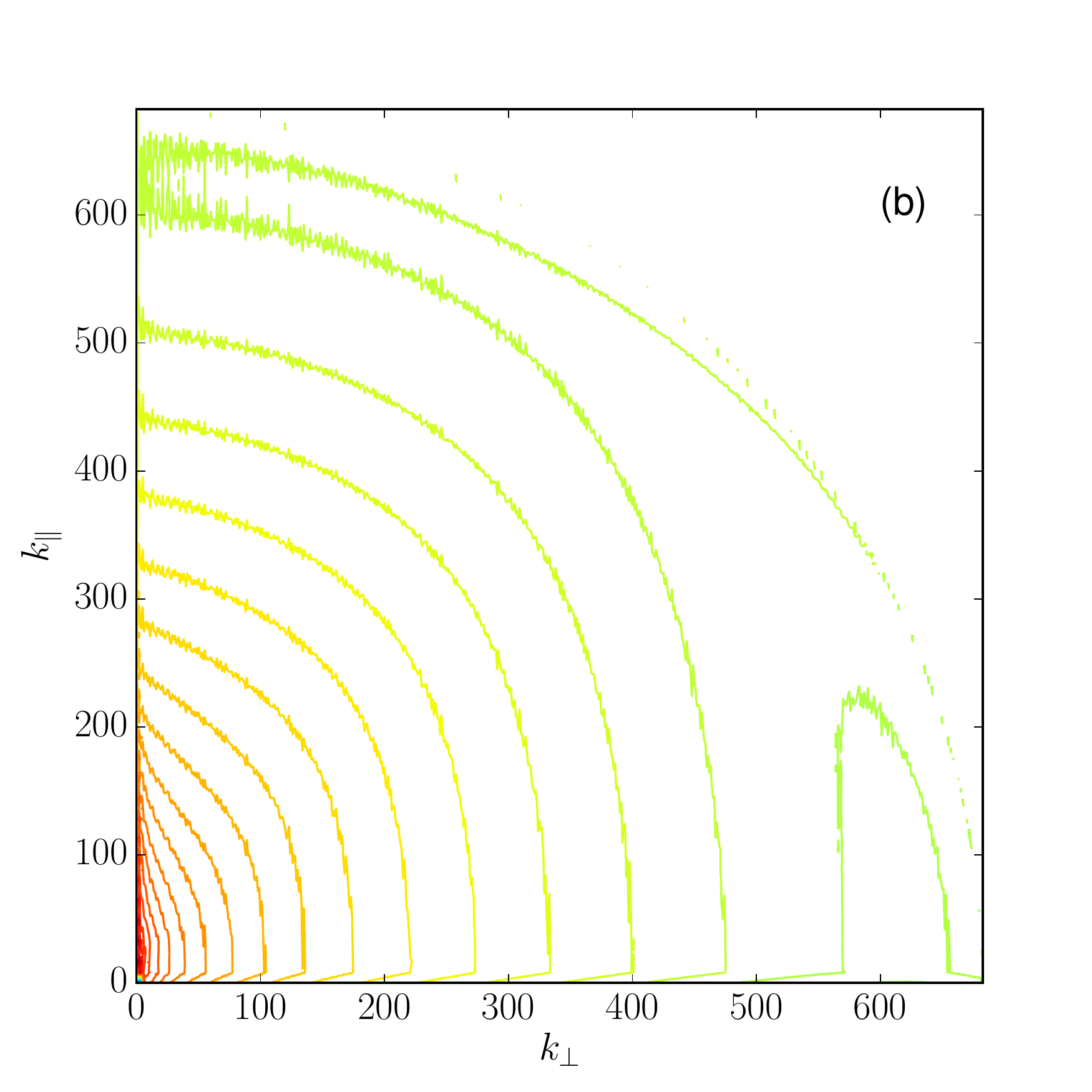}
\caption{Contour plots \ADD{for run A8} of (a) the kinetic and (b) the potential axisymmetric energy spectra. Contour levels are separated in logarithmic scale. Note contours are elongated for small wave numbers, but become closer to circles for large wave numbers, indicative of a recovery of isotropy at small scale.}
\label{f:2D}
\end{figure}

\ADD{\section{Spectral behavior for run A8} \label{S:FOURIER}}

We finally examine the properties of run A8 in Fourier space, to study the turbulence that develops after the peak of enstrophy. Thus, all spectra shown in this section correspond to averages in time over the turbulent steady state (i.e., here, from $t\approx 7$ to 15). Because the discretization in the horizontal and vertical directions is not the same in terms of wave numbers, we display first in Fig.~\ref{f:F}(a) the isotropic energy spectrum (for the kinetic and potential energy) as defined in Eq.~(\ref{eq:iso}). In the discrete case the integral is replaced by a sum, and all modes in spherical shells of width $\Delta k$ are summed up. We show these spectra using two summations over Fourier shells, of respective width $\Delta k=1$ (thin lines) and $\Delta k=8$ (thick lines). The oscillations for the spectra with $\Delta k=1$ are due to the fact that all shells with $k_\parallel \mod{8} \neq 0$ are depleted because of the aspect ratio of the box and of the different density of modes in parallel and perpendicular directions in Fourier space.

The intensity of the peaks for $\Delta k=1$ is quite strong, although they become much less visible  as we move on to higher wave numbers, since the density of wave numbers in shells with high $k$ is substantially larger \ADD{($\sim k^2$).} However, these peaks are also related to, on the one hand, the lack of resonances in the elongated box, so that the energy accumulates at the large-scale flow, and on the other hand, the growth in stratified flows of the zero mode at $k_\perp=0$ leading to strong mean flows \cite{smith_02, rorai_15} that are known to dominate the dynamics when one performs for example a measurement of the wave dispersion as done in \cite{clark_17}. Smoothing out these harmonics (e.g., for $\Delta k=8$), a slope is discernible; it is steeper at large scales, and compatible with a Kolmogorov spectrum at wave numbers larger than the Ozmidov wave number.

\begin{figure}
\includegraphics[width=8.7cm,trim=10 10 5 20,clip]{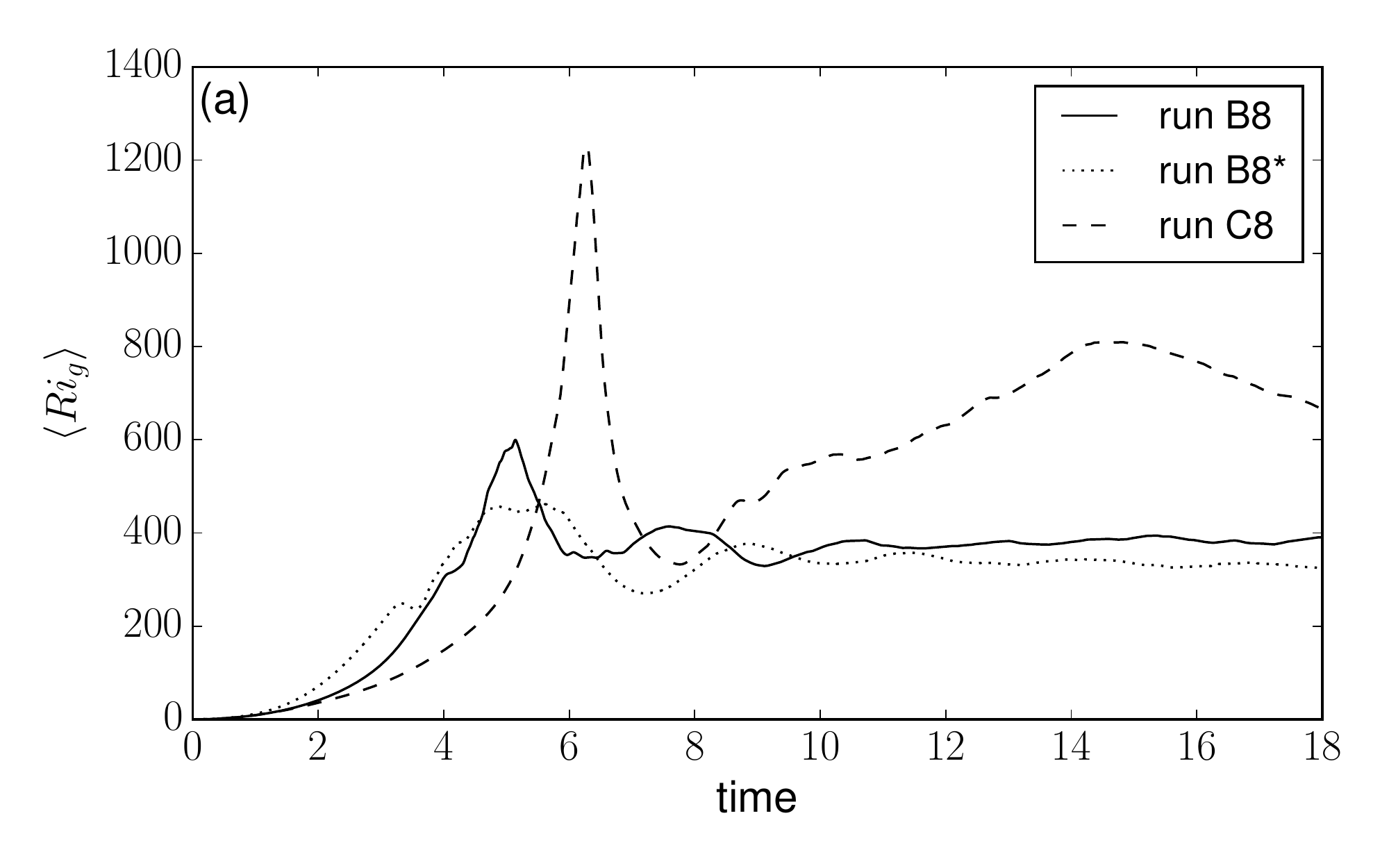}
\includegraphics[width=8.7cm,trim=0 10 0 20,clip]{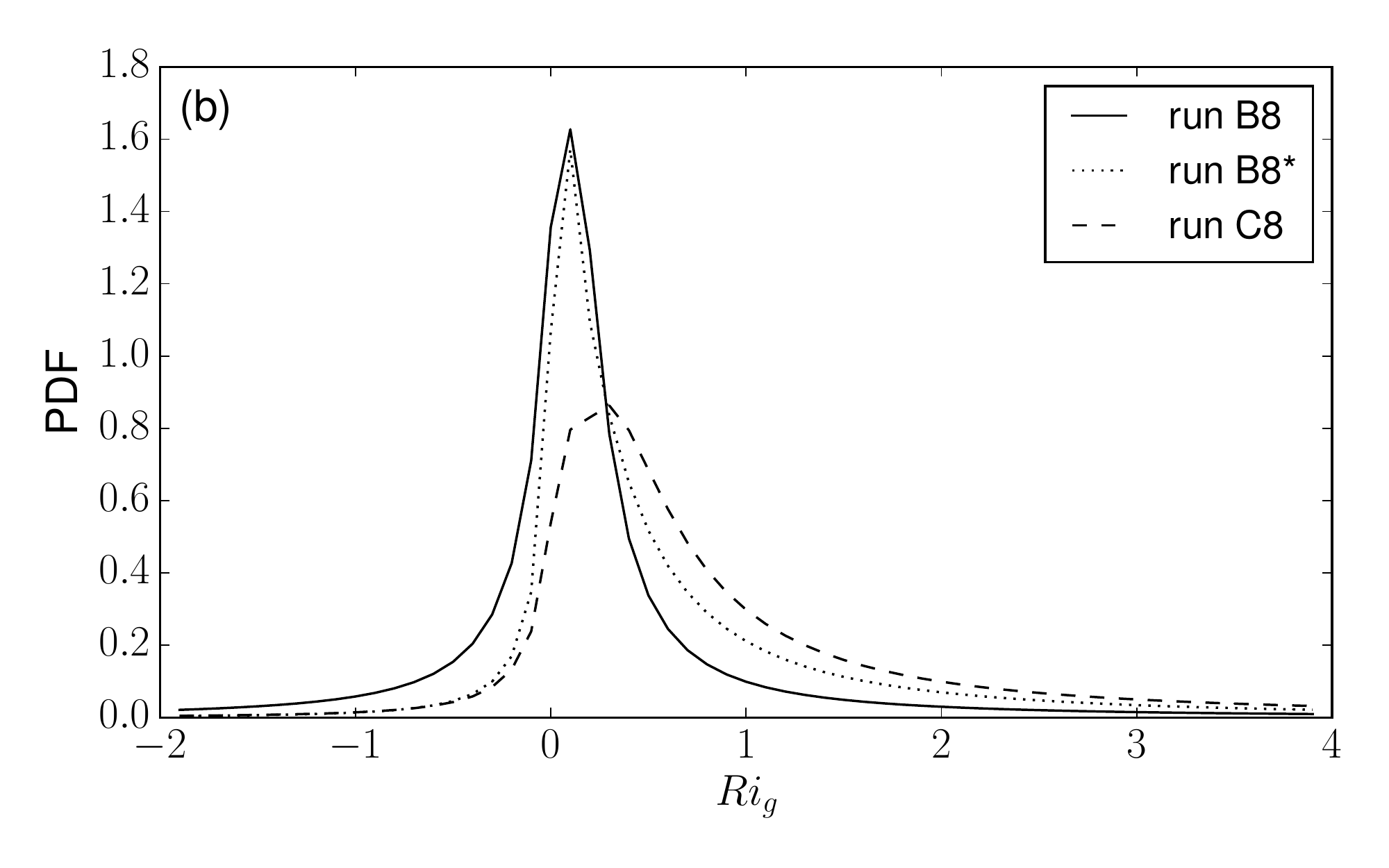}
\caption{\ADD{(a) Spatially averaged gradient Richardson number $\left<Ri_g\right>$ (over the entire domain) as a function of time for the set of runs with aspect ratio $A_r = 1/8$: runs B8, ${\bf B8^\ast}$, and C8 (see Table \ref{t:runs}). (b) PDFs of the local gradient Richardson number $Ri_g$ for the same runs during the developed turbulent regime.}} 
\label{f:rig}
\end{figure}

Even with $\Delta k=8$, an oscillation remains in the spectrum of the potential energy; the first peak is at $k=16$, the next peak at $k=2\cdot 16$, and the next peaks have period $16$. It corresponds to the spatial frequency of the dominant temperature fluctuations being created by the flow circulation. This is compatible with the argument derived in Eq.~(\ref{eq:parity}), where we showed that to the lowest order in a short-time expansion, the TG flow acting at $k_x=k_y=1$ and $k_z=8$ excites temperature fluctuations at $k_x=k_y=2$ and $k_z=16$ (corresponding to the isotropic wave number $k = [16^2+2 \cdot 2^2]^{1/2} \approx 16.2$).

The total energy flux $\Pi$ for run A8 is shown in Fig.~\ref{f:F}(b). Just as was done for the energy in Eqs.~(\ref{eq:iso})-(\ref{eq:per}), by integrating over isotropic wave numbers or perpendicular wave numbers, we can obtain the reduced isotropic energy flux $\Pi(k)$, or the reduced perpendicular energy flux $\Pi(k_\perp)$. There is no sign of inverse cascade (the isotropic energy flux is zero at large scales, and the perpendicular energy flux is positive at all wave numbers), and they are relatively constant in the inertial range. In Fig.~\ref{f:F}, the two vertical dashed lines always indicate the buoyancy and Ozmidov wave numbers. Their ratio is proportional to $\textrm{Fr}^{-1/2}\approx 5.8$. Note that $L_B$ is barely resolved in the computation, given the choice of parameters.

The kinetic and potential energy spectra for run A8, now reduced in terms of the perpendicular and parallel wave numbers as defined by Eqs.~(\ref{eq:para}) and (\ref{eq:per}), are given in Fig.~\ref{f:SS}. Data points are denser in terms of $k_\perp$, and the behavior of the spectrum at large scales is more visible here. All modes with $k_\perp < 8$ have no contribution from $k_z$ except for the mode $k_z=0$. The large-scale kinetic energy spectrum follows a $k_\perp^{-3}$ law, in agreement with a large-scale flow close to balance \ADD{for these small Froude numbers,} and at scales smaller than the Ozmidov scale a Kolmogorov spectrum, $\sim k_\perp^{-5/3}$ is plausible although not well resolved. Between the buoyancy and the Ozmidov scale, the spectrum is shallower. Moreover, in the spectra in terms of $k_\parallel$ the break at $k\sim k_B$ is more clear, with at large scales an approximate law $\sim k_\parallel^{-5/2}$ shown as a reference.

\begin{figure}
\includegraphics[width=8.7cm]{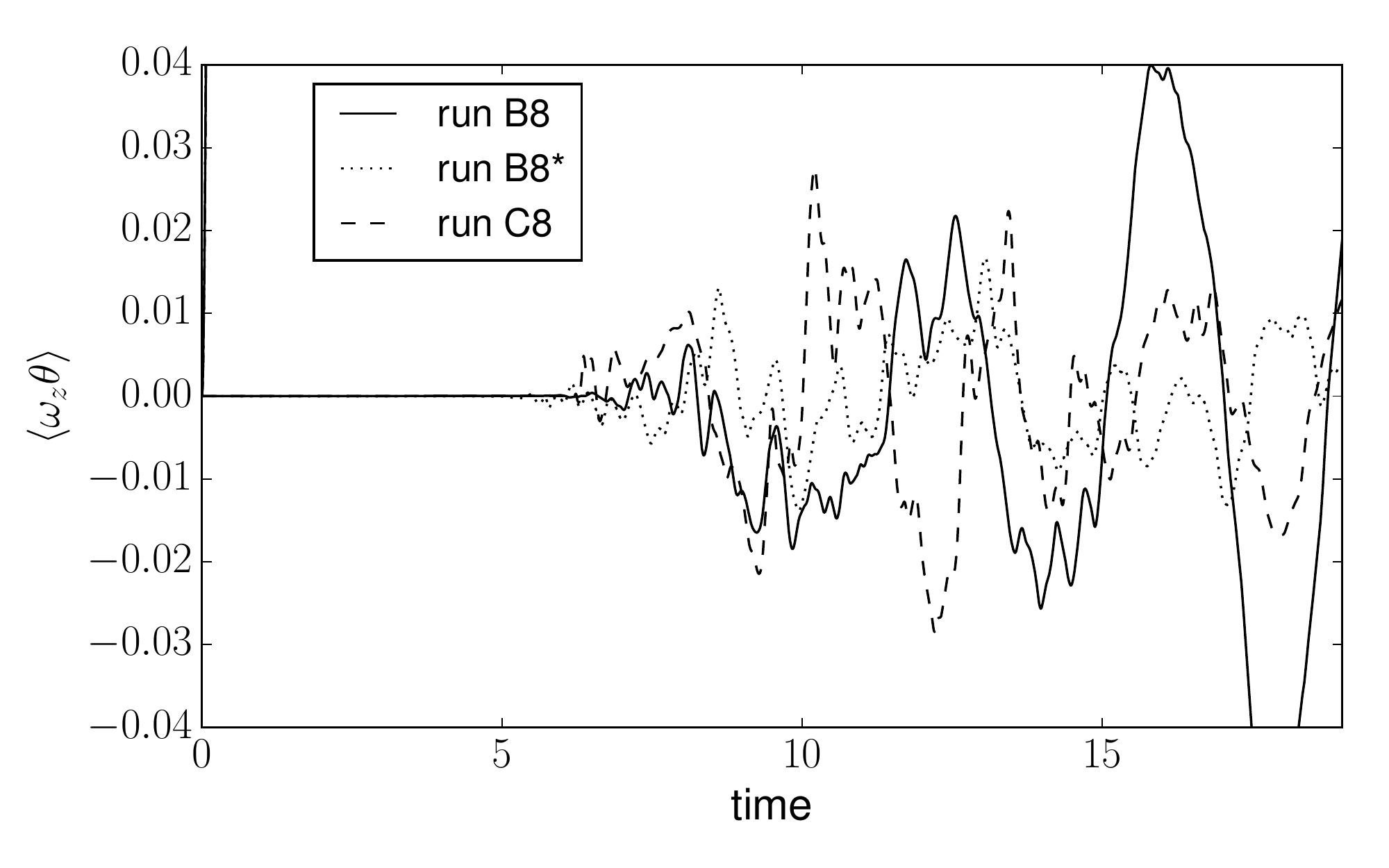}
\caption{\ADD{Mean correlation between vertical vorticity and temperature fluctuations, $\langle \omega_z \theta \rangle$, as a function of time and for runs with $A_r=1/8$: B8 (with $N=8$, solid line), ${\bf B8^\ast}$ (with $N=8$ and thermal forcing, dotted line), and C8 (with $N=16$, dashed line). Note that the large oscillations are significantly slower than the period of internal gravity waves, and in fact, independent of $N$. Faster fluctuations, associated with the internal gravity waves, can also be seen.}}
\label{f:WCO}
\end{figure}

Although reduced parallel and perpendicular spectra give some information of the anisotropy (or isotropy) of the flow, a better quantification of the scale-by-scale anisotropy can be obtained from the axisymmetric spectrum as defined in Eq.~(\ref{etheta}). In Fig.~\ref{f:2D} we thus show the two-dimensional axisymmetric spectrum as a function of $k_\parallel$ and $k_\perp$ for the kinetic and potential energy in run A8. Ellipsoidal at large scales, and with strong accumulation of energy in modes with $k_\perp \approx 0$ and $k_\parallel \gg k_\perp$ (i.e., on modes corresponding to strong vertical shear), the isocontours of these angular spectra tend to be more spherical as the wave number increases, indicating that small scales are more isotropic. However, note this trend is slower for $E_P$, suggesting that the isotropization process at small scales is slower for the temperature fluctuations.

Overall, all these spectra, which are consistent with spectra reported for stably stratified turbulence in previous studies, and which become more isotropic and closer to Kolmogorov scaling at small scales, confirm the generation of strong turbulence by the fronts in the flow.

\ADD{\section{Variation of aspect ratio and of the \BV} \label{S:PARAM}}

\begin{figure}
\includegraphics[width=8.7cm,clip]{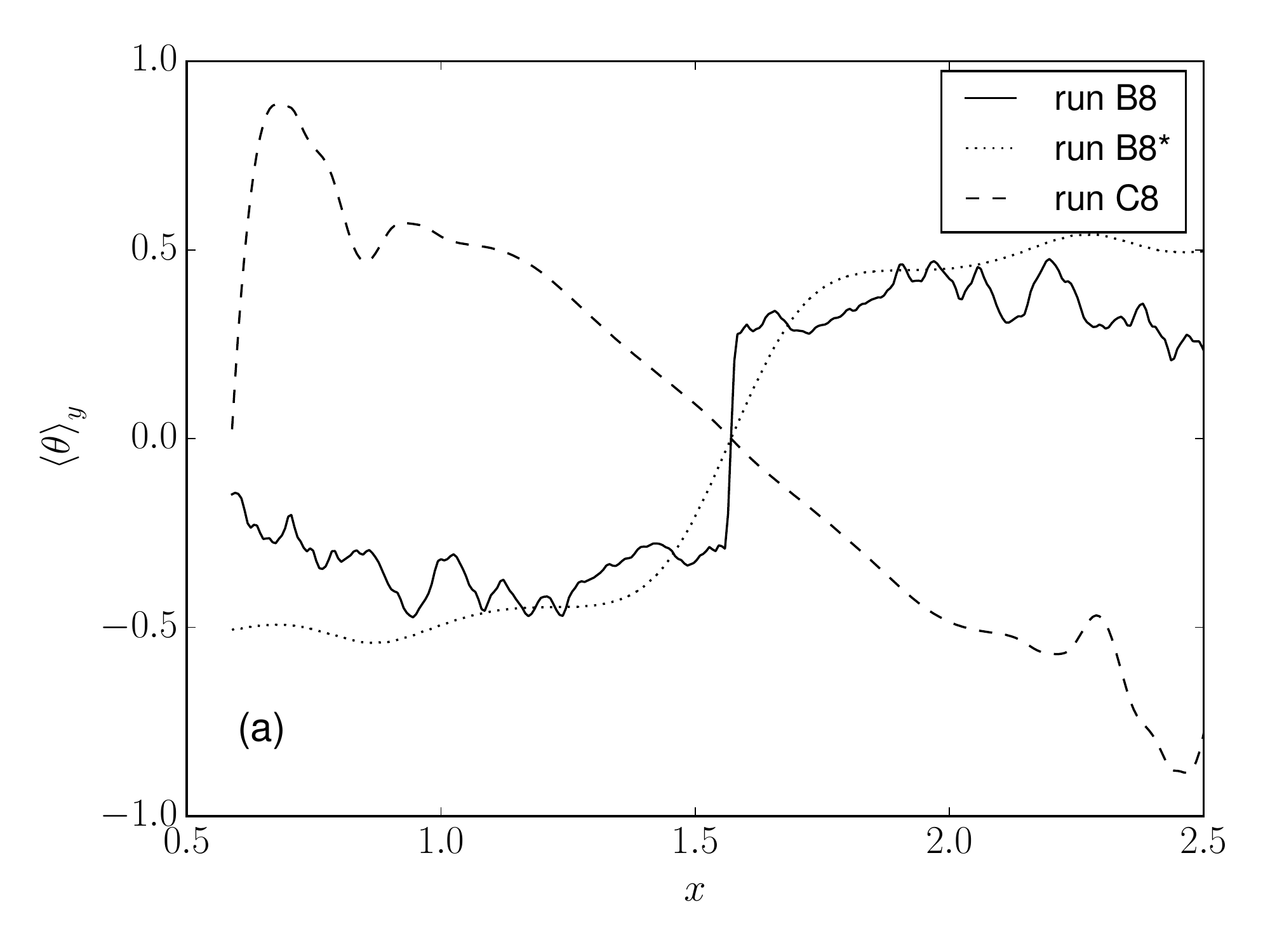} 
\includegraphics[width=8.7cm,clip]{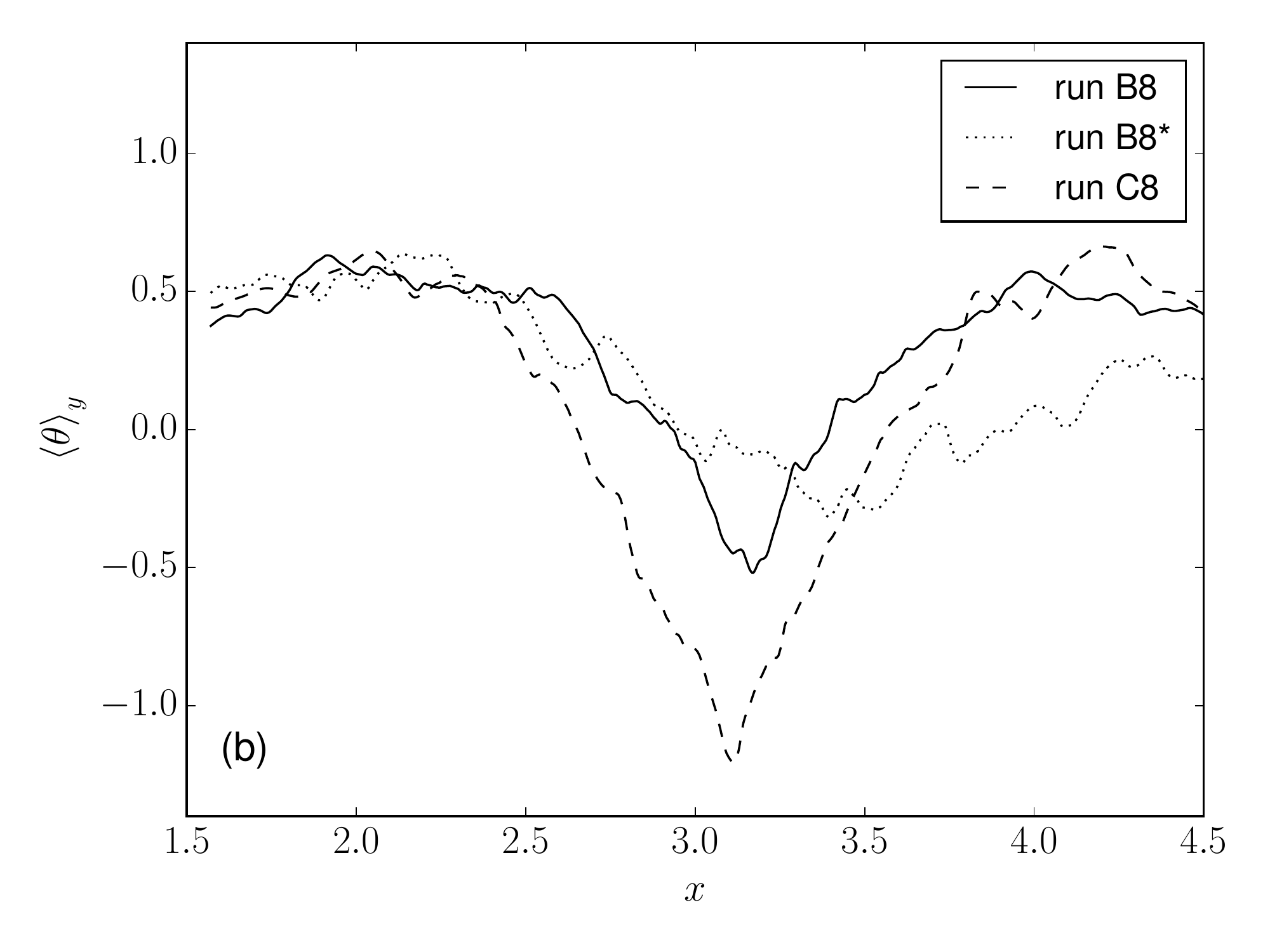} 
\caption{\ADD{(a) Instantaneous temperature profiles, averaged in the horizontal ($y$) direction in the vicinity of a front-like structure at early time ($t=5.0$) in runs B8, ${\bf B8^\ast}$, and C8 (solid, dotted and dashed lines respectively). (b) Same as in (a) for a cold filament-like structure at a later time. Compare with Fig.~\ref{f:P} which shows the temperature profiles in run A8.}}
\label{f:efr}
\end{figure}

\ADD{We now examine how the results described in the preceding sections are affected by a change in the \BV, and thus in the Froude number, as well as by a change in the aspect ratio of the box (which also results in a change in the vertical scale of the shear). Six simulations were done for this parametric study (see Table \ref{t:runs}): A first set of three runs comprises runs B8, ${\bf B8^\ast}$, and C8, which have the same aspect ratio as run A8 (i.e., $A_r=1/8$) but with a spatial resolution of $1024^2\times128$ grid points. Except for the change in resolution (and thus of viscosity and diffusivity, and thus different $Re$ and ${\cal R}_B$), run B8 is identical to run A8. Run C8 has the same TGz8 forcing with $N=16$, and thus, the buoyancy length scale in this run is approximately half that in runs A8 and B8. Finally, run ${\bf B8^\ast}$ has the same parameters as run B8 but is forced also in the temperature, with a thermal source that opposes in sign the thermal fluctuations induced by the TG forcing (see Eq.~\ref{eq:parity}), in an attempt to generate a more balanced flow; this is intended to be contrasted with the other runs. In the second set, runs D4, E4, and F4 have an aspect ratio $A_r=1/4$ (and thus are forced with a TGz4 forcing), and a resolution of $768^2\times192$ grid points. In this set of runs the \BV\ is varied from $N=4$ to $16$, and as a result the buoyancy length scale varies from the box height (which is also the forcing scale) to less than $1/3$ of the box height. Note that in all runs the \BV\ is always chosen to have the buoyancy length scale equal or smaller than the box height; in other runs, we do not perform simulations with $N$ smaller than the minimum required to have $L_B \le L_z$.}

\subsection{\ADD{Variation of the \BV\ at fixed aspect ratio}}

\begin{figure}
\includegraphics[width=8.7cm]{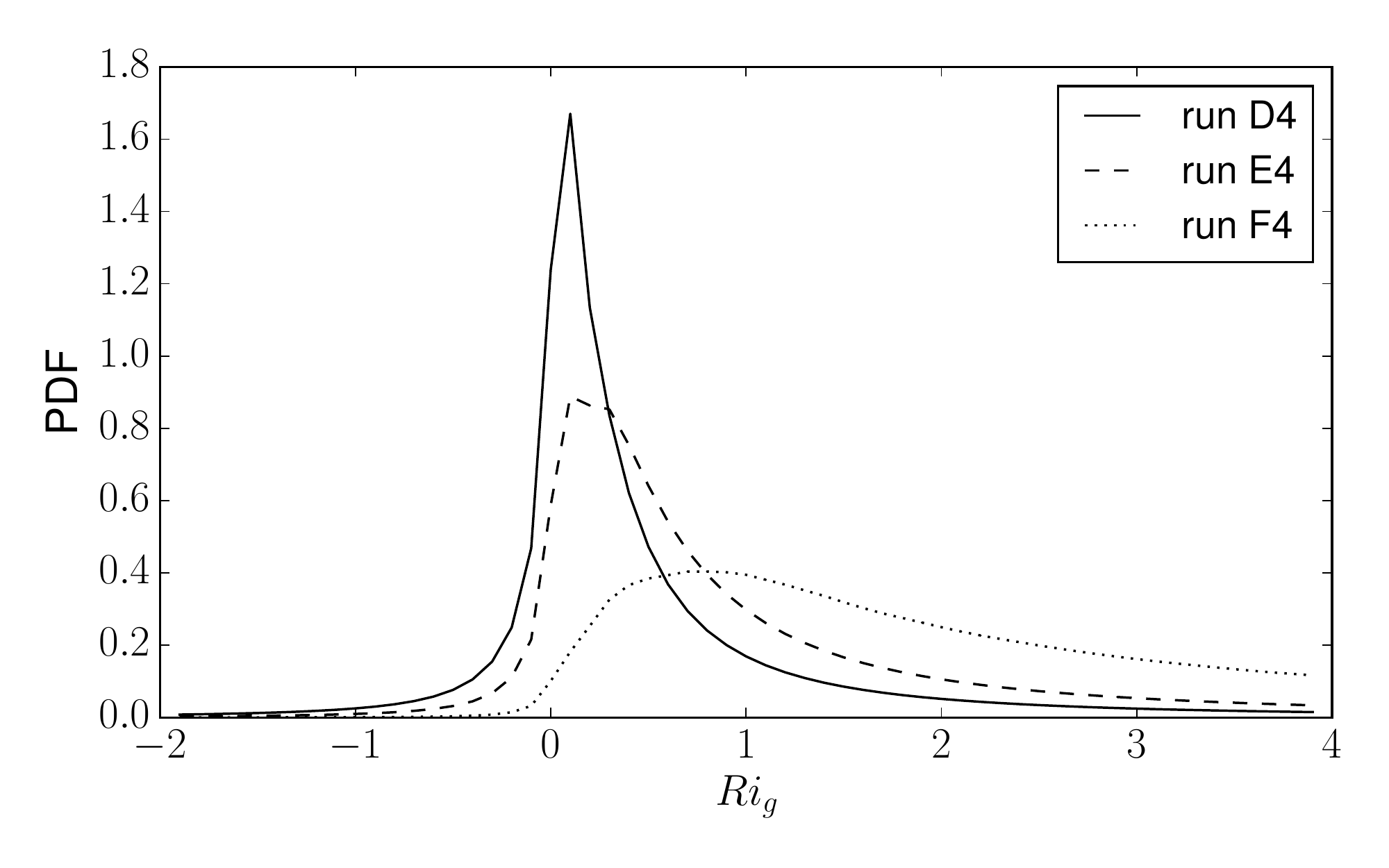} 
\caption{\ADD{PDFs of the local gradient Richardson number $Ri_g$ for runs D4, E4, and F4, with aspect ratio $A_r=1/4$ and respectively with $N=4$, $8$, and $16$.}}
\label{f:ri4}
\end{figure}

\ADD{We start discussing the runs with the same aspect ratio as run A8, i.e., with $A_r$ of =$1:8$. In Fig.~\ref{f:rig} we show the spatially averaged gradient Richardson number $\left< Ri_g \right>$ for runs B8, ${\bf B8^\ast}$, and C8. Both runs with $N=8$ (B8 and ${\bf B8^\ast}$) show the same behavior (qualitatively similar to run A8), while for run C8 (with $N=16$) the averaged gradient Richardson number keeps increasing for longer times and saturates at a larger value. Figure \ref{f:rig} also shows the PDFs of the local gradient Richardson number for the same runs. The thermal forcing, intended to keep run ${\bf B8^\ast}$ more balanced, reduces the probability of finding points with $Ri_g<0$ when compared with run B8 (i.e., the probability of overturning is reduced). However, the peak and the probability near $Ri_g\approx 1/4$ (associated with the upper threshold for local shear instabilities) remain relatively unchanged. Thus, overall, the runs studied in this work have points that can suffer local shear or overturning instabilities. In comparison, run C8 (with stronger stratification) has significantly lower probabilities of having points with $Ri_g<1/4$ or $Ri_g<0$.}

\ADD{Figure \ref{f:WCO} shows the correlation between vertical vorticity and temperature fluctuations for runs  B8, ${\bf B8^\ast}$, and C8. Runs B8 and C8, with different \BV, display slow fluctuations (similar to those in run A8) on a time scale that is much larger than (and seems independent of) $1/N$, and which is proportional to the eddy turnover time in each run. Superimposed to these slow oscillations, fast fluctuations at a time scale inversely proportional to the buoyancy frequency can be also observed. However, run ${\bf B8^\ast}$, which has a forcing intended to counteract temperature fluctuations, display much smaller slow oscillations, and thus the fast wave fluctuations are more prominent in the dynamics of $\left< \omega_z \theta \right>$.}

\ADD{In Fig.~\ref{f:efr}(a) we show the average temperature profile $\left< \theta \right>_{y}$ (averaged over $y$ in the vicinity of the structure) at early time for runs B8, ${\bf B8^\ast}$, and C8, and which can be directly compared with Fig.~\ref{f:P}(a) (for clarity, only the temperature is shown in this case). Run B8 displays a clear sharp front similar to the one observed in run A8, while run ${\bf B8^\ast}$ displays a smoother gradient, as expected since the thermal forcing opposes the temperature profile excited by the mechanical TG forcing. However, a gradient in the temperature is still visible. Run C8 displays a different behavior, with no front. The front at early times develops when the buoyancy length scale and the length scale of large-scale shear (associated with the mechanical forcing) are comparable, a result which is also confirmed by runs D4, E4, and F4 with a different aspect ratio (see below). Figure ~\ref{f:efr}(b) shows again the temperature profile $\left< \theta \right>_{y}$ but now at later times and in the region where filament-like structures develop. The width of the temperature drop is controlled by the large-scale forcing, but it has a dependence on the \BV: for larger values of $N$ the drop in the temperature becomes wider. Also, in run ${\bf B8^\ast}$ the drop is smoother, as expected for the choice of the thermal forcing that tries to balance the TG forcing.}

\begin{figure}
\includegraphics[width=8.7cm]{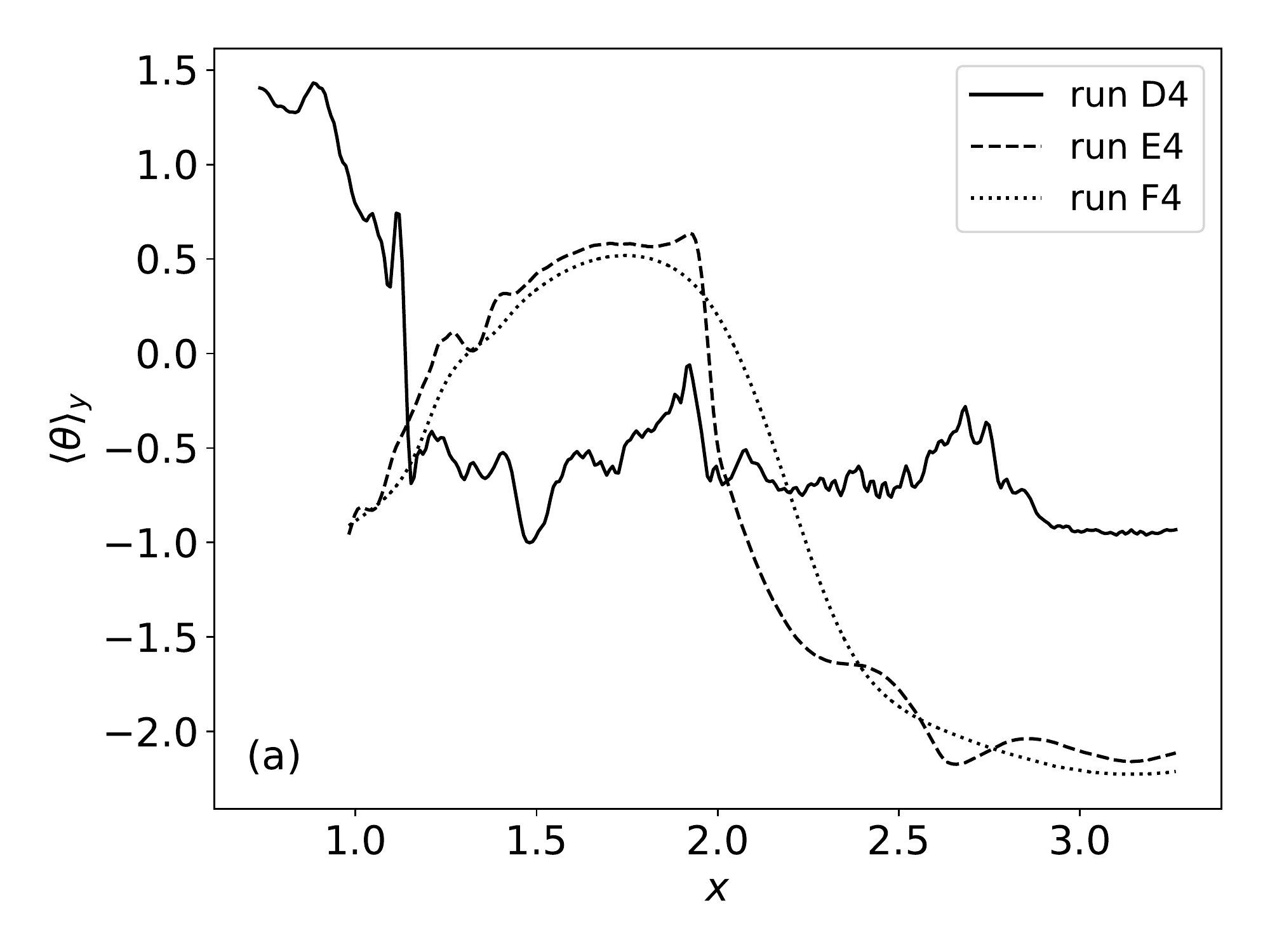}   \includegraphics[width=8.7cm]{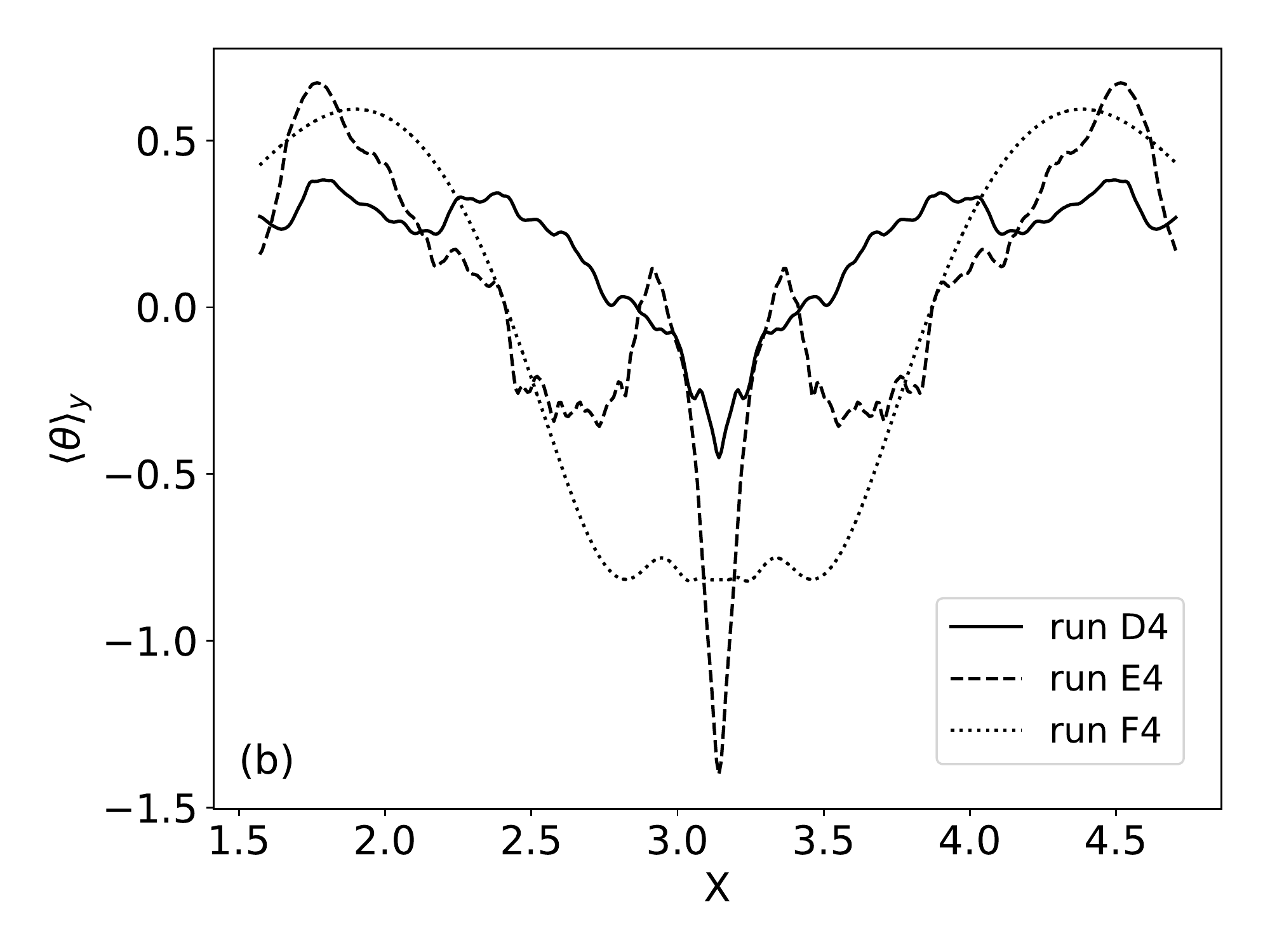} 
\caption{\ADD{Instantaneous temperature profiles, averaged in the horizontal ($y$) direction in the vicinity of the structure, for the filament-like features observed in the simulations (a) at early and (b) at late times, for Runs D4, E4 and F4 (see Table \ref{t:runs}).}}
\label{f:temp4} \end{figure}

\subsection{\ADD{Variation of the aspect ratio}}

\ADD{We finally briefly discuss runs D4, E4, and F4, which have aspect ratio of $1:4$ with mechanical TGz4 forcing, and without thermal forcing. In these runs, the \BV\ is $N=4$, $8$, and $16$ respectively. The overall behavior of run D4 is similar to that of runs A8 and B8. Figure \ref{f:ri4} shows the PDFs of the local gradient Richardson number $Ri_g$ for all these runs. Run D4 displays non-negligible probabilities of points with $Ri_g<1/4$ and $Ri_g<0$ (with a sharp peak of the PDF in between these two values), while run E4 and C4 have lower probabilities of local shear instabilities and overturning as the \BV\ is increased (compare the PDFs with those in Fig.~\ref{f:rig} for runs with aspect ratio of $1:8$).}

\ADD{The sharp structures observed in the previous sections are also affected by the change in the aspect ratio and in the \BV. As an illustration, Fig.~\ref{f:temp4} shows the instantaneous temperature profiles averaged in $y$ in the vicinity of the structures, $\left<\theta\right>_y$, for the fronts at early times in Fig.~\ref{f:temp4}(a), and for the filament-like features observed at late times in the simulations in Fig.~\ref{f:temp4}(b). In run D4 (which has $L_B \approx L_z$, as runs A8 and B8 but with a different aspect ratio), the early-time front is sharp and with length scales again given by the dissipation, although its position is shifted with respect to the other simulations, a result of the turbulent fluctuations in the flow. And as before, see runs E4 and F4 in Fig.~\ref{f:temp4}(a), increasing $N$ results in a smoother temperature gradient in the same region. At late times, see Fig.~\ref{f:temp4}(b), the drop in the temperature is clear for runs D4 and E4, and significantly smoother for run F4 (with the largest value of $N$), as also observed in the simulations with $A_r=1/8$.}

\section{Conclusion} \label{S:CONCLU}

We have shown in this paper that a classical configuration in turbulence studies, the Taylor-Green flow, suitably adapted to have an aspect ratio that mimics that of geophysical flows such as the atmosphere or the ocean (although much less extreme), is able to create front- and filament-like structures that further destabilize, for example through a local shear instability process, to produce fully developed turbulence in the bulk of the flow. Moreover, we presented evidence that, \ADD{when the shear stemming from the TG forcing and the background stratification are comparable,} the stratified TG flow develops turbulence following this procedure: the sharp gradients form first, then destabilize, and then quasi-isotropic turbulence ensues with a Kolmogorov spectrum and vortex filaments, the origin of which, however, is not the classical destabilization of a vortex sheet through a self-similar process \cite{brachet_92}, but the formation of the sharp temperature gradients that become unstable. Once turbulence is generated, \ADD{sharp gradients} and turbulence are regenerated in a cyclic behavior governed by the turnover time of the eddies at the largest available scale. One of the mechanisms behind this cycle may be the restratification of fronts observed in previous studies (see, e.g., \cite{lapeyre_06}).

The forcing configuration corresponds to a large-scale two-dimensional field, with no vertical velocity and with a modulation in the vertical that creates locally strong shear. \ADD{A case also forced in the temperature to try to counteract temperature fluctuations induced by the TG forcing, and intended to develop a more balanced configuration, was also considered.} Since the Ozmidov scale is resolved, with ${\cal R}_B\approx 36$ \ADD{for the highest Reynolds number considered (run A8), unbalanced} dynamics, quasi-isotropic turbulence, and the formation of vortex filaments can all take place at small scales. This allows for strong gradients which prevail in front- and filament-like structures to excite small scale structures, carving the road to dissipate energy through a nonlinear process, with a  direct energy cascade which is clearly observed. The sharp gradients in our simple set up, although lacking several important effects in oceanic configurations (see, e.g., \cite{dasaro_11}), give enhanced dissipation and realistic values for the energy dissipation rate, which is also of the order of dissipation rates in isotropic and homogeneous turbulence.

Of the many open issues left in the understanding of sub-mesoscale structures in stratified turbulence such as those observed in the ocean, as mentioned in the conclusion in \cite{mcwilliams_16}, a few may be addressed with the type of studies we present here. For example, \ADD{one could add rotation which is rather strong in the oceans, with $N/f\approx 5$, contrary to the case of the atmosphere for which $N/f\approx 100$, with $f=2\Omega$ twice the strength of the imposed rotation. Another example is that,} as already mentioned, the rate of energy cascade as measured for TGz8 forcing is quite close to its dimensional (Kolmogorov) evaluation. Together with the multitude of vortex tubes that are visible in the flow, this indicates that the  generation of fully developed turbulence by the strong stirring linked with large-scale vertical shear can be studied in this simplified set up. The ratio of dissipation of kinetic to potential energy is roughly 3 \ADD{for the high-resolution run A8; thus, $r_\epsilon\approx 1/4$,} comparable to what was found in \cite{rosenberg_16, rosenberg_17, pouquet_17p} for an ensemble of rotating stratified flows in the absence of forcing. Similarly, when measuring the so-called mixing efficiency, $\Gamma=B_V/\epsilon_V$, with $B_V=N\left<w\theta \right>$ the properly dimensionalised vertical heat flux, we find $\Gamma\approx 0.4$ \ADD{for run A8,} again comparable with previous studies and observations. However, a marked difference between the results presented here and those from some previous studies \cite{bartello_95, kurien_14, rorai_15, maffioli_16, rosenberg_16, pouquet_17j} is the aforementioned level of dissipation. This indicates again that the specific configuration employed here, namely that of a strong vertical shear, plays an important role in energizing the flow towards the small scales \ADD{through the formation of strong gradients and shear-induced instabilities.}

Other processes, such as the arrest in the growth of the sharp gradients, which may also be linked to enhanced turbulent dissipation, and the coupling of these structures with gravity waves and nonlinear eddies, can be considered using \ADD{these TG flows.} Finally, von K\'arm\'an cells have helicity (although the TG flow has zero net helicity), and little is known of its role in stratified flows. It has been observed that the decay of energy can be substantially slowed down in the presence of strong helicity \cite{teitelbaum_09,rorai_13}, and that its presence may be associated with flat spectra, \ADD{as} observed in the strongly stratified nocturnal planetary boundary layer \cite{rorai_15,koprov_05}. However, most of the \ADD{numerical} studies considered flows in isotropic boxes, and thus \ADD{the role of helicity} in the specific context of fluids with a \ADD{small} aspect ratio remains to be examined. It will be interesting to see if the presence of net helicity affects the creation and further development of fronts, and other \ADD{phenomena,} such as the dispersion of Lagrangian particles by the flow.

\begin{acknowledgments}
Computations were performed using resources at NCAR \ADD{and at UBA, which are both} gratefully acknowledged. Visualizations were done with the NCAR-developed VAPOR software \cite{clyne_07}. Useful discussions with Peter Sullivan and John Clyne were instrumental at the onset of this work. Support for AP, from LASP and in particular from Bob Ergun, is gratefully acknowledged. PDM acknowledges support from the CISL visitor program at NCAR, and from grants UBACYT No.~20020130100738BA, and PICT Nos.~2011-1529 and 2015-3530.
\end{acknowledgments}

\bibliography{ms}

\begin{thebibliography}{92}
\expandafter\ifx\csname natexlab\endcsname\relax\def\natexlab#1{#1}\fi
\expandafter\ifx\csname bibnamefont\endcsname\relax
  \def\bibnamefont#1{#1}\fi
\expandafter\ifx\csname bibfnamefont\endcsname\relax
  \def\bibfnamefont#1{#1}\fi
\expandafter\ifx\csname citenamefont\endcsname\relax
  \def\citenamefont#1{#1}\fi
\expandafter\ifx\csname url\endcsname\relax
  \def\url#1{\texttt{#1}}\fi
\expandafter\ifx\csname urlprefix\endcsname\relax\def\urlprefix{URL }\fi
\providecommand{\bibinfo}[2]{#2}
\providecommand{\eprint}[2][]{\url{#2}}

\bibitem[{\citenamefont{Charney}(1971)}]{charney_71}
\bibinfo{author}{\bibfnamefont{J.}~\bibnamefont{Charney}},
  \bibinfo{journal}{J.\ Atmos.\ Sci.} \textbf{\bibinfo{volume}{28}},
  \bibinfo{pages}{1087} (\bibinfo{year}{1971}).

\bibitem[{\citenamefont{Kartashova et~al.}(2008)\citenamefont{Kartashova,
  Nazarenko, and Rudenko}}]{kartashova_08}
\bibinfo{author}{\bibfnamefont{E.}~\bibnamefont{Kartashova}},
  \bibinfo{author}{\bibfnamefont{S.}~\bibnamefont{Nazarenko}},
  \bibnamefont{and} \bibinfo{author}{\bibfnamefont{O.}~\bibnamefont{Rudenko}},
  \bibinfo{journal}{Phys. Rev. E} \textbf{\bibinfo{volume}{78}},
  \bibinfo{pages}{016304} (\bibinfo{year}{2008}).

\bibitem[{\citenamefont{Kafiabad and Bartello}(2016)}]{kafiabad_16}
\bibinfo{author}{\bibfnamefont{H.}~\bibnamefont{Kafiabad}} \bibnamefont{and}
  \bibinfo{author}{\bibfnamefont{P.}~\bibnamefont{Bartello}},
  \bibinfo{journal}{J. Fluid Mech.} \textbf{\bibinfo{volume}{795}},
  \bibinfo{pages}{914} (\bibinfo{year}{2016}).

\bibitem[{\citenamefont{Deusebio et~al.}(2013)\citenamefont{Deusebio, Vallgren,
  and Lindborg}}]{deusebio_13}
\bibinfo{author}{\bibfnamefont{E.}~\bibnamefont{Deusebio}},
  \bibinfo{author}{\bibfnamefont{A.}~\bibnamefont{Vallgren}}, \bibnamefont{and}
  \bibinfo{author}{\bibfnamefont{E.}~\bibnamefont{Lindborg}},
  \bibinfo{journal}{J. Fluid Mech.} \textbf{\bibinfo{volume}{720}},
  \bibinfo{pages}{66} (\bibinfo{year}{2013}).

\bibitem[{\citenamefont{Deusebio
  et~al.}(2014{\natexlab{a}})\citenamefont{Deusebio, Augier, and
  Lindborg}}]{deusebio_14b}
\bibinfo{author}{\bibfnamefont{E.}~\bibnamefont{Deusebio}},
  \bibinfo{author}{\bibfnamefont{P.}~\bibnamefont{Augier}}, \bibnamefont{and}
  \bibinfo{author}{\bibfnamefont{E.}~\bibnamefont{Lindborg}},
  \bibinfo{journal}{J . Fluid Mech.} \textbf{\bibinfo{volume}{755}},
  \bibinfo{pages}{294} (\bibinfo{year}{2014}{\natexlab{a}}).

\bibitem[{\citenamefont{King et~al.}(2015)\citenamefont{King, Vogelzang, and
  Stoffelen}}]{king_15a}
\bibinfo{author}{\bibfnamefont{G.~P.} \bibnamefont{King}},
  \bibinfo{author}{\bibfnamefont{J.}~\bibnamefont{Vogelzang}},
  \bibnamefont{and}
  \bibinfo{author}{\bibfnamefont{A.}~\bibnamefont{Stoffelen}},
  \bibinfo{journal}{J. Geophys. Res.} \textbf{\bibinfo{volume}{120}},
  \bibinfo{pages}{346} (\bibinfo{year}{2015}).

\bibitem[{\citenamefont{Pouquet
  et~al.}(2017{\natexlab{a}})\citenamefont{Pouquet, Rosenberg, Marino, and
  Herbert}}]{pouquet_17j}
\bibinfo{author}{\bibfnamefont{A.}~\bibnamefont{Pouquet}},
  \bibinfo{author}{\bibfnamefont{D.}~\bibnamefont{Rosenberg}},
  \bibinfo{author}{\bibfnamefont{R.}~\bibnamefont{Marino}}, \bibnamefont{and}
  \bibinfo{author}{\bibfnamefont{C.}~\bibnamefont{Herbert}},
  \bibinfo{journal}{J. Fluid Mech., Submitted}
  (\bibinfo{year}{2017}{\natexlab{a}}).

\bibitem[{\citenamefont{Davis and Yan}(2004)}]{davis_04}
\bibinfo{author}{\bibfnamefont{A.}~\bibnamefont{Davis}} \bibnamefont{and}
  \bibinfo{author}{\bibfnamefont{X.-H.} \bibnamefont{Yan}},
  \bibinfo{journal}{Geophys. Res. Lett.} \textbf{\bibinfo{volume}{31}},
  \bibinfo{pages}{L17304} (\bibinfo{year}{2004}).

\bibitem[{\citenamefont{Rossi et~al.}(2009)\citenamefont{Rossi, L\'opez,
  Hern\'andez-Garci\'a, Sudre, Gar{\c c}on, and Morel}}]{rossi_09}
\bibinfo{author}{\bibfnamefont{V.}~\bibnamefont{Rossi}},
  \bibinfo{author}{\bibfnamefont{C.}~\bibnamefont{L\'opez}},
  \bibinfo{author}{\bibfnamefont{E.}~\bibnamefont{Hern\'andez-Garci\'a}},
  \bibinfo{author}{\bibfnamefont{J.}~\bibnamefont{Sudre}},
  \bibinfo{author}{\bibfnamefont{V.}~\bibnamefont{Gar{\c c}on}},
  \bibnamefont{and} \bibinfo{author}{\bibfnamefont{Y.}~\bibnamefont{Morel}},
  \bibinfo{journal}{Nonlin. Processes Geophys.} \textbf{\bibinfo{volume}{16}},
  \bibinfo{pages}{557} (\bibinfo{year}{2009}).

\bibitem[{\citenamefont{Gruber et~al.}(2011)\citenamefont{Gruber, Lachkar,
  Frenzel, Marchesiello, M\"unnich, McWilliams, Nagai, and
  Plattner}}]{gruber_11}
\bibinfo{author}{\bibfnamefont{N.}~\bibnamefont{Gruber}},
  \bibinfo{author}{\bibfnamefont{Z.}~\bibnamefont{Lachkar}},
  \bibinfo{author}{\bibfnamefont{H.}~\bibnamefont{Frenzel}},
  \bibinfo{author}{\bibfnamefont{P.}~\bibnamefont{Marchesiello}},
  \bibinfo{author}{\bibfnamefont{M.}~\bibnamefont{M\"unnich}},
  \bibinfo{author}{\bibfnamefont{J.~C.} \bibnamefont{McWilliams}},
  \bibinfo{author}{\bibfnamefont{T.}~\bibnamefont{Nagai}}, \bibnamefont{and}
  \bibinfo{author}{\bibfnamefont{G.-K.} \bibnamefont{Plattner}},
  \bibinfo{journal}{Nature Geosc.} \textbf{\bibinfo{volume}{4}},
  \bibinfo{pages}{787} (\bibinfo{year}{2011}).

\bibitem[{\citenamefont{Shulman et~al.}(2015)\citenamefont{Shulman, Penta,
  Richman, Jacobs, Anderson, and Sakalaukus}}]{shulman_15}
\bibinfo{author}{\bibfnamefont{I.}~\bibnamefont{Shulman}},
  \bibinfo{author}{\bibfnamefont{B.}~\bibnamefont{Penta}},
  \bibinfo{author}{\bibfnamefont{J.}~\bibnamefont{Richman}},
  \bibinfo{author}{\bibfnamefont{G.}~\bibnamefont{Jacobs}},
  \bibinfo{author}{\bibfnamefont{S.}~\bibnamefont{Anderson}}, \bibnamefont{and}
  \bibinfo{author}{\bibfnamefont{P.}~\bibnamefont{Sakalaukus}},
  \bibinfo{journal}{J. Geophys. Res.} \textbf{\bibinfo{volume}{120}},
  \bibinfo{pages}{2050} (\bibinfo{year}{2015}).

\bibitem[{\citenamefont{Smith et~al.}(2016)\citenamefont{Smith, Hamlington, and
  Fox-Kemper}}]{smith_16}
\bibinfo{author}{\bibfnamefont{K.}~\bibnamefont{Smith}},
  \bibinfo{author}{\bibfnamefont{P.}~\bibnamefont{Hamlington}},
  \bibnamefont{and}
  \bibinfo{author}{\bibfnamefont{B.}~\bibnamefont{Fox-Kemper}},
  \bibinfo{journal}{J. Geophys. Res.} \textbf{\bibinfo{volume}{121}},
  \bibinfo{pages}{908} (\bibinfo{year}{2016}).

\bibitem[{\citenamefont{Papenberg et~al.}(2010)\citenamefont{Papenberg,
  Klaeschen, Krahmann, and Hobbs}}]{papenberg_10}
\bibinfo{author}{\bibfnamefont{C.}~\bibnamefont{Papenberg}},
  \bibinfo{author}{\bibfnamefont{D.}~\bibnamefont{Klaeschen}},
  \bibinfo{author}{\bibfnamefont{G.}~\bibnamefont{Krahmann}}, \bibnamefont{and}
  \bibinfo{author}{\bibfnamefont{R.~W.} \bibnamefont{Hobbs}},
  \bibinfo{journal}{Geophys. Res. Lett.} \textbf{\bibinfo{volume}{37}}
  (\bibinfo{year}{2010}).

\bibitem[{\citenamefont{McWilliams}(2016)}]{mcwilliams_16}
\bibinfo{author}{\bibfnamefont{J.}~\bibnamefont{McWilliams}},
  \bibinfo{journal}{Proc. Roy. Soc. A} \textbf{\bibinfo{volume}{472}},
  \bibinfo{pages}{2016.0117} (\bibinfo{year}{2016}).

\bibitem[{\citenamefont{Celani et~al.}(2004)\citenamefont{Celani, Cencini,
  Mazzino, and Vergassola}}]{celani_04}
\bibinfo{author}{\bibfnamefont{A.}~\bibnamefont{Celani}},
  \bibinfo{author}{\bibfnamefont{M.}~\bibnamefont{Cencini}},
  \bibinfo{author}{\bibfnamefont{A.}~\bibnamefont{Mazzino}}, \bibnamefont{and}
  \bibinfo{author}{\bibfnamefont{M.}~\bibnamefont{Vergassola}},
  \bibinfo{journal}{New J. Phys.} \textbf{\bibinfo{volume}{6}}
  (\bibinfo{year}{2004}).

\bibitem[{\citenamefont{Ishihara et~al.}(2009)\citenamefont{Ishihara, Gotoh,
  and Kaneda}}]{ishihara_09}
\bibinfo{author}{\bibfnamefont{T.}~\bibnamefont{Ishihara}},
  \bibinfo{author}{\bibfnamefont{T.}~\bibnamefont{Gotoh}}, \bibnamefont{and}
  \bibinfo{author}{\bibfnamefont{Y.}~\bibnamefont{Kaneda}},
  \bibinfo{journal}{Ann. Rev. Fluid Mech.} \textbf{\bibinfo{volume}{41}},
  \bibinfo{pages}{165} (\bibinfo{year}{2009}).

\bibitem[{\citenamefont{Oieroset et~al.}(2011)\citenamefont{Oieroset, Phan,
  Eastwood, Fujimoto, Daughton, Shay, Angelopoulos, Mozer, McFadden, Larson
  et~al.}}]{oieroset_11}
\bibinfo{author}{\bibfnamefont{M.}~\bibnamefont{Oieroset}},
  \bibinfo{author}{\bibfnamefont{T.}~\bibnamefont{Phan}},
  \bibinfo{author}{\bibfnamefont{J.~P.} \bibnamefont{Eastwood}},
  \bibinfo{author}{\bibfnamefont{M.}~\bibnamefont{Fujimoto}},
  \bibinfo{author}{\bibfnamefont{W.}~\bibnamefont{Daughton}},
  \bibinfo{author}{\bibfnamefont{M.~A.} \bibnamefont{Shay}},
  \bibinfo{author}{\bibfnamefont{V.}~\bibnamefont{Angelopoulos}},
  \bibinfo{author}{\bibfnamefont{F.~S.} \bibnamefont{Mozer}},
  \bibinfo{author}{\bibfnamefont{J.~P.} \bibnamefont{McFadden}},
  \bibinfo{author}{\bibfnamefont{D.~E.} \bibnamefont{Larson}},
  \bibnamefont{et~al.}, \bibinfo{journal}{Phys. Rev. Lett.}
  \textbf{\bibinfo{volume}{107}}, \bibinfo{pages}{165007}
  (\bibinfo{year}{2011}).

\bibitem[{\citenamefont{Mininni et~al.}(2006)\citenamefont{Mininni, Pouquet,
  and Montgomery}}]{mininni_06b}
\bibinfo{author}{\bibfnamefont{P.}~\bibnamefont{Mininni}},
  \bibinfo{author}{\bibfnamefont{A.}~\bibnamefont{Pouquet}}, \bibnamefont{and}
  \bibinfo{author}{\bibfnamefont{D.}~\bibnamefont{Montgomery}},
  \bibinfo{journal}{Phys. Rev. Lett.} \textbf{\bibinfo{volume}{97}},
  \bibinfo{pages}{244503} (\bibinfo{year}{2006}).

\bibitem[{\citenamefont{Hoskins}(1982)}]{hoskins_82}
\bibinfo{author}{\bibfnamefont{B.}~\bibnamefont{Hoskins}},
  \bibinfo{journal}{Ann. Rev. Fluid Mech.} \textbf{\bibinfo{volume}{14}},
  \bibinfo{pages}{131} (\bibinfo{year}{1982}).

\bibitem[{\citenamefont{Grabowski and Wang}(2013)}]{grabowski_13}
\bibinfo{author}{\bibfnamefont{W.}~\bibnamefont{Grabowski}} \bibnamefont{and}
  \bibinfo{author}{\bibfnamefont{L.-P.} \bibnamefont{Wang}},
  \bibinfo{journal}{Ann. Rev. Fluid Mech.} \textbf{\bibinfo{volume}{45}},
  \bibinfo{pages}{293} (\bibinfo{year}{2013}).

\bibitem[{\citenamefont{Rorai et~al.}(2014)\citenamefont{Rorai, Mininni, and
  Pouquet}}]{rorai_14}
\bibinfo{author}{\bibfnamefont{C.}~\bibnamefont{Rorai}},
  \bibinfo{author}{\bibfnamefont{P.}~\bibnamefont{Mininni}}, \bibnamefont{and}
  \bibinfo{author}{\bibfnamefont{A.}~\bibnamefont{Pouquet}},
  \bibinfo{journal}{Phys. Rev. E} \textbf{\bibinfo{volume}{89}},
  \bibinfo{pages}{043002} (\bibinfo{year}{2014}).

\bibitem[{\citenamefont{Maffioli et~al.}(2014)\citenamefont{Maffioli, Davidson,
  Dalziel, and Swaminathan}}]{maffioli_14}
\bibinfo{author}{\bibfnamefont{A.}~\bibnamefont{Maffioli}},
  \bibinfo{author}{\bibfnamefont{P.}~\bibnamefont{Davidson}},
  \bibinfo{author}{\bibfnamefont{S.}~\bibnamefont{Dalziel}}, \bibnamefont{and}
  \bibinfo{author}{\bibfnamefont{N.}~\bibnamefont{Swaminathan}},
  \bibinfo{journal}{J. Fluid Mech.} \textbf{\bibinfo{volume}{739}},
  \bibinfo{pages}{229} (\bibinfo{year}{2014}).

\bibitem[{\citenamefont{Liang et~al.}(2016)\citenamefont{Liang, Zareei, and
  {R}eza Alam}}]{liang_17}
\bibinfo{author}{\bibfnamefont{Y.}~\bibnamefont{Liang}},
  \bibinfo{author}{\bibfnamefont{A.}~\bibnamefont{Zareei}}, \bibnamefont{and}
  \bibinfo{author}{\bibfnamefont{M.}~\bibnamefont{{R}eza Alam}},
  \bibinfo{journal}{J. Fluid Mech.} \textbf{\bibinfo{volume}{811}},
  \bibinfo{pages}{400} (\bibinfo{year}{2016}).

\bibitem[{\citenamefont{Brouzet et~al.}(2017)\citenamefont{Brouzet, Ermanyuk,
  Joubaud, Pillet, and Dauxois}}]{brouzet_17}
\bibinfo{author}{\bibfnamefont{C.}~\bibnamefont{Brouzet}},
  \bibinfo{author}{\bibfnamefont{E.}~\bibnamefont{Ermanyuk}},
  \bibinfo{author}{\bibfnamefont{S.}~\bibnamefont{Joubaud}},
  \bibinfo{author}{\bibfnamefont{G.}~\bibnamefont{Pillet}}, \bibnamefont{and}
  \bibinfo{author}{\bibfnamefont{T.}~\bibnamefont{Dauxois}},
  \bibinfo{journal}{J. Fluid Mech.} \textbf{\bibinfo{volume}{811}},
  \bibinfo{pages}{544} (\bibinfo{year}{2017}).

\bibitem[{\citenamefont{M\'etais and Herring}(1989)}]{metais_89}
\bibinfo{author}{\bibfnamefont{O.}~\bibnamefont{M\'etais}} \bibnamefont{and}
  \bibinfo{author}{\bibfnamefont{J.}~\bibnamefont{Herring}},
  \bibinfo{journal}{J. Fluid Mech.} \textbf{\bibinfo{volume}{202}},
  \bibinfo{pages}{117} (\bibinfo{year}{1989}).

\bibitem[{\citenamefont{M\'etais et~al.}(1994)\citenamefont{M\'etais, Riley,
  and Lesieur}}]{metais_94}
\bibinfo{author}{\bibfnamefont{O.}~\bibnamefont{M\'etais}},
  \bibinfo{author}{\bibfnamefont{J.}~\bibnamefont{Riley}}, \bibnamefont{and}
  \bibinfo{author}{\bibfnamefont{M.}~\bibnamefont{Lesieur}}, in
  \emph{\bibinfo{booktitle}{Stably-Stratified Flows: Flow and Dispersion over
  Topography}} (\bibinfo{year}{1994}), pp. \bibinfo{pages}{139--151}.

\bibitem[{\citenamefont{Lilly}(1983)}]{lilly_83}
\bibinfo{author}{\bibfnamefont{D.}~\bibnamefont{Lilly}}, \bibinfo{journal}{J.
  Atm. Sci.} \textbf{\bibinfo{volume}{40}}, \bibinfo{pages}{749}
  (\bibinfo{year}{1983}).

\bibitem[{\citenamefont{Cambon et~al.}(2004)\citenamefont{Cambon, Godeferd,
  Nicolleau, and Vassilicos}}]{cambon_04}
\bibinfo{author}{\bibfnamefont{C.}~\bibnamefont{Cambon}},
  \bibinfo{author}{\bibfnamefont{F.~S.} \bibnamefont{Godeferd}},
  \bibinfo{author}{\bibfnamefont{F.}~\bibnamefont{Nicolleau}},
  \bibnamefont{and} \bibinfo{author}{\bibfnamefont{J.~C.}
  \bibnamefont{Vassilicos}}, \bibinfo{journal}{J. Fluid Mech.}
  \textbf{\bibinfo{volume}{499}}, \bibinfo{pages}{231} (\bibinfo{year}{2004}).

\bibitem[{\citenamefont{Lindborg}(2005)}]{lindborg_05}
\bibinfo{author}{\bibfnamefont{E.}~\bibnamefont{Lindborg}},
  \bibinfo{journal}{Geophys. Res. Lett.} \textbf{\bibinfo{volume}{32}},
  \bibinfo{pages}{1} (\bibinfo{year}{2005}).

\bibitem[{\citenamefont{Lindborg}(2006)}]{lindborg_06}
\bibinfo{author}{\bibfnamefont{E.}~\bibnamefont{Lindborg}},
  \bibinfo{journal}{J. Fluid Mech.} \textbf{\bibinfo{volume}{550}},
  \bibinfo{pages}{207} (\bibinfo{year}{2006}).

\bibitem[{\citenamefont{Waite and Snyder}(2009)}]{waite_09}
\bibinfo{author}{\bibfnamefont{M.}~\bibnamefont{Waite}} \bibnamefont{and}
  \bibinfo{author}{\bibfnamefont{C.}~\bibnamefont{Snyder}},
  \bibinfo{journal}{J. Atmos. Sci.} \textbf{\bibinfo{volume}{66}},
  \bibinfo{pages}{883} (\bibinfo{year}{2009}).

\bibitem[{\citenamefont{Rorai et~al.}(2015)\citenamefont{Rorai, Mininni, and
  Pouquet}}]{rorai_15}
\bibinfo{author}{\bibfnamefont{C.}~\bibnamefont{Rorai}},
  \bibinfo{author}{\bibfnamefont{P.}~\bibnamefont{Mininni}}, \bibnamefont{and}
  \bibinfo{author}{\bibfnamefont{A.}~\bibnamefont{Pouquet}},
  \bibinfo{journal}{Phys. Rev. E} \textbf{\bibinfo{volume}{92}},
  \bibinfo{pages}{013003} (\bibinfo{year}{2015}).

\bibitem[{\citenamefont{Maffioli and Davidson}(2016)}]{maffioli_16}
\bibinfo{author}{\bibfnamefont{A.}~\bibnamefont{Maffioli}} \bibnamefont{and}
  \bibinfo{author}{\bibfnamefont{P.}~\bibnamefont{Davidson}},
  \bibinfo{journal}{J. Fluid Mech.} \textbf{\bibinfo{volume}{786}},
  \bibinfo{pages}{210} (\bibinfo{year}{2016}).

\bibitem[{\citenamefont{Augier et~al.}(2012)\citenamefont{Augier, Chomaz, and
  Billant}}]{augier_12}
\bibinfo{author}{\bibfnamefont{P.}~\bibnamefont{Augier}},
  \bibinfo{author}{\bibfnamefont{J.-M.} \bibnamefont{Chomaz}},
  \bibnamefont{and} \bibinfo{author}{\bibfnamefont{P.}~\bibnamefont{Billant}},
  \bibinfo{journal}{J. Fluid Mech.} \textbf{\bibinfo{volume}{713}},
  \bibinfo{pages}{86} (\bibinfo{year}{2012}).

\bibitem[{\citenamefont{Billant and Chomaz}(2001)}]{billant_01}
\bibinfo{author}{\bibfnamefont{P.}~\bibnamefont{Billant}} \bibnamefont{and}
  \bibinfo{author}{\bibfnamefont{J.-M.} \bibnamefont{Chomaz}},
  \bibinfo{journal}{Phys. Fluids} \textbf{\bibinfo{volume}{13}},
  \bibinfo{pages}{1645} (\bibinfo{year}{2001}).

\bibitem[{\citenamefont{Mininni et~al.}(2012)\citenamefont{Mininni, Rosenberg,
  and Pouquet}}]{3072}
\bibinfo{author}{\bibfnamefont{P.}~\bibnamefont{Mininni}},
  \bibinfo{author}{\bibfnamefont{D.}~\bibnamefont{Rosenberg}},
  \bibnamefont{and} \bibinfo{author}{\bibfnamefont{A.}~\bibnamefont{Pouquet}},
  \bibinfo{journal}{J. Fluid Mech.} \textbf{\bibinfo{volume}{699}},
  \bibinfo{pages}{263} (\bibinfo{year}{2012}).

\bibitem[{\citenamefont{Brethouwer et~al.}(2007)\citenamefont{Brethouwer,
  Billant, Lindborg, and Chomaz}}]{brethouwer_07}
\bibinfo{author}{\bibfnamefont{G.}~\bibnamefont{Brethouwer}},
  \bibinfo{author}{\bibfnamefont{P.}~\bibnamefont{Billant}},
  \bibinfo{author}{\bibfnamefont{E.}~\bibnamefont{Lindborg}}, \bibnamefont{and}
  \bibinfo{author}{\bibfnamefont{J.-M.} \bibnamefont{Chomaz}},
  \bibinfo{journal}{J. Fluid Mech.} \textbf{\bibinfo{volume}{585}},
  \bibinfo{pages}{343} (\bibinfo{year}{2007}).

\bibitem[{\citenamefont{Ivey et~al.}(2008)\citenamefont{Ivey, Winters, and
  Koseff}}]{ivey_08}
\bibinfo{author}{\bibfnamefont{G.}~\bibnamefont{Ivey}},
  \bibinfo{author}{\bibfnamefont{K.}~\bibnamefont{Winters}}, \bibnamefont{and}
  \bibinfo{author}{\bibfnamefont{J.}~\bibnamefont{Koseff}},
  \bibinfo{journal}{Ann. Rev. Fluid Mech.} \textbf{\bibinfo{volume}{{\bf 40}}},
  \bibinfo{pages}{169} (\bibinfo{year}{2008}).

\bibitem[{\citenamefont{Waite}(2011)}]{waite_11}
\bibinfo{author}{\bibfnamefont{M.~L.} \bibnamefont{Waite}},
  \bibinfo{journal}{Phys. Fluids} \textbf{\bibinfo{volume}{23}},
  \bibinfo{pages}{066602} (\bibinfo{year}{2011}).

\bibitem[{\citenamefont{Venayagamoorthy and Koseff}(2016)}]{venayagamoorthy_16}
\bibinfo{author}{\bibfnamefont{S.}~\bibnamefont{Venayagamoorthy}}
  \bibnamefont{and} \bibinfo{author}{\bibfnamefont{J.}~\bibnamefont{Koseff}},
  \bibinfo{journal}{J. Fluid Mech.} \textbf{\bibinfo{volume}{798}},
  \bibinfo{pages}{R1} (\bibinfo{year}{2016}).

\bibitem[{\citenamefont{Maffioli et~al.}(2016)\citenamefont{Maffioli,
  Brethouwer, and Lindborg}}]{maffioli_16b}
\bibinfo{author}{\bibfnamefont{A.}~\bibnamefont{Maffioli}},
  \bibinfo{author}{\bibfnamefont{G.}~\bibnamefont{Brethouwer}},
  \bibnamefont{and} \bibinfo{author}{\bibfnamefont{E.}~\bibnamefont{Lindborg}},
  \bibinfo{journal}{J . Fluid Mech.} \textbf{\bibinfo{volume}{794}},
  \bibinfo{pages}{R3} (\bibinfo{year}{2016}).

\bibitem[{\citenamefont{Marino et~al.}(2015{\natexlab{a}})\citenamefont{Marino,
  Pouquet, and Rosenberg}}]{marino_15p}
\bibinfo{author}{\bibfnamefont{R.}~\bibnamefont{Marino}},
  \bibinfo{author}{\bibfnamefont{A.}~\bibnamefont{Pouquet}}, \bibnamefont{and}
  \bibinfo{author}{\bibfnamefont{D.}~\bibnamefont{Rosenberg}},
  \bibinfo{journal}{Phys. Rev. Lett.} \textbf{\bibinfo{volume}{114}},
  \bibinfo{pages}{114504} (\bibinfo{year}{2015}{\natexlab{a}}).

\bibitem[{\citenamefont{Pouquet
  et~al.}(2017{\natexlab{b}})\citenamefont{Pouquet, Marino, Mininni, and
  Rosenberg}}]{pouquet_17p}
\bibinfo{author}{\bibfnamefont{A.}~\bibnamefont{Pouquet}},
  \bibinfo{author}{\bibfnamefont{R.}~\bibnamefont{Marino}},
  \bibinfo{author}{\bibfnamefont{P.~D.} \bibnamefont{Mininni}},
  \bibnamefont{and}
  \bibinfo{author}{\bibfnamefont{D.}~\bibnamefont{Rosenberg}},
  \bibinfo{journal}{Phys. Fluids} \textbf{\bibinfo{volume}{29}}
  (\bibinfo{year}{2017}{\natexlab{b}}).

\bibitem[{\citenamefont{Venaille et~al.}(2017)\citenamefont{Venaille, Gostiaux,
  and Sommeria}}]{venaille_17}
\bibinfo{author}{\bibfnamefont{A.}~\bibnamefont{Venaille}},
  \bibinfo{author}{\bibfnamefont{L.}~\bibnamefont{Gostiaux}}, \bibnamefont{and}
  \bibinfo{author}{\bibfnamefont{J.}~\bibnamefont{Sommeria}},
  \bibinfo{journal}{J. Fluid Mech.} \textbf{\bibinfo{volume}{810}},
  \bibinfo{pages}{554} (\bibinfo{year}{2017}).

\bibitem[{\citenamefont{Klymak et~al.}(2008)\citenamefont{Klymak, Pinkel, and
  Rainville}}]{klymak_08}
\bibinfo{author}{\bibfnamefont{J.}~\bibnamefont{Klymak}},
  \bibinfo{author}{\bibfnamefont{R.}~\bibnamefont{Pinkel}}, \bibnamefont{and}
  \bibinfo{author}{\bibfnamefont{L.}~\bibnamefont{Rainville}},
  \bibinfo{journal}{J. Phys. Oceano.} \textbf{\bibinfo{volume}{38}},
  \bibinfo{pages}{380} (\bibinfo{year}{2008}).

\bibitem[{\citenamefont{van Haren and Gostiaux}(2016)}]{vanharen_16}
\bibinfo{author}{\bibfnamefont{H.}~\bibnamefont{van Haren}} \bibnamefont{and}
  \bibinfo{author}{\bibfnamefont{L.}~\bibnamefont{Gostiaux}},
  \bibinfo{journal}{J. Mar. Res.} \textbf{\bibinfo{volume}{74}},
  \bibinfo{pages}{161} (\bibinfo{year}{2016}).

\bibitem[{\citenamefont{Cl\'ement et~al.}(2017)\citenamefont{Cl\'ement,
  Thurnherr, and Laurent}}]{clement_17}
\bibinfo{author}{\bibfnamefont{L.}~\bibnamefont{Cl\'ement}},
  \bibinfo{author}{\bibfnamefont{A.~M.} \bibnamefont{Thurnherr}},
  \bibnamefont{and} \bibinfo{author}{\bibfnamefont{L.~C.~S.}
  \bibnamefont{Laurent}}, \bibinfo{journal}{J. \ Phys.\ Ocean.}
  \textbf{\bibinfo{volume}{47}}, \bibinfo{pages}{1873} (\bibinfo{year}{2017}).

\bibitem[{\citenamefont{David et~al.}(2017)\citenamefont{David, Marshall, and
  Zanna}}]{david_17}
\bibinfo{author}{\bibfnamefont{T.~W.} \bibnamefont{David}},
  \bibinfo{author}{\bibfnamefont{D.~P.} \bibnamefont{Marshall}},
  \bibnamefont{and} \bibinfo{author}{\bibfnamefont{L.}~\bibnamefont{Zanna}},
  \bibinfo{journal}{Ocean Modelling} \textbf{\bibinfo{volume}{113}},
  \bibinfo{pages}{34} (\bibinfo{year}{2017}).

\bibitem[{\citenamefont{Mater and
  Venayagamoorthy}(2014{\natexlab{a}})}]{mater_14}
\bibinfo{author}{\bibfnamefont{B.}~\bibnamefont{Mater}} \bibnamefont{and}
  \bibinfo{author}{\bibfnamefont{S.}~\bibnamefont{Venayagamoorthy}},
  \bibinfo{journal}{Geophys. Res. Lett.} \textbf{\bibinfo{volume}{41}},
  \bibinfo{pages}{4646} (\bibinfo{year}{2014}{\natexlab{a}}).

\bibitem[{\citenamefont{Mater and
  Venayagamoorthy}(2014{\natexlab{b}})}]{mater_14b}
\bibinfo{author}{\bibfnamefont{B.}~\bibnamefont{Mater}} \bibnamefont{and}
  \bibinfo{author}{\bibfnamefont{S.}~\bibnamefont{Venayagamoorthy}},
  \bibinfo{journal}{Phys. Fluids} \textbf{\bibinfo{volume}{26}},
  \bibinfo{pages}{036601} (\bibinfo{year}{2014}{\natexlab{b}}).

\bibitem[{\citenamefont{Karimpour and Venayagamoorthy}(2015)}]{karimpour_15}
\bibinfo{author}{\bibfnamefont{F.}~\bibnamefont{Karimpour}} \bibnamefont{and}
  \bibinfo{author}{\bibfnamefont{S.}~\bibnamefont{Venayagamoorthy}},
  \bibinfo{journal}{Phys. Fluids} \textbf{\bibinfo{volume}{27}},
  \bibinfo{pages}{046603} (\bibinfo{year}{2015}).

\bibitem[{\citenamefont{Bachman et~al.}(2017)\citenamefont{Bachman, Fox-Kemper,
  Taylor, and Thomas}}]{bachman_17}
\bibinfo{author}{\bibfnamefont{S.}~\bibnamefont{Bachman}},
  \bibinfo{author}{\bibfnamefont{B.}~\bibnamefont{Fox-Kemper}},
  \bibinfo{author}{\bibfnamefont{J.}~\bibnamefont{Taylor}}, \bibnamefont{and}
  \bibinfo{author}{\bibfnamefont{L.}~\bibnamefont{Thomas}},
  \bibinfo{journal}{Ocean Modelling} \textbf{\bibinfo{volume}{109}},
  \bibinfo{pages}{72} (\bibinfo{year}{2017}).

\bibitem[{\citenamefont{Smyth and Moum}(2000)}]{smyth_00b}
\bibinfo{author}{\bibfnamefont{W.}~\bibnamefont{Smyth}} \bibnamefont{and}
  \bibinfo{author}{\bibfnamefont{J.}~\bibnamefont{Moum}},
  \bibinfo{journal}{Phys. Fluids} \textbf{\bibinfo{volume}{12}},
  \bibinfo{pages}{1343} (\bibinfo{year}{2000}).

\bibitem[{\citenamefont{Taylor and Green}(1937)}]{taylor_37}
\bibinfo{author}{\bibfnamefont{G.}~\bibnamefont{Taylor}} \bibnamefont{and}
  \bibinfo{author}{\bibfnamefont{A.}~\bibnamefont{Green}},
  \bibinfo{journal}{Proc. Roy. Soc Lond. Series A}
  \textbf{\bibinfo{volume}{158}}, \bibinfo{pages}{499} (\bibinfo{year}{1937}).

\bibitem[{\citenamefont{Brachet et~al.}(1983)\citenamefont{Brachet, Meiron,
  Orszag, Nickel, Morf, and Frisch}}]{brachet_83}
\bibinfo{author}{\bibfnamefont{M.}~\bibnamefont{Brachet}},
  \bibinfo{author}{\bibfnamefont{D.}~\bibnamefont{Meiron}},
  \bibinfo{author}{\bibfnamefont{S.}~\bibnamefont{Orszag}},
  \bibinfo{author}{\bibfnamefont{B.}~\bibnamefont{Nickel}},
  \bibinfo{author}{\bibfnamefont{R.}~\bibnamefont{Morf}}, \bibnamefont{and}
  \bibinfo{author}{\bibfnamefont{U.}~\bibnamefont{Frisch}}, \bibinfo{journal}{J
  . Fluid Mech.} \textbf{\bibinfo{volume}{130}}, \bibinfo{pages}{411}
  (\bibinfo{year}{1983}).

\bibitem[{\citenamefont{Cichowlas et~al.}(2005)\citenamefont{Cichowlas,
  Bona\"iti, Debbasch, and Brachet}}]{brachet_05}
\bibinfo{author}{\bibfnamefont{C.}~\bibnamefont{Cichowlas}},
  \bibinfo{author}{\bibfnamefont{P.}~\bibnamefont{Bona\"iti}},
  \bibinfo{author}{\bibfnamefont{F.}~\bibnamefont{Debbasch}}, \bibnamefont{and}
  \bibinfo{author}{\bibfnamefont{M.}~\bibnamefont{Brachet}},
  \bibinfo{journal}{Phys Rev. Lett.} \textbf{\bibinfo{volume}{95}}
  (\bibinfo{year}{2005}).

\bibitem[{\citenamefont{Brachet et~al.}(2013)\citenamefont{Brachet, Bustamante,
  Krstulovic, Mininni, Pouquet, and Rosenberg}}]{brachet_13}
\bibinfo{author}{\bibfnamefont{M.}~\bibnamefont{Brachet}},
  \bibinfo{author}{\bibfnamefont{M.}~\bibnamefont{Bustamante}},
  \bibinfo{author}{\bibfnamefont{G.}~\bibnamefont{Krstulovic}},
  \bibinfo{author}{\bibfnamefont{P.}~\bibnamefont{Mininni}},
  \bibinfo{author}{\bibfnamefont{A.}~\bibnamefont{Pouquet}}, \bibnamefont{and}
  \bibinfo{author}{\bibfnamefont{D.}~\bibnamefont{Rosenberg}},
  \bibinfo{journal}{Phys. Rev. E} \textbf{\bibinfo{volume}{87}},
  \bibinfo{pages}{013110} (\bibinfo{year}{2013}).

\bibitem[{\citenamefont{Nore et~al.}(1997)\citenamefont{Nore, Politano, and
  Pouquet}}]{nore_97}
\bibinfo{author}{\bibfnamefont{C.}~\bibnamefont{Nore}},
  \bibinfo{author}{\bibfnamefont{M.~B.~H.} \bibnamefont{Politano}},
  \bibnamefont{and} \bibinfo{author}{\bibfnamefont{A.}~\bibnamefont{Pouquet}},
  \bibinfo{journal}{Phys. Plasmas Lett.} \textbf{\bibinfo{volume}{4}},
  \bibinfo{pages}{1} (\bibinfo{year}{1997}).

\bibitem[{\citenamefont{Ponty et~al.}(2005)\citenamefont{Ponty, Mininni,
  Montgomery, Pinton, Politano, and Pouquet}}]{ponty_05}
\bibinfo{author}{\bibfnamefont{Y.}~\bibnamefont{Ponty}},
  \bibinfo{author}{\bibfnamefont{P.~D.} \bibnamefont{Mininni}},
  \bibinfo{author}{\bibfnamefont{D.}~\bibnamefont{Montgomery}},
  \bibinfo{author}{\bibfnamefont{J.-F.} \bibnamefont{Pinton}},
  \bibinfo{author}{\bibfnamefont{H.}~\bibnamefont{Politano}}, \bibnamefont{and}
  \bibinfo{author}{\bibfnamefont{A.}~\bibnamefont{Pouquet}},
  \bibinfo{journal}{Phys. Rev. Lett.} \textbf{\bibinfo{volume}{94}},
  \bibinfo{pages}{164502} (\bibinfo{year}{2005}).

\bibitem[{\citenamefont{Riley and deBruynKops}(2003)}]{riley_03}
\bibinfo{author}{\bibfnamefont{J.}~\bibnamefont{Riley}} \bibnamefont{and}
  \bibinfo{author}{\bibfnamefont{S.}~\bibnamefont{deBruynKops}},
  \bibinfo{journal}{Phys. Fluids} \textbf{\bibinfo{volume}{15}},
  \bibinfo{pages}{2047} (\bibinfo{year}{2003}).

\bibitem[{\citenamefont{Douady et~al.}(1991)\citenamefont{Douady, Couder, and
  Brachet}}]{douady_91}
\bibinfo{author}{\bibfnamefont{S.}~\bibnamefont{Douady}},
  \bibinfo{author}{\bibfnamefont{Y.}~\bibnamefont{Couder}}, \bibnamefont{and}
  \bibinfo{author}{\bibfnamefont{M.-E.} \bibnamefont{Brachet}},
  \bibinfo{journal}{Phys. Rev. Lett.} \textbf{\bibinfo{volume}{67}},
  \bibinfo{pages}{983} (\bibinfo{year}{1991}).

\bibitem[{\citenamefont{Odier et~al.}(1998)\citenamefont{Odier, Pinton, and
  Fauve}}]{odier_98}
\bibinfo{author}{\bibfnamefont{P.}~\bibnamefont{Odier}},
  \bibinfo{author}{\bibfnamefont{J.-F.} \bibnamefont{Pinton}},
  \bibnamefont{and} \bibinfo{author}{\bibfnamefont{S.}~\bibnamefont{Fauve}},
  \bibinfo{journal}{Phys. Rev. E} \textbf{\bibinfo{volume}{58}},
  \bibinfo{pages}{7397} (\bibinfo{year}{1998}).

\bibitem[{\citenamefont{Lee et~al.}(2010)\citenamefont{Lee, Brachet, Pouquet,
  Mininni, and Rosenberg}}]{lee_10}
\bibinfo{author}{\bibfnamefont{E.}~\bibnamefont{Lee}},
  \bibinfo{author}{\bibfnamefont{M.}~\bibnamefont{Brachet}},
  \bibinfo{author}{\bibfnamefont{A.}~\bibnamefont{Pouquet}},
  \bibinfo{author}{\bibfnamefont{P.}~\bibnamefont{Mininni}}, \bibnamefont{and}
  \bibinfo{author}{\bibfnamefont{D.}~\bibnamefont{Rosenberg}},
  \bibinfo{journal}{Phys. Rev. E} \textbf{\bibinfo{volume}{81}},
  \bibinfo{pages}{016318} (\bibinfo{year}{2010}).

\bibitem[{\citenamefont{Rosenberg et~al.}(2015)\citenamefont{Rosenberg,
  Pouquet, Marino, and Mininni}}]{rosenberg_15}
\bibinfo{author}{\bibfnamefont{D.}~\bibnamefont{Rosenberg}},
  \bibinfo{author}{\bibfnamefont{A.}~\bibnamefont{Pouquet}},
  \bibinfo{author}{\bibfnamefont{R.}~\bibnamefont{Marino}}, \bibnamefont{and}
  \bibinfo{author}{\bibfnamefont{P.}~\bibnamefont{Mininni}},
  \bibinfo{journal}{Phys. Fluids} \textbf{\bibinfo{volume}{27}},
  \bibinfo{pages}{055105} (\bibinfo{year}{2015}).

\bibitem[{\citenamefont{G\'omez et~al.}(2005)\citenamefont{G\'omez, Mininni,
  and Dmitruk}}]{gomez_05}
\bibinfo{author}{\bibfnamefont{D.~O.} \bibnamefont{G\'omez}},
  \bibinfo{author}{\bibfnamefont{P.~D.} \bibnamefont{Mininni}},
  \bibnamefont{and} \bibinfo{author}{\bibfnamefont{P.}~\bibnamefont{Dmitruk}},
  \bibinfo{journal}{Physica Scripta} \textbf{\bibinfo{volume}{T116}},
  \bibinfo{pages}{123} (\bibinfo{year}{2005}).

\bibitem[{\citenamefont{di~Leoni et~al.}(2016)\citenamefont{di~Leoni, Mininni,
  and Brachet}}]{clark_16}
\bibinfo{author}{\bibfnamefont{P.~C.} \bibnamefont{di~Leoni}},
  \bibinfo{author}{\bibfnamefont{P.~D.} \bibnamefont{Mininni}},
  \bibnamefont{and} \bibinfo{author}{\bibfnamefont{M.~E.}
  \bibnamefont{Brachet}}, \bibinfo{journal}{Phys. Rev. A}
  \textbf{\bibinfo{volume}{94}} (\bibinfo{year}{2016}).

\bibitem[{\citenamefont{Mininni et~al.}(2011)\citenamefont{Mininni, Rosenberg,
  Reddy, and Pouquet}}]{hybrid_11}
\bibinfo{author}{\bibfnamefont{P.}~\bibnamefont{Mininni}},
  \bibinfo{author}{\bibfnamefont{D.}~\bibnamefont{Rosenberg}},
  \bibinfo{author}{\bibfnamefont{R.}~\bibnamefont{Reddy}}, \bibnamefont{and}
  \bibinfo{author}{\bibfnamefont{A.}~\bibnamefont{Pouquet}},
  \bibinfo{journal}{Parallel Computing} \textbf{\bibinfo{volume}{37}},
  \bibinfo{pages}{316} (\bibinfo{year}{2011}).

\bibitem[{\citenamefont{Waite and Bartello}(2004)}]{waite_04}
\bibinfo{author}{\bibfnamefont{M.}~\bibnamefont{Waite}} \bibnamefont{and}
  \bibinfo{author}{\bibfnamefont{P.}~\bibnamefont{Bartello}},
  \bibinfo{journal}{J. Fluid Mech.} \textbf{\bibinfo{volume}{517}},
  \bibinfo{pages}{281} (\bibinfo{year}{2004}).

\bibitem[{\citenamefont{Smith and Waleffe}(2002)}]{smith_02}
\bibinfo{author}{\bibfnamefont{L.~M.} \bibnamefont{Smith}} \bibnamefont{and}
  \bibinfo{author}{\bibfnamefont{F.}~\bibnamefont{Waleffe}},
  \bibinfo{journal}{J. Fluid Mech.} \textbf{\bibinfo{volume}{451}},
  \bibinfo{pages}{145} (\bibinfo{year}{2002}).

\bibitem[{\citenamefont{Marino et~al.}(2013)\citenamefont{Marino, Mininni,
  Rosenberg, and Pouquet}}]{marino_13i}
\bibinfo{author}{\bibfnamefont{R.}~\bibnamefont{Marino}},
  \bibinfo{author}{\bibfnamefont{P.}~\bibnamefont{Mininni}},
  \bibinfo{author}{\bibfnamefont{D.}~\bibnamefont{Rosenberg}},
  \bibnamefont{and} \bibinfo{author}{\bibfnamefont{A.}~\bibnamefont{Pouquet}},
  \bibinfo{journal}{EuroPhys. Lett.} \textbf{\bibinfo{volume}{102}},
  \bibinfo{pages}{44006} (\bibinfo{year}{2013}).

\bibitem[{\citenamefont{Marino et~al.}(2015{\natexlab{b}})\citenamefont{Marino,
  Rosenberg, Herbert, and Pouquet}}]{marino_15w}
\bibinfo{author}{\bibfnamefont{R.}~\bibnamefont{Marino}},
  \bibinfo{author}{\bibfnamefont{D.}~\bibnamefont{Rosenberg}},
  \bibinfo{author}{\bibfnamefont{C.}~\bibnamefont{Herbert}}, \bibnamefont{and}
  \bibinfo{author}{\bibfnamefont{A.}~\bibnamefont{Pouquet}},
  \bibinfo{journal}{EuroPhys. Lett.} \textbf{\bibinfo{volume}{112}},
  \bibinfo{pages}{49001} (\bibinfo{year}{2015}{\natexlab{b}}).

\bibitem[{\citenamefont{Celani et~al.}(2010)\citenamefont{Celani, Musacchio,
  and Vincenzi}}]{celani_10}
\bibinfo{author}{\bibfnamefont{A.}~\bibnamefont{Celani}},
  \bibinfo{author}{\bibfnamefont{S.}~\bibnamefont{Musacchio}},
  \bibnamefont{and} \bibinfo{author}{\bibfnamefont{D.}~\bibnamefont{Vincenzi}},
  \bibinfo{journal}{Phys. Rev. Lett.} \textbf{\bibinfo{volume}{104}},
  \bibinfo{pages}{184506} (\bibinfo{year}{2010}).

\bibitem[{\citenamefont{Deusebio
  et~al.}(2014{\natexlab{b}})\citenamefont{Deusebio, Boffetta, Lindborg, and
  Musacchio}}]{deusebio_14}
\bibinfo{author}{\bibfnamefont{E.}~\bibnamefont{Deusebio}},
  \bibinfo{author}{\bibfnamefont{G.}~\bibnamefont{Boffetta}},
  \bibinfo{author}{\bibfnamefont{E.}~\bibnamefont{Lindborg}}, \bibnamefont{and}
  \bibinfo{author}{\bibfnamefont{S.}~\bibnamefont{Musacchio}},
  \bibinfo{journal}{Phys. Rev. E} \textbf{\bibinfo{volume}{90}},
  \bibinfo{pages}{023005} (\bibinfo{year}{2014}{\natexlab{b}}).

\bibitem[{\citenamefont{Sozza et~al.}(2015)\citenamefont{Sozza, Boffetta,
  Muratore-Ginanneschi, and Musacchio}}]{sozza_15}
\bibinfo{author}{\bibfnamefont{A.}~\bibnamefont{Sozza}},
  \bibinfo{author}{\bibfnamefont{G.}~\bibnamefont{Boffetta}},
  \bibinfo{author}{\bibfnamefont{P.}~\bibnamefont{Muratore-Ginanneschi}},
  \bibnamefont{and}
  \bibinfo{author}{\bibfnamefont{S.}~\bibnamefont{Musacchio}},
  \bibinfo{journal}{Phys. Fluids} \textbf{\bibinfo{volume}{27}},
  \bibinfo{pages}{035112} (\bibinfo{year}{2015}).

\bibitem[{\citenamefont{Babin et~al.}(1997)\citenamefont{Babin, Mahalov, and
  Nicolaenko}}]{babin_97}
\bibinfo{author}{\bibfnamefont{A.}~\bibnamefont{Babin}},
  \bibinfo{author}{\bibfnamefont{A.}~\bibnamefont{Mahalov}}, \bibnamefont{and}
  \bibinfo{author}{\bibfnamefont{B.}~\bibnamefont{Nicolaenko}},
  \bibinfo{journal}{Theor. Comput. Fluid Dyn.} \textbf{\bibinfo{volume}{9}},
  \bibinfo{pages}{223} (\bibinfo{year}{1997}).

\bibitem[{\citenamefont{D'{A}saro et~al.}(2011)\citenamefont{D'{A}saro, Lee,
  Rainville, Harcourt, and Thomas}}]{dasaro_11}
\bibinfo{author}{\bibfnamefont{E.}~\bibnamefont{D'{A}saro}},
  \bibinfo{author}{\bibfnamefont{C.}~\bibnamefont{Lee}},
  \bibinfo{author}{\bibfnamefont{L.}~\bibnamefont{Rainville}},
  \bibinfo{author}{\bibfnamefont{R.}~\bibnamefont{Harcourt}}, \bibnamefont{and}
  \bibinfo{author}{\bibfnamefont{L.}~\bibnamefont{Thomas}},
  \bibinfo{journal}{Science} \textbf{\bibinfo{volume}{332}},
  \bibinfo{pages}{318} (\bibinfo{year}{2011}).

\bibitem[{\citenamefont{Zuo et~al.}(2012)\citenamefont{Zuo, Zhang, Xu, Mu, Li,
  and Chen}}]{Zuo_12}
\bibinfo{author}{\bibfnamefont{J.-C.} \bibnamefont{Zuo}},
  \bibinfo{author}{\bibfnamefont{M.}~\bibnamefont{Zhang}},
  \bibinfo{author}{\bibfnamefont{Q.}~\bibnamefont{Xu}},
  \bibinfo{author}{\bibfnamefont{L.}~\bibnamefont{Mu}},
  \bibinfo{author}{\bibfnamefont{J.}~\bibnamefont{Li}}, \bibnamefont{and}
  \bibinfo{author}{\bibfnamefont{M.-X.} \bibnamefont{Chen}},
  \bibinfo{journal}{Water Sc. and Eng.} \textbf{\bibinfo{volume}{5}},
  \bibinfo{pages}{428 } (\bibinfo{year}{2012}).

\bibitem[{\citenamefont{Hoskins and Bretherton}(1972)}]{hoskins_72}
\bibinfo{author}{\bibfnamefont{B.}~\bibnamefont{Hoskins}} \bibnamefont{and}
  \bibinfo{author}{\bibfnamefont{F.}~\bibnamefont{Bretherton}},
  \bibinfo{journal}{J. Atmos. Sci.} \textbf{\bibinfo{volume}{29}},
  \bibinfo{pages}{11} (\bibinfo{year}{1972}).

\bibitem[{\citenamefont{Molemaker et~al.}(2010)\citenamefont{Molemaker,
  McWilliams, and Capet}}]{molemaker_10b}
\bibinfo{author}{\bibfnamefont{M.}~\bibnamefont{Molemaker}},
  \bibinfo{author}{\bibfnamefont{J.}~\bibnamefont{McWilliams}},
  \bibnamefont{and} \bibinfo{author}{\bibfnamefont{X.}~\bibnamefont{Capet}},
  \bibinfo{journal}{J. Fluid Mech.} \textbf{\bibinfo{volume}{{\bf 654}}},
  \bibinfo{pages}{35} (\bibinfo{year}{2010}).

\bibitem[{\citenamefont{Kimura and Herring}(2012)}]{kimura_12}
\bibinfo{author}{\bibfnamefont{Y.}~\bibnamefont{Kimura}} \bibnamefont{and}
  \bibinfo{author}{\bibfnamefont{J.~R.} \bibnamefont{Herring}},
  \bibinfo{journal}{J. Fluid Mech.} \textbf{\bibinfo{volume}{698}},
  \bibinfo{pages}{19} (\bibinfo{year}{2012}).

\bibitem[{\citenamefont{de~Bruyn~Kops}(2015)}]{debruynkops_15}
\bibinfo{author}{\bibfnamefont{S.}~\bibnamefont{de~Bruyn~Kops}},
  \bibinfo{journal}{J. Fluid Mech.} \textbf{\bibinfo{volume}{775}},
  \bibinfo{pages}{436} (\bibinfo{year}{2015}).

\bibitem[{\citenamefont{Clyne et~al.}(2007)\citenamefont{Clyne, Norton, and
  Rast}}]{clyne_07}
\bibinfo{author}{\bibfnamefont{J.}~\bibnamefont{Clyne}},
  \bibinfo{author}{\bibfnamefont{P.~D. M.~A.} \bibnamefont{Norton}},
  \bibnamefont{and} \bibinfo{author}{\bibfnamefont{M.}~\bibnamefont{Rast}},
  \bibinfo{journal}{New J. of Physics} \textbf{\bibinfo{volume}{9}},
  \bibinfo{pages}{301} (\bibinfo{year}{2007}).

\bibitem[{\citenamefont{di~Leoni et~al.}(2017)\citenamefont{di~Leoni, Cobelli,
  and Mininni}}]{clark_17}
\bibinfo{author}{\bibfnamefont{P.~C.} \bibnamefont{di~Leoni}},
  \bibinfo{author}{\bibfnamefont{P.}~\bibnamefont{Cobelli}}, \bibnamefont{and}
  \bibinfo{author}{\bibfnamefont{P.~D.} \bibnamefont{Mininni}},
  \bibinfo{journal}{Eur. Phys. J. E}  (\bibinfo{year}{2017}).

\bibitem[{\citenamefont{Brachet et~al.}(1992)\citenamefont{Brachet, Meneguzzi,
  Vincent, Politano, and Sulem}}]{brachet_92}
\bibinfo{author}{\bibfnamefont{M.}~\bibnamefont{Brachet}},
  \bibinfo{author}{\bibfnamefont{M.}~\bibnamefont{Meneguzzi}},
  \bibinfo{author}{\bibfnamefont{A.}~\bibnamefont{Vincent}},
  \bibinfo{author}{\bibfnamefont{H.}~\bibnamefont{Politano}}, \bibnamefont{and}
  \bibinfo{author}{\bibfnamefont{P.}~\bibnamefont{Sulem}},
  \bibinfo{journal}{Phys. Fluids} \textbf{\bibinfo{volume}{A4}},
  \bibinfo{pages}{2845} (\bibinfo{year}{1992}).

\bibitem[{\citenamefont{Lapeyre et~al.}(2006)\citenamefont{Lapeyre, Klein, and
  Hua}}]{lapeyre_06}
\bibinfo{author}{\bibfnamefont{G.}~\bibnamefont{Lapeyre}},
  \bibinfo{author}{\bibfnamefont{P.}~\bibnamefont{Klein}}, \bibnamefont{and}
  \bibinfo{author}{\bibfnamefont{B.~L.} \bibnamefont{Hua}},
  \bibinfo{journal}{J. Phys. Ocean.} \textbf{\bibinfo{volume}{36}},
  \bibinfo{pages}{1577} (\bibinfo{year}{2006}).

\bibitem[{\citenamefont{Rosenberg et~al.}(2016)\citenamefont{Rosenberg, Marino,
  Herbert, and Pouquet}}]{rosenberg_16}
\bibinfo{author}{\bibfnamefont{D.}~\bibnamefont{Rosenberg}},
  \bibinfo{author}{\bibfnamefont{R.}~\bibnamefont{Marino}},
  \bibinfo{author}{\bibfnamefont{C.}~\bibnamefont{Herbert}}, \bibnamefont{and}
  \bibinfo{author}{\bibfnamefont{A.}~\bibnamefont{Pouquet}},
  \bibinfo{journal}{Eur. Phys. J. E} \textbf{\bibinfo{volume}{39}},
  \bibinfo{pages}{8} (\bibinfo{year}{2016}).

\bibitem[{\citenamefont{Rosenberg et~al.}(2017)\citenamefont{Rosenberg, Marino,
  Herbert, and Pouquet}}]{rosenberg_17}
\bibinfo{author}{\bibfnamefont{D.}~\bibnamefont{Rosenberg}},
  \bibinfo{author}{\bibfnamefont{R.}~\bibnamefont{Marino}},
  \bibinfo{author}{\bibfnamefont{C.}~\bibnamefont{Herbert}}, \bibnamefont{and}
  \bibinfo{author}{\bibfnamefont{A.}~\bibnamefont{Pouquet}},
  \bibinfo{journal}{Eur. Phys. J. E} \textbf{\bibinfo{volume}{40}}
  (\bibinfo{year}{2017}).

\bibitem[{\citenamefont{Bartello}(1995)}]{bartello_95}
\bibinfo{author}{\bibfnamefont{P.}~\bibnamefont{Bartello}},
  \bibinfo{journal}{J. Atmos. Sci.} \textbf{\bibinfo{volume}{52}},
  \bibinfo{pages}{4410} (\bibinfo{year}{1995}).

\bibitem[{\citenamefont{Kurien and Smith}(2014)}]{kurien_14}
\bibinfo{author}{\bibfnamefont{S.}~\bibnamefont{Kurien}} \bibnamefont{and}
  \bibinfo{author}{\bibfnamefont{L.~M.} \bibnamefont{Smith}},
  \bibinfo{journal}{J. of Turb.} \textbf{\bibinfo{volume}{15}},
  \bibinfo{pages}{241} (\bibinfo{year}{2014}).

\bibitem[{\citenamefont{Teitelbaum and Mininni}(2009)}]{teitelbaum_09}
\bibinfo{author}{\bibfnamefont{T.}~\bibnamefont{Teitelbaum}} \bibnamefont{and}
  \bibinfo{author}{\bibfnamefont{P.}~\bibnamefont{Mininni}},
  \bibinfo{journal}{Phys. Rev. Lett.} \textbf{\bibinfo{volume}{103}},
  \bibinfo{pages}{014501} (\bibinfo{year}{2009}).

\bibitem[{\citenamefont{Rorai et~al.}(2013)\citenamefont{Rorai, Rosenberg,
  Pouquet, and Mininni}}]{rorai_13}
\bibinfo{author}{\bibfnamefont{C.}~\bibnamefont{Rorai}},
  \bibinfo{author}{\bibfnamefont{D.}~\bibnamefont{Rosenberg}},
  \bibinfo{author}{\bibfnamefont{A.}~\bibnamefont{Pouquet}}, \bibnamefont{and}
  \bibinfo{author}{\bibfnamefont{P.}~\bibnamefont{Mininni}},
  \bibinfo{journal}{Phys. Rev. E} \textbf{\bibinfo{volume}{87}},
  \bibinfo{pages}{063007} (\bibinfo{year}{2013}).

\bibitem[{\citenamefont{Koprov et~al.}(2005)\citenamefont{Koprov, Koprov,
  Ponomarev, and Chkhetiani}}]{koprov_05}
\bibinfo{author}{\bibfnamefont{B.}~\bibnamefont{Koprov}},
  \bibinfo{author}{\bibfnamefont{V.}~\bibnamefont{Koprov}},
  \bibinfo{author}{\bibfnamefont{V.}~\bibnamefont{Ponomarev}},
  \bibnamefont{and}
  \bibinfo{author}{\bibfnamefont{O.}~\bibnamefont{Chkhetiani}},
  \bibinfo{journal}{Dokl. Phys.} \textbf{\bibinfo{volume}{50}},
  \bibinfo{pages}{419} (\bibinfo{year}{2005}).

\end{thebibliography}

\end{document}